\documentclass[prd,twocolumn,twoside,preprintnumbers,superscriptaddress,nofootinbib]{revtex4}

\usepackage{amsmath,slashed}
\usepackage{graphicx,graphics,color}
\usepackage{dcolumn}
\usepackage[hyperfootnotes=false]{hyperref}
\usepackage{xspace}
\usepackage{enumerate}
\usepackage{varwidth}
\usepackage{comment}

\usepackage{amssymb}
\usepackage{mathtools}
\usepackage{dsfont}
\usepackage{multirow}

\newcommand{\GeV}{\,\text{GeV}}
\newcommand{\MeV}{\,\text{MeV}}
\newcommand{\keV}{\,\text{keV}}
\newcommand{\fm}{\,\text{fm}}
\newcommand{\eV}{\,\text{eV}}

\newcommand{\mpi}{M_\pi}

\newcommand{\mW}{M_W}
\newcommand{\mN}{m_N}
\newcommand{\eps}{\epsilon}

\newcommand{\Order}{\mathcal{O}}

\newcommand{\diff}{\text{d}}
\newcommand{\cc}{\mathbf{c}}
\newcommand{\rr}{\mathbf{r}}
\newcommand{\qq}{\mathbf{q}}
\newcommand{\pp}{\mathbf{p}}
\newcommand{\xx}{\mathbf{x}}
\newcommand{\yy}{\mathbf{y}}
\newcommand{\zz}{\mathbf{z}}
\newcommand{\PP}{\mathbf{P}}
\newcommand{\LL}{\mathbf{L}}

\newcommand{\qg}{\mathbf{q}_\gamma}

\newcommand{\qe}{q_\text{ext}}

\newcommand{\beq}{\begin{equation}}
\newcommand{\eeq}{\end{equation}}

\newcommand{\epspi}{\epsilon_{\slashed{\pi}}}

\newcommand{\bsigma}{\boldsymbol{\sigma}}

\newcommand{\mue}{\mu_\text{ext}}
\newcommand{\mupi}{\mu_\pi}
\newcommand{\muchi}{\mu_\chi}
\newcommand{\muW}{\mu_W}

\allowdisplaybreaks[1]

\begin{document}

\preprint{INT-PUB-24-021, LA-UR-24-25160}

\title{Ab-initio electroweak corrections to superallowed $\boldsymbol{\beta}$ decays and their impact on  $\boldsymbol{V_{ud}}$}

\author{Vincenzo Cirigliano}
\affiliation{Institute for Nuclear Theory, University of Washington, Seattle WA 91195-1550, USA}
\author{Wouter Dekens}
\affiliation{Institute for Nuclear Theory, University of Washington, Seattle WA 91195-1550, USA}
\author{Jordy de Vries}
\affiliation{Institute for Theoretical Physics Amsterdam and Delta Institute for Theoretical
Physics,University of Amsterdam, Science Park 904, 1098 XH Amsterdam, The Netherlands}
\affiliation{Nikhef, Theory Group, Science Park 105, 1098 XG, Amsterdam, The Netherlands}
\author{\\Stefano Gandolfi}
\affiliation{Theoretical Division, Los Alamos National Laboratory, Los Alamos, NM 87545, USA}
\author{Martin Hoferichter}
\affiliation{Albert Einstein Center for Fundamental Physics, Institute for Theoretical Physics, University of Bern, Sidlerstrasse 5, 3012 Bern, Switzerland}
\author{Emanuele Mereghetti}
\affiliation{Theoretical Division, Los Alamos National Laboratory, Los Alamos, NM 87545, USA}

\begin{abstract}
Radiative corrections are essential for an accurate determination of $V_{ud}$ from superallowed $\beta$ decays. In view of recent progress in the single-nucleon sector, the uncertainty is dominated by the theoretical description of nucleus-dependent effects, limiting the precision that can currently  be achieved for $V_{ud}$. In this work, we provide a detailed account of the electroweak corrections to superallowed $\beta$ decays in effective field theory (EFT), including the power counting, potential and ultrasoft contributions, and factorization in the decay rate. 
We present a first numerical evaluation of the dominant corrections in light nuclei 
based on Quantum Monte Carlo methods, 
confirming the expectations from the EFT power counting. 
Finally, we  discuss strategies how to extract from data 
the low-energy constants that parameterize short-distance contributions and whose values are not predicted by the EFT. 
Combined with advances in ab-initio nuclear-structure calculations, this EFT framework allows one to systematically address the dominant uncertainty in $V_{ud}$, as illustrated in detail for the $^{14}$O $\to$ $^{14}$N transition.   
\end{abstract}

\maketitle

\section{Introduction}

Superallowed $\beta$ decays constitute the prime source of information on $V_{ud}$, the first element of the Cabibbo--Kobayashi--Maskawa (CKM) matrix~\cite{Cabibbo:1963yz,Kobayashi:1973fv}.  That is, by measuring the decay half-life $t$, the traditional master formula~\cite{Hardy:2020qwl,Gorchtein:2023naa}
\begin{equation}
    \frac{1}{t}  =    \frac{G_F^2 |V_{ud}|^2  m_e^5}{\pi^3 \log 2} (1 + \Delta^V_R)  (1 + \delta_R^\prime) (1 + \delta_\text{NS} - \delta_C)  \times f 
    \label{eq:master0}
\end{equation}
in principle allows one to extract $V_{ud}$ at high precision, provided that the various radiative corrections (RC) can be controlled at a sufficient level. In the traditional decomposition~\eqref{eq:master0}, 
 $f$ denotes a phase-space factor that includes the Fermi function, which captures the main effect of the Coulomb interaction of the outgoing electron  in the nuclear field. This 
factor 
 depends on the nuclear electroweak (EW) form factor, and involves corrections related to nuclear recoil,  atomic electron screening, and  atomic overlap~\cite{Hardy:2020qwl,Gorchtein:2023naa,Hayen:2017pwg}. Next, $\delta_C$ is defined by $M_\text{F} = \langle f | \tau^+ |i \rangle = M_\text{F}^{(0)} (1 - \delta_C/2)$, i.e., it measures the deviation of the Fermi matrix element from $M_\text{F}^{(0)} = \sqrt{2}$ as expected in the isospin limit. Further RC  are contained in 
 the so-called outer correction $\delta_R^\prime$, comprising infrared (IR)-sensitive effects not included in the Fermi function, while the remaining, inner RC are separated into a universal, single-nucleon correction $\Delta^V_R$ and nucleus-dependent RC $\delta_\text{NS}$~\cite{Gorchtein:2018fxl,Seng:2022cnq}. In this paper, we provide a detailed description of an approach to superallowed $\beta$ decays in effective field theory (EFT)~\cite{Cirigliano:2024rfk}, including the factorization assumptions inherent in Eq.~\eqref{eq:master0} and a first numerical evaluation of the dominant RC in light nuclei. 
 
 Revisiting the formalism for superallowed $\beta$ decays in this manner is highly motivated by precision tests of the Standard Model, most notably the unitarity of the first row of the CKM matrix
 \begin{align}
 \label{CKM}
 |V_{ud}|^2+|V_{us}|^2+|V_{ub}|^2=1.  
\end{align}
First, a global fit of all available constraints on $V_{ud}$ and $V_{us}$, with the $V_{ub}$ contribution being numerically irrelevant at present, suggests a deficit of  
$2.8\sigma$~\cite{Cirigliano:2022yyo}. 
Despite a separate tension in $V_{us}$ originating from determinations of $K_{\ell 3}$ and $K_{\ell 2}$ decays, requiring experimental clarification~\cite{Cirigliano:2022yyo,Anzivino:2023bhp}, $V_{ud}$
has attracted renewed interest
following a reevaluation of the universal RC associated with $\Delta_R^V$~\cite{Marciano:2005ec,Seng:2018yzq,Seng:2018qru,Czarnecki:2019mwq,Seng:2020wjq,Hayen:2020cxh,Shiells:2020fqp}, given the significant increase in the possible deficit in Eq.~\eqref{CKM}. Implications for beyond-the-Standard-Model scenarios~\cite{Belfatto:2019swo,Coutinho:2019aiy} have been investigated studying vector-like quarks~\cite{Cheung:2020vqm,Belfatto:2021jhf,Branco:2021vhs,Crivellin:2021bkd} and leptons~\cite{Crivellin:2020ebi,Kirk:2020wdk}, modifications of the Fermi constant~\cite{Marciano:1999ih,Crivellin:2021njn}, the violation of lepton flavor universality~\cite{Crivellin:2020lzu,Crivellin:2020klg,Capdevila:2020rrl,Crivellin:2021sff,Crivellin:2020oup,Marzocca:2021azj}, as well as in the context of Standard-Model EFT~\cite{Alok:2021ydy,Cirigliano:2022qdm,Cirigliano:2023nol,Dawid:2024wmp}. The significance of all these conclusions ultimately depends on the reliability of RC in superallowed $0^+\to 0^+$ transitions~\cite{Hardy:2020qwl}, 
which currently provide the most precise value for $V_{ud}$.

The experimental component of the resulting uncertainty, obtained after an average over a large number of isotopes, is currently subleading compared to the theory uncertainties from the RC, in stark contrast to alternative probes. In neutron decay, uncertainties in the experimental input still dominate the uncertainty in $V_{ud}$. These arise from the 
 lifetime $\tau_n$~\cite{UCNt:2021pcg} and especially the decay parameter $\lambda$~\cite{Markisch:2018ndu}, which currently limits the precision of the $V_{ud}$ determination from the neutron, also in view of  Ref.~\cite{Beck:2019xye}. Pion $\beta$ decay would permit an even cleaner  determination of $V_{ud}$ in a purely mesonic system~\cite{Cirigliano:2002ng,Czarnecki:2019iwz,Feng:2020zdc}, yet the experimental challenges are substantial~\cite{Pocanic:2003pf,PIONEER:2022yag}. In this situation, improvements in the RC for superallowed $\beta$ decays would have a direct impact on the unitarity test~\eqref{CKM}.
 
 For the single-nucleon RC contained in $\Delta_R^V$, recent improvements include a comprehensive EFT analysis~\cite{Cirigliano:2023fnz} and a first lattice-QCD calculation~\cite{Ma:2023kfr}, but the reliability of nucleus-dependent corrections remains a serious concern, both for $\delta_C$~\cite{Miller:2008my,Miller:2009cg,Martin:2021bud,Condren:2022dji,Seng:2022epj,Crawford:2022yhi,Seng:2023cvt,Seng:2023cgl} and $\delta_\text{NS}$~\cite{Gorchtein:2018fxl,Seng:2022cnq,Gorchtein:2023naa}, motivating the development of an EFT framework also for the nuclear corrections.  To this end, we start with a discussion of the general EFT approach, the power counting, and the relevant momentum regions in Sec.~\ref{sec:EFT}. We will show that the dominant contributions can be expressed in terms of two-body (2b) currents~\cite{Cirigliano:2024rfk}, which  are discussed in detail in Sec.~\ref{sec:potentials}. We find that, at the required precision, also contact terms need to be included. Similarly to the case of neutrinoless double-$\beta$ decay~\cite{Cirigliano:2018hja,Cirigliano:2019vdj,Cirigliano:2020dmx,Cirigliano:2021qko,Wirth:2021pij}, these come with unknown coefficients, but here their values can be determined by a simultaneous fit to different isotopes.  
 Ultrasoft contributions, generated by photons with very small momenta, are addressed in Sec.~\ref{sec:ultrasoft}, in the context of which we also make the connection to the dispersive approach of Refs.~\cite{Seng:2022cnq,Gorchtein:2023naa}. In particular, such a comparison is useful to clarify whether the expected EFT scalings hold in the presence of 
 low-lying intermediate states, such as the $3^+$ and $1^+$ levels of $^{10}$B  in the $^{10}\text{C}\to{}^{10}\text{B}$ $0^+\to 0^+$ transition~\cite{Gennari:2024sbn}.  
In Sec.~\ref{sec:decay_rate} we combine all the ingredients into a master formula for the decay rate, with particular attention to the question to what extent the factorization assumptions in  Eq.~\eqref{eq:master0} can be justified from the EFT perspective. First numerical evaluations are presented in 
Sec.~\ref{sec:light_nuclei}, to see whether the expectations from the EFT power counting are realized in practice. Based on these results, we discuss the application to $^{14}\text{O}\to{}^{14}\text{N}$ in Sec.~\ref{sec:example}, as a concrete numerical example to illustrate the formalism. Strategies for the determination of the contact terms are discussed in Sec.~\ref{sec:LECs}, before summarizing our findings and sketching future work in
Sec.~\ref{sec:conclusions}.

To keep this paper readable, we have put many technical, but crucial, discussions into several appendices. In App.~\ref{app:energy} we discuss the role of energy-dependent potentials, in particular, subtleties that arise for the zero component of the momentum transfer, while Apps.~\ref{app:coordinate} and~\ref{app:sub} provide the potentials in coordinate space as well as the required subtraction of ultrasoft contributions. Appendices~\ref{app:usoftRGE} and~\ref{App:RGE}  discuss the renormalization group (RG) evolution to low-energy scales, App.~\ref{app:toy} details about the comparison to the dispersive approach used in the literature, and App.~\ref{app:Cfactor} various corrections to the phase-space integrals that are not the focus of this work.

\begin{figure*}[t]
\includegraphics[width=\textwidth]{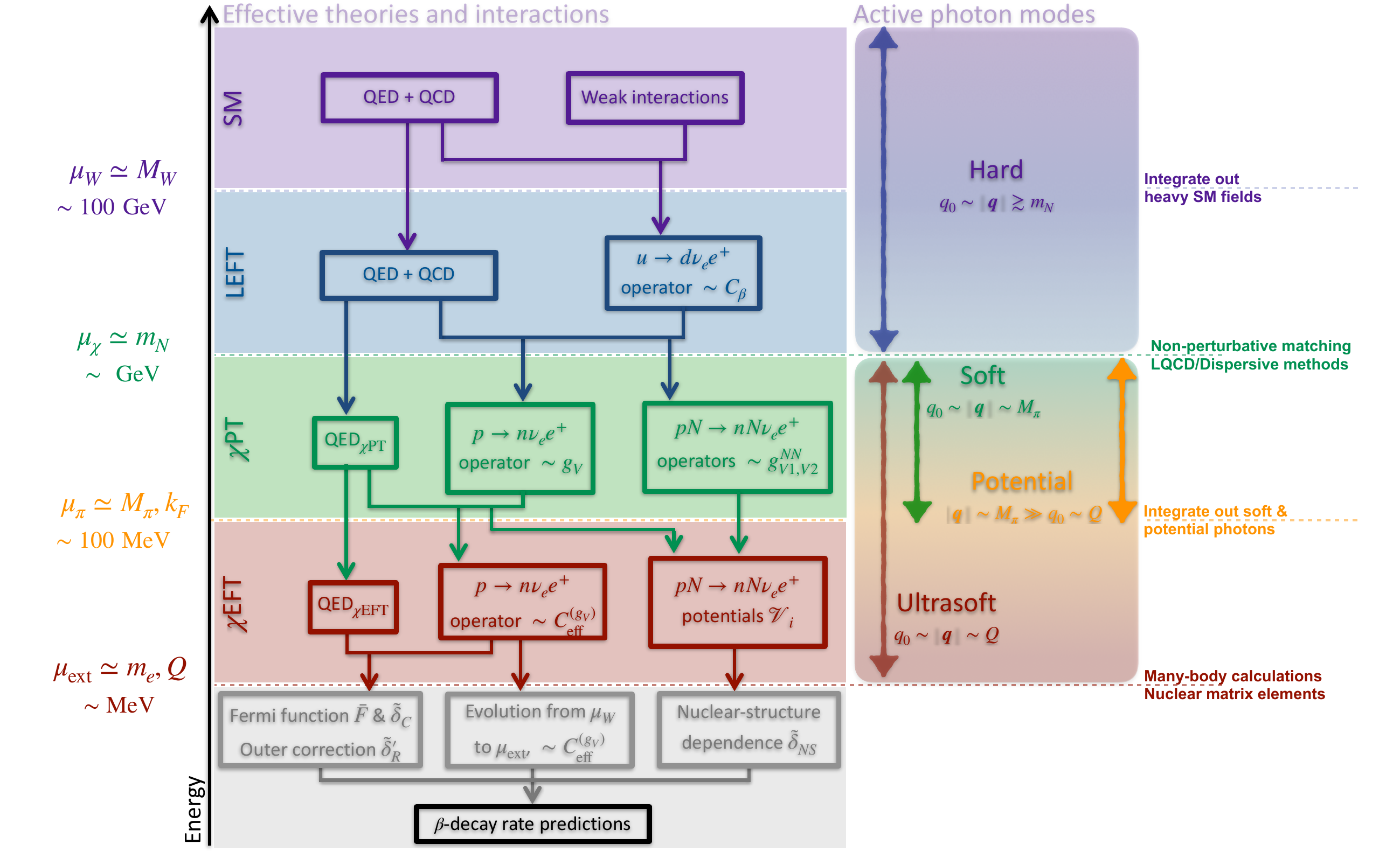}
\caption{Hierarchy of scales in the EFT. The left panel summarizes the different EFTs and their interactions, the right panel the associated photon modes. }
\label{fig:scales}
\end{figure*}

\section{Effective field theory}
\label{sec:EFT}

The main advantage of an EFT approach to the evaluation of RC 
to nuclear $\beta$ decay is that the different scales inherent to the problem can be taken into account in a systematic manner. While the overall scale is set by $G_F q^2_\text{ext}$, where $G_F$ denotes the Fermi constant~\cite{MuLan:2012sih} and  $\qe$ the low scale of the order of the $\mathcal Q_\text{EC}$ value of the reaction, the relevant energies for RC range from  
the EW scale over hadronic scales down to $\qe$. The different regimes are as follows:
\begin{enumerate}
 \item Low-energy scales: $\qe\simeq m_e\simeq E_0$, with the electron endpoint energy $E_0=\mathcal Q_\text{EC}-m_e$. 
\item Nuclear scales: $\gamma\simeq R^{-1} \simeq \mpi \simeq k_F = \mathcal O(100\MeV)$, with pion mass $\mpi$, nucleon binding momentum $\gamma$, inverse nuclear radius $R^{-1}$, and the Fermi momentum $k_F$. 
\item Chiral/hadronic scales: $\Lambda_\chi\simeq 4\pi F_\pi\simeq\mN\simeq 1 \GeV$, where $\Lambda_\chi$ denotes the cutoff scale of chiral perturbation theory ($\chi$PT) (the pion decay constant is taken in the conventions $F_\pi=92.3\MeV$), coinciding with typical hadronic scales of the order of the nucleon mass $\mN$.
\item EW scale: $\mW\simeq 100$ GeV.
\end{enumerate}
The matching scales in the EFT for the different regions are denoted by $\mue$, $\mupi$, $\muchi$, and $\muW$, respectively, see Fig.~\ref{fig:scales} for an illustration of the different scales.  
They satisfy the hierarchy 
\begin{equation}
  \qe \ll \mpi \ll  \Lambda_{\chi}  \ll \mW,
\end{equation}
leading us to define expansion parameters   
\begin{equation}
    \epsilon_\text{recoil} = \mathcal O\bigg(\frac{\qe}{\Lambda_\chi}\bigg), \quad  \epsilon_{\slashed{\pi}} = 
    \mathcal O\bigg(\frac{\qe}{\mpi}\bigg), \quad
    \epsilon_{\chi} = 
    \mathcal O\bigg(\frac{\mpi}{\Lambda_\chi}\bigg), 
 \end{equation}
in terms of which we will organize the RC, together with the respective scaling in the fine-structure constant $\alpha=e^2/(4\pi)$.  To reach the required precision at the $10^{-4}$ level, one needs the RC at ${\mathcal O}(\alpha)$, ${\mathcal O}(\alpha\epsilon_\chi)$, ${\mathcal O}(\alpha \epsilon_{\slashed{\pi}})$, but also some leading ${\mathcal O}(\alpha^2)$ contributions, including 
   leading logarithms (LL)  and next-to-leading logarithms (NLL), ${\mathcal O}(\alpha^2 L^2)$ and ${\mathcal O}(\alpha^2 L)$ with $L\simeq \log \mW/\Lambda_\chi$ or $\log \Lambda_\chi/\qe $, respectively,  
as well as  Coulomb- and $Z$-enhanced corrections.

To capture all these effects related to multiple different scales, one needs to use 
a tower of  EFTs, as done for meson decays~\cite{Descotes-Genon:2005wrq,Cirigliano:2008wn} and 
neutron decay~\cite{Cirigliano:2022hob,Cirigliano:2023fnz}. 
In this section, we provide a detailed account of the various EFTs and the power counting, starting with the contributions from hard photons.

\subsection{Hard photon contributions}

We begin by discussing the important contributions that arise from the exchange of hard photons, 
i.e., photons with  virtuality   
$\Lambda_\chi^2 \lesssim Q^2  \lesssim \mW^2$.
Between the EW scale and the hadronic scale, the relevant EFT is  the Fermi theory 
obtained by integrating out the heavy Standard-Model particles, commonly referred to as low-energy EFT (LEFT)~\cite{Jenkins:2017jig}. 
The relevant part of the LEFT effective Lagrangian reads
\begin{equation}
    {\mathcal L}_\text{LEFT}  =   - 2 \sqrt{2} G_F  V_{ud} \ C^r_\beta (\mu)  \ \bar{e}_L \gamma_\mu \nu_{L}  \,  \bar{u}_{L} \gamma^\mu d_L + \ \text{h.c.}\label{eq:LLEFT}
\end{equation}
Here, $G_F$ is implicitly understood to be defined from muon decay~\cite{MuLan:2012sih}, absorbing a set of RC,
and the   Wilson coefficient $C^r_\beta (\mu)$ encodes the RC 
due to hard photons. 
The corresponding anomalous dimension is known 
to ${\mathcal O}(\alpha)$~\cite{Sirlin:1981ie}, ${\mathcal O}(\alpha^2)$~\cite{Cirigliano:2023fnz}, and ${\mathcal O}(\alpha \alpha_s)$~\cite{Sirlin:1981ie}, and the  
RG equations allow one to evaluate 
$C^r_\beta (\mu)$ at the hadronic scale $\mu \simeq \muchi$, 
thus resumming the LL and NLL of the 
ratio $\mW/\Lambda_\chi$. 
This correction is universal for all hadronic $\beta$ decay processes.

At the hadronic scale, we next match  onto  an EFT  written in terms of  nucleons, 
pions, light leptons, and photons~\cite{Ando:2004rk,Cirigliano:2022hob}, 
according to the exact and broken symmetries of low-energy EW interactions, QED, and QCD. We give more details of this chiral EFT below, but here we already present a few key interactions 
that will be necessary. 
We focus on the effects induced by hard photons in  single-nucleon ($N$), mesonic, 
and nucleon--nucleon ($N\!N$) interactions. 

First of all, the leading-order (LO) EW one-body (1b) Lagrangian is 
\beq
   \mathcal L_W^\text{1b}   =
 - \sqrt{2} G_F V_{ud} \,   \bar e_L \gamma_\mu \nu_L
  \bar N   (g_V  v^\mu   - 2 g_A S^\mu ) \tau^+ N  + \cdots, 
\label{eq:Lagrangian_at_leading_order}
\eeq
in terms of the nucleon $N^T=(p,n)$  isodoublet,  the nucleon four-velocity $v_\mu$ and spin $S_\mu$,  and isospin Pauli matrices $\tau^a$. In the nucleon rest frame, $v^{\mu} = (1, {\bf{0}})$, and $S^\mu = (0, \bsigma/2)$, with $\sigma$
the spin Pauli matrices.
The ellipsis denotes omitted terms involving pion fields or of  higher order in  $\epsilon_\chi$. 
At this level, the effects of hard photons are captured in the deviation of the (scale-dependent) vector coupling $g_V (\mu)$ from one (and $g_A (\mu)$ from $g_A^\text{QCD}$~\cite{Cirigliano:2022hob}). 
The vector coupling  $g_V (\mu)$  can be represented as follows~\cite{Cirigliano:2023fnz}:
\begin{align}
g_V (\mu) &= \tilde U (\mu, \muchi) 
\notag \\
& \times  \left[ 1 +  \overline \Box^V_\text{had} (\mu_0) -  \frac{\alpha (\muchi)}{2 \pi}  \, 
\kappa \left( \frac{\mu}{\mu_0},\frac{\mu_0}{\muchi} \right)   \right]
\notag \\
& \times \left( 1+\frac{\alpha(\muchi)}{\pi}B(a)\right)^{-1}  U(\muchi, \muW)  \  C^r_\beta (\muW).
\label{eq:gVM}
\end{align}
From right to left, the terms appearing in the above expression represent contributions of virtual photons
of decreasing virtuality. 
$C^r_\beta (\muW)$ is the LEFT Wilson coefficient defined in Eq.~\eqref{eq:LLEFT}, 
evaluated at the weak scale $\muW \simeq \mW$.
The  function $U(\muchi, \muW)$ encodes the RG evolution 
from $\muW$ down to $\muchi$ and sums the LL and NLL
of $\mW/\Lambda_\chi$. The term involving $B(a)$ is a scheme-dependent quantity that enters the matching onto $\chi$PT \cite{Cirigliano:2023fnz}. Similarly, both $U (\muchi,\muW)$ and $C^r_\beta(\muW)$ depend on the arbitrary parameter $a$, while the product of these three factors  
is scheme independent.
The terms in  square bracket in Eq.~\eqref{eq:gVM} represent the contributions to $g_V$ from 
 matching LEFT onto chiral EFT. 
This involves a perturbative term 
\beq
\kappa \left( \frac{\mu}{\mu_0},\frac{\mu_0}{\muchi} \right)  = \frac{5}{8} + \frac{3}{4}\log \frac{\mu^2}{\mu_0^2} + \Big( 1 - \frac{\alpha_s(\mu_0^2)}{4\pi}\Big) 
\log\frac{ \mu_0^2}{\muchi^2}
\eeq
and 
a non-perturbative contribution 
$\overline{\Box}_\text{had}^V (\mu_0)$, which is a subtracted version of  the standard
$\gamma W$ box 
$\Box_{\gamma W}^V$ of Refs.~\cite{Seng:2018qru,Seng:2018yzq} 
and can be expressed in terms of the unpolarized structure function $T_3 (\nu, Q^2)$ ($\nu \equiv q^0$, $Q^2 \equiv - q^2$) as follows: 
\begin{align}
    \overline{\Box}_\text{had}^V (\mu_0)  &= \frac{e^2}{i} \int \frac{d^4 q}{\left( 2 \pi \right)^4} \frac{ \nu^2 + Q^2}{Q^4} 
    \Bigg[ \frac{{ T}_3 (\nu, Q^2)}{2 \mN \nu}
    \notag \\
    & -\frac{2}{3} \frac{1}{Q^2 + \mu^2_0}\, \left(1 - \frac{\alpha_s (\mu_0^2)}{\pi}   \right) \Bigg].\label{eq:boxbar}    
\end{align}
The leftmost factor $\tilde U(\mu, \muchi)$ in Eq.~\eqref{eq:gVM} encodes the running of 
$g_V (\mu)$ in chiral EFT, whose anomalous dimension is known  to ${\mathcal O}(\alpha^2)$~\cite{Cirigliano:2023fnz}.
Note that to NLL accuracy $g_V$ does not depend on the scales $\muW$, $\mu_0$, and $\muchi$, see Ref.~\cite{Cirigliano:2023fnz} for further details.
In this work, we will need as input for the nuclear-level EFT the value 
\beq
\label{eq:gVmu0}
g_V(\mu = M_{\pi^\pm})= 1.01494(12), 
\eeq  
where the error is dominated by the
non-perturbative contribution
$\overline{\Box}_\text{had}^V (\mu_0)$~\cite{Cirigliano:2023fnz}, which was evaluated with input from Refs.~\cite{Seng:2018qru,Seng:2018yzq,Czarnecki:2019mwq,Shiells:2020fqp,Hayen:2020cxh,Seng:2020wjq,Cirigliano:2022yyo}.
It is also interesting to give $g_V$ at the nucleon mass scale, which is related to $\Delta_R^V$ in the traditional approach,
\begin{equation}\label{eq:gVmN}
    g_V(\mu = \mN) = 1.01153(12).
\end{equation}

A matching formula similar to Eq.~\eqref{eq:gVM} 
holds for the axial effective coupling $g_A (\mu)$ in 
Eq.~\eqref{eq:Lagrangian_at_leading_order}. 
While details will be given in Ref.~\cite{gAmatching}, for the purposes of this analysis 
we note that  the short-distance ($\muW \to \muchi$) and long-distance 
($\mupi \to \mue$)  RG evolution factors are the same for $g_V$ and $g_A$, 
so that $g_A/g_V$ is scale independent  and contains non-perturbative 
information from  matching at the scale $\mu \simeq \muchi$ and $\mu \simeq \mupi$.

Next, hard photons generate contributions to the pion chiral Lagrangian
 \beq
 \label{pion_mass_splitting}
\mathcal L_{\pi} = 2 e^2 F_\pi^2 Z_\pi\,\pi^+\pi^-+\dots,
 \eeq
where $Z_\pi$ is a low-energy constant (LEC) determined from $M^2_{\pi^\pm}-M^2_{\pi^0}= 2 e^2 F_\pi^2 Z_\pi$. Diagrams involving $Z_\pi$ lead to isospin-breaking corrections to $g_A$ \cite{Cirigliano:2022hob} and, as we will see below, to RC to nuclear $\beta$ decay. 
In this work, we define the isospin limit by $\mpi=M_{\pi^0}$, including corrections from 
the pion mass splitting as generated by hard photons via the chiral Lagrangian.
For the nucleon, we did not find any relevant isospin-breaking effects, for the numerics we use $\mN= (m_n + m_p)/2 =0.939 \GeV$.

Hard photons also generate EW 2b contact operators between nucleons at $\mathcal O(G_F \alpha)$. 
The interactions with the lowest number of derivatives act in an $S$ wave.
There are two $^1S_0$ operators, with isospin $T=1,2$, and one 
spin-dependent operator connecting $^1S_0$ and $^3 S_1$ waves. Omitting terms involving pions, we can write
\begin{align}
\label{eq:2nuc1}
      \mathcal L_W^\text{2b}  & = -\sqrt{2} e^2 G_F V_{ud}  \bar e_L \gamma_\mu \nu_L  \bigg[  v^\mu 
    g^{N\!N}_{V1}   N^\dagger \tau^+ N \, N^\dagger N  
     \notag\\
     &+ v^\mu g^{N\!N}_{V2}  N^\dagger \tau^+ N \, N^\dagger \tau^3 N  
     + 2 g_{V3}^{N\!N} N^\dagger \tau^+ N \, N^\dagger S^\mu N  \bigg] \notag\\
    &+ \cdots
\end{align}
Naive dimensional analysis would indicate that $g^{N\!N}_{V1,V2,V3} = \mathcal O (\Lambda_\chi^{-3})$, but as we will discuss in more detail below, the RG equations require the 
two $^1S_0$ 
LECs to scale as
\begin{equation}\label{gVLEC}
g^{N\!N}_{V1,V2} = \mathcal O \left(\frac{1}{\Lambda_\chi F_\pi^2} \right).
\end{equation} 
The values of these LECs are not known at present, but could be determined in a global analysis of superallowed $\beta$ decays together with $V_{ud}$, see Sec.~\ref{sec:LECs}.   
Finally, hard photons also lead to isospin-breaking corrections to $N\!N$ strong interactions~\cite{vanKolck:1997fu,Walzl:2000cx}, which play a role  
in the evaluation of $\delta_C$.

\begin{figure*}
\includegraphics[width=\textwidth]{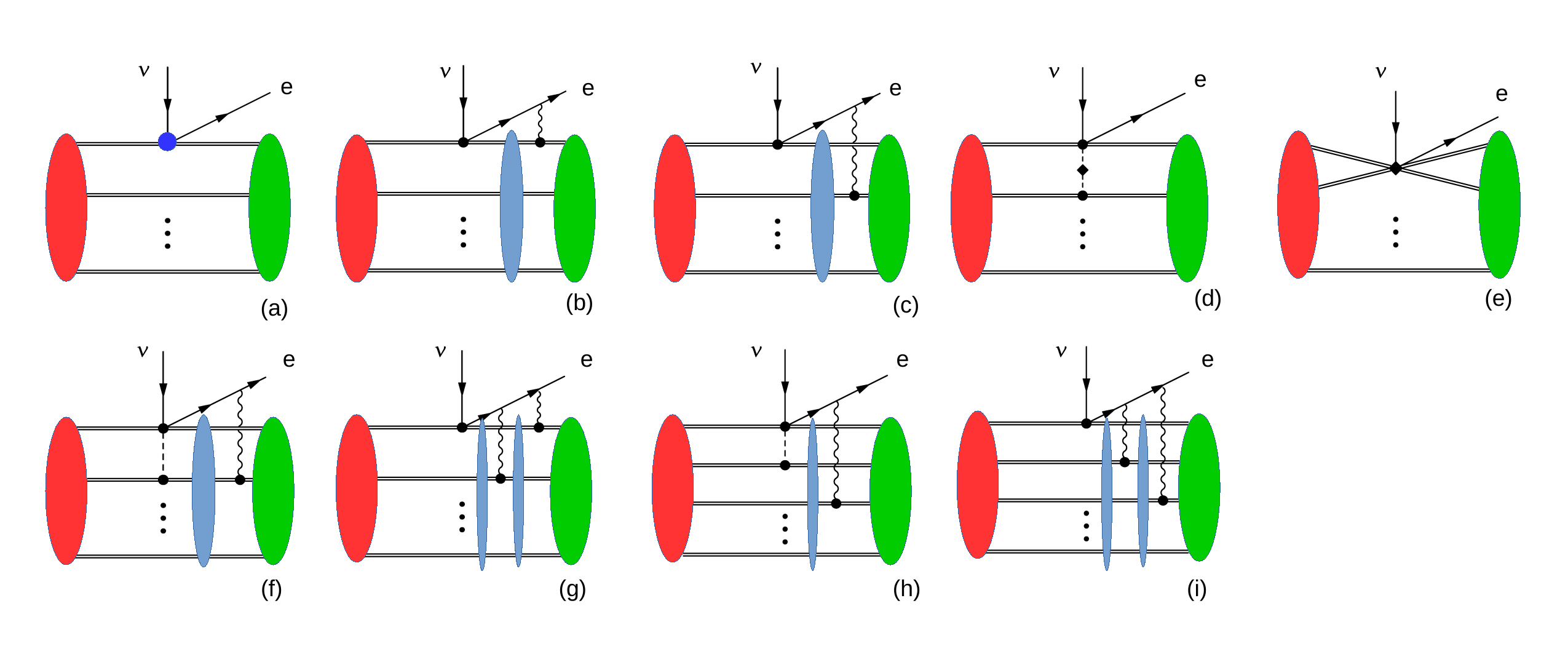}
\caption{Representative diagrams for RC to superallowed decays in EFT. 
Leptons, nucleons, photons, and pions are denoted by plain, double, wavy, and dashed lines, respectively. A blue circle denotes the insertion of the EW current, including $\mathcal O(\alpha)$ corrections from hard photon exchange, see Eq.~\eqref{eq:Lagrangian_at_leading_order}. Black circles denotes 1b EW and EM currents and pion--nucleon vertices from the chiral Lagrangian.  The red and green ovals denote the wave functions of the initial and final nuclei, the blue oval the nuclear Green's function.
    }
\label{fig:diagrams}
\end{figure*}

\subsection{Power counting in the hadronic EFT}
\label{sec:PC}

Having integrated out hard photons, we can now investigate various RC in chiral EFT with dynamical photons and leptons. 
Before doing any actual calculations we would like to identify the diagrams that give the most important contributions by formulating a power counting (PC). This is somewhat complicated by the fact that we encounter diagrams involving loops with virtual pions, nucleons, and photons. In the presence of more than one nucleon, we can identify three regions for the loop momentum $q$:
\begin{enumerate}
    \item soft: $q^0 \simeq |\qq| \simeq \mpi$,
    \item\label{potential} potential: $q^0 \simeq \qq^2/\mN \simeq \qe$, $ |\qq| \simeq \mpi$. 
        \item ultrasoft: $q^0 \simeq |\qq| \simeq \qe\simeq \mpi^2/\mN$.
\end{enumerate}

The most common loops in chiral EFT involve virtual pions corresponding to a  soft scaling for which one has to track powers of $Q\simeq \mpi \simeq \gamma \simeq k_F$. Diagrams with soft loops can be estimated by the following PC rules
\begin{itemize}
    \item Soft: each loop integration picks up a factor $Q^4/(4\pi)^2$. Each pion or photon propagator scales as $1/Q^2$. Each heavy-baryon nucleon propagator or electron propagator scales as $1/Q$. 
\end{itemize}

Diagrams with two nucleons in the intermediate state become sensitive to a different momentum scaling. In such diagrams the contour integration over the zeroth component of the loop integral cannot be performed in a way to avoid all nucleon poles so that $q^0\simeq Q^2/\mN$. The nucleon propagators then scale as $\mN/Q^2 \simeq 1/\qe$. In addition, these loops also pick up an enhancement of $4\pi$~\cite{Kaplan:1998tg, vanKolck:2020plz}. These potential diagrams can be counted with the PC rules
\begin{itemize}
    \item Potential: each loop integration picks up a factor $Q^5/(4\pi \mN)$. Each pion or photon propagator scales as $1/Q^2$. Electron propagators scale as $1/Q$, but nucleon propagators are associated with a factor $\mN/Q^2$. 
\end{itemize}
As an example, let us consider an insertion of a LO pion exchange in a diagram. It gives rise to an additional loop $Q^5/(4\pi \mN)$, two extra nucleon propagators $\simeq (\mN/Q^2)^2$, one extra pion propagator $\simeq 1/Q^2$, and two LO pion--nucleon vertices $\simeq (g_A Q/F_\pi)^2$. Altogether, this amounts to $ g_A^2 Q \mN/(4\pi F_\pi^2)$ and after identifying $g_A \simeq 1$, $Q\simeq F_\pi$, and $\mN \simeq \Lambda_\chi \simeq 4\pi F_\pi$ we obtain $g_A^2 Q \mN/(4\pi F_\pi^2) \simeq \mathcal O(1)$. This counting implies that insertions of the LO strong $N\!N$ potential are not suppressed and must be resummed leading to nuclear bound states and intermediate excited states. These iterations lead to the red, green, and blue ovals in Fig.~\ref{fig:diagrams}.

Finally, we have diagrams in which the only external scales involved are of $\mathcal O(\qe)$ (such loops do not involve virtual pions to  ${\mathcal O}(\epsilon_{\slashed{\pi}}^1)$). These ultrasoft loops scale similarly to soft loops upon replacing $Q \rightarrow \qe$:
\begin{itemize}
    \item Ultrasoft: each loop integration picks up a factor $\qe^4/(4\pi)^2$. Each photon propagator scales as $1/\qe^2$. Each heavy-baryon nucleon propagator or electron propagator scales as $1/\qe$. 
\end{itemize}

Let us now apply these PC rules to the diagrams in Fig.~\ref{fig:diagrams} starting with diagram \ref{fig:diagrams}(a). This diagram involves at LO just the single nucleon $\beta$-decay vertex proportional to $G_F$. In addition, there appear $A+1$ intermediate nucleon propagators and $A-1$ loop integrations but these are common to all diagrams and can be omitted when estimating their relative importance. We thus estimate
\begin{equation}
 \mathcal A_a \simeq \mathcal O(G_F).
\end{equation}
Diagram~\ref{fig:diagrams}(b) involves (apart from the blue oval which counts as $\mathcal O(1)$, see above) one ultrasoft loop because the loop momenta can always be routed in such a way that the electron, the photon, and one nucleon propagator only become sensitive to the external scale $\qe$. With respect to $\mathcal A_a$, this diagram then picks up one ultrasoft loop $\qe^4/(4\pi)^2$, two insertions of the charge $\simeq e^2$, and the combinations of one ultrasoft electron, one photon, and on nucleon propagator that become $1/\qe^4$. Altogether we obtain 
\begin{equation}
 \mathcal A_b \simeq  \mathcal O\left(G_F\,\frac{\alpha}{\pi}\right).
\end{equation}
However, explicit calculation shows that part of the diagram is actually enhanced by a factor $\pi^2$ leading to ${\mathcal O}\left(G_F\,\alpha\,\pi\right)$ contributions. These $\pi^2$-enhanced terms are usually collected in the Fermi function~\cite{Fermi:1934sk}, while the terms following the PC estimates are collected in the Sirlin function~\cite{Sirlin:1967zza}, see Sec.~\ref{sec:decay_rate} for the matching to the traditional notation.

We emphasize that trying to account for numerical factors in the PC is only possible in case there are universal features of certain topologies, e.g., the factors of $4\pi$ that can be associated with $N\!N$ loops~\cite{Kaplan:1998tg, vanKolck:2020plz}, but, in general, the PC cannot be expected to capture numerical enhancements of dimensionless integrals.\footnote{In some cases, e.g., triangle diagrams for isospin-breaking corrections to pion--nucleon scattering~\cite{Gasser:2002am,Hoferichter:2009ez,Hoferichter:2009gn}, $\pi$ enhancements that one might be able to guess from the topology of the diagram can be further accompanied by large numerical prefactors, which can only be found by an explicit calculation.} Another example for the intricacies of such $\pi$-enhanced contributions concerns the multiple scattering series in pion--deuteron and $N\!N$ scattering~\cite{Beane:2002wk,Liebig:2010ki,Baru:2010xn,Baru:2011bw,Baru:2012iv}, for which Coulombic pion propagators produce $\pi^2$-enhanced contribution that do not correspond to a special momentum scaling. For that reason, we only consider the universal $4\pi$ factors mentioned above, while other enhanced contributions, such as the numerical enhancement in the Fermi function, require explicit calculations.

Next, in diagram~\ref{fig:diagrams}(c) the additional loop can be either ultrasoft or potential. Let us first consider the ultrasoft scaling, in which case the extra loop gives $\qe^4/(4\pi)^2$, the vertices again $e^2$, the electron and photon propagator are both ultrasoft and give rise to $1/\qe^3$. The extra nucleon propagator, however, has potential scaling and picks up $\mN/Q^2$. This implies 
\begin{equation}
 \mathcal A^{\text{us}}_c \simeq \mathcal O\left(G_F\, \frac{\alpha}{\pi}  \frac {\qe \mN}{Q^2}\right) = \mathcal O\left(G_F\,\frac{\alpha}{\pi}\right),
\end{equation}
where we again identified $\qe =Q^2/\mN$. Accordingly, the ultrasoft part of diagram~\ref{fig:diagrams}(c) thus appears 
at the same order as diagram~\ref{fig:diagrams}(b), and we will show that the sum of these diagrams amounts to the Fermi and Sirlin functions. Assuming potential scaling instead, the extra loop in 
diagram~\ref{fig:diagrams}(c)  gives $Q^5/(4\pi \mN)$, the vertices $e^2$, the photon and electron propagator combined $1/Q^3$, and the nucleon propagator $\mN/Q^2$. This would combine to a PC scaling $\mathcal O(G_F \alpha)$ as well, but it turns out that the actual diagram vanishes unless one loop momentum picks up an external scale, which costs a power $\qe/Q = \epspi$. This implies the non-vanishing part of the diagram becomes
\begin{equation}
 \mathcal A^{\text{pot}, \epspi}_c \simeq  \mathcal O\left(G_F\,\alpha\,\epspi\right).
\end{equation}
Instead of using an external scale, we can also use a next-to-leading (NLO) EM vertex, which brings in a power $Q/\Lambda_\chi = \epsilon_\chi$. In that case the diagram scales as 
\begin{equation}
 \mathcal A^{\text{pot}, \epsilon_\chi}_c \simeq  \mathcal O\left(G_F\,\alpha\,\epsilon_\chi\right).
\end{equation}
Both assigned scalings of $\mathcal A^{\text{pot}}_c$ contribute at the order in which we are interested  and we will compute the effects of these diagrams explicitly below.  
We can finally consider the contribution to diagram~\ref{fig:diagrams}(c) from 
ultrasoft photons coupling to NLO EM and weak vertices, e.g., to the nucleon magnetic moment. 
These NLO vertices scale as $\qe/\mN\simeq \qe/\Lambda_\chi$ so that
\begin{equation}
 \mathcal A^{\text{us}, \text{NLO}}_c \simeq \mathcal O\left(G_F\, \frac{\alpha}{\pi}  \frac {\qe }{\Lambda_\chi}\right) = \mathcal O\left(G_F\,\frac{\alpha}{\pi} \epsilon_\text{recoil}\right),
\end{equation}
which are thus beyond the order of this calculation.

Turning to diagram~\ref{fig:diagrams}(d), we encounter one additional potential loop. There is no ultrasoft contribution due to the pion propagator. The diagram involves an EM contribution to the pion-mass splitting that scales as $\Delta \mpi^2 = \mathcal O(e^2 \Lambda_\chi^2/(4\pi)^2)$. In addition, we have one potential loop $\simeq Q^5/(4\pi \mN)$, one nucleon propagator $\mN/Q^2$, the pion propagator $\simeq \Delta \mpi^2/Q^4$, and the combination of weak and strong pion vertices $G_F Q/F_\pi^2$. As in diagram~\ref{fig:diagrams}(c), the contribution vanishes unless we consider one external momenta, or a subleading vertex from the chiral Lagrangian. Together, we then obtain 
\begin{equation}
 \mathcal A_d \simeq  \mathcal O\left(G_F\,\alpha\,\epspi,\, G_F\,\alpha\,\epsilon_\chi\right),
\end{equation}
and thus the same scaling as $\mathcal A^{\text{pot}}_c$. Moreover, the parts of diagrams~\ref{fig:diagrams}(c) that scale as $\mathcal O\left( G_F\,\alpha\,\epsilon_\chi\right)$ lead to divergences that must be absorbed by diagram~\ref{fig:diagrams}(e). With the scaling of the LECs in Eq.~\eqref{gVLEC}, one obtains  
\begin{equation}
 \mathcal A_e \simeq  \mathcal O\left(G_F\,\alpha\,\epsilon_\chi\right),
\end{equation}
of exactly the right size to be able to absorb the divergence. 

We now turn to the diagrams on the second line of Fig.~\ref{fig:diagrams}. Diagram~\ref{fig:diagrams}(f)  involves two additional loops. The first loop is a potential loop, but in the second loop the nucleon pole can always be avoided and thus this loop acquires soft scaling. Putting all factors together we obtain 
\begin{equation}
 \mathcal A_f \simeq  \mathcal O\left(G_F\,\alpha\,\epsilon^2_\chi\right),
\end{equation}
beyond the accuracy we consider. 

Diagram~\ref{fig:diagrams}(g) involves two photon exchanges. If both loops have potential scaling we obtain corrections that scale as 
\begin{equation}
 \mathcal A_g \simeq \mathcal O(G_F\,\alpha^2).
 \end{equation}
 Since, numerically, $\alpha \simeq \epspi$, we have to consider such corrections. In the ultrasoft limit one would naively obtain an additional suppression by $(4\pi)^2$, but again we find enhanced terms that will contribute to the Fermi function. In fact, only the combination of potential and ultrasoft contributions will lead to regulator-independent results. 

Diagrams~\ref{fig:diagrams}(h) and \ref{fig:diagrams}(i) involve three-body (3b) corrections. Assuming two potential loops in diagram (h) leads to an assigned scaling $\mathcal A_h \simeq \mathcal O(G_F\,\alpha \eps_\chi)$ and $\mathcal A_i \simeq \mathcal O(G_F\,\alpha^2)$ and thus potentially relevant. We will see that, similarly to $\mathcal A_g $, $\mathcal A_i$ is connected to the Fermi function. While $\mathcal A_h$ seems potentially relevant as well, it must be emphasized that the PC for three-nucleon processes is not very well tested. Our PC follows Ref.~\cite{Friar:1996zw}, but using the rules of Refs.~\cite{Weinberg:1990rz, Weinberg:1991um} would demote $\mathcal A_h \simeq \mathcal O(G_F\,\alpha \eps_\chi^2)$, and the latter scaling was borne out explicitly in calculations of 3b corrections to nuclear electric dipole moments~\cite{deVries:2012ab,Bsaisou:2014zwa}. For this reason, we will not explicitly compute the 3b corrections in this work, but stress that it would be interesting and important to verify their sizes. 

Finally, we also remark on PC estimates for $\delta_C$. Generalizing the theorems from Refs.~\cite{Behrends:1960nf,Ademollo:1964sr,Brown:1969lsx}, it was shown in Ref.~\cite{Miller:2008my} that there are no first-order corrections, and therefore $\delta_C$ scales with $\mathcal O(\alpha^2)$. A diagram with two Coulomb photon exchanges would have two potential loops, four nucleon propagators, and $e^4/Q^4$ from the photons combining to $\mathcal O(G_F \alpha^2 \mN^2/k_F^2)$, and thus be sizable. NLO correction in which the Coulomb exchange $e^2/Q^2$ is replaced by $e^2/\Lambda_\chi^2$~\cite{{vanKolck:1997fu,Walzl:2000cx}} would then appear at $\mathcal O(G_F \alpha^2)$ and could thus still be relevant. Ultimately, 
the counting of such corrections to Coulomb photon exchanges depends on the way in which $\delta_C$ is evaluated in practice, in particular, which corrections are included in the employed nuclear wave functions. 

We conclude our discussion of the PC with a summary of the main observations, see Ref.~\cite{Cirigliano:2024rfk}: 
\begin{enumerate}
    \item\label{ultrasoft_modes} Ultrasoft modes in diagrams~\ref{fig:diagrams}(b,c)
    contribute to the Fermi and Sirlin functions, while corrections beyond these functions are suppressed by $\mathcal O(\alpha \epsilon_\text{recoil})$, and therefore do not have to be considered. 
\item Potential modes in diagram~\ref{fig:diagrams}(c)
scale like  $\mathcal O(\alpha \epsilon_{\slashed{\pi}})$
and $\mathcal O(\alpha \epsilon_{\chi})$ relative to LO, and therefore need to be included.
\item  Soft modes first contribute suppressed by $\mathcal O(\alpha \epsilon^2_\chi )$, and thus will not be considered. 
\item  Hard modes generate several relevant contributions: (i) $\mathcal O(\alpha)$ corrections to $g_V$; (ii) 
$\mathcal O(\alpha \epsilon_\chi )$ 
two-nucleon contact terms $g_{V1,V2}^{N\!N}$ needed to absorb divergences induced by potential modes; (iii)
$\mathcal O(\alpha \epsilon_{\slashed{\pi}}, \alpha \epsilon_\chi)$ effects via the pion mass splitting.
\item\label{alpha2_corrections} There are sizable two-photon-exchange diagrams that scale as ${\mathcal O}(\alpha^2)$ compared to the LO contribution, and thus have to be considered. Potential, soft, and ultrasoft scalings are relevant for these contributions. 
\end{enumerate}
Accordingly, the dominant contributions to be combined into 
$\delta_\text{NS}$ can be evaluated
as the matrix element of  EW potentials between the initial and final nuclear states, and these effects will be described in detail in Sec.~\ref{sec:potentials}, while the role of ultrasoft contributions will be discussed further in Sec.~\ref{sec:ultrasoft}. ${\mathcal O}(\alpha^2)$ corrections are particularly important to justify the factorization of the decay rate, see Sec.~\ref{sec:decay_rate}. Moreover, the interplay of potential and ultrasoft modes becomes crucial to obtain regulator-independent results.

\subsection{Nuclear \texorpdfstring{$\boldsymbol{\beta}$}{\textbeta} decay  in EFT}\label{sec:decay}

In chiral EFT with dynamical photons and leptons, the starting point for the calculations of nuclear decay amplitudes 
is the  Hamiltonian obtained after integrating out pions and photons with momenta that have soft and potential scaling, only ultrasoft photons are left as dynamical degrees of freedom. 
The Hamiltonian takes the schematic form
\begin{align}
H_\text{eff} &=   H_\text{nucl}  \ + 
H_\text{EM}  + H_\text{EW}.
\end{align}
$H_\text{nucl}$ contains the nucleon kinetic terms and  the strong interaction potentials, up to a given chiral order. 
$H_\text{EM}$ contains EM interactions
\begin{align}\label{eq:Hem}
H_\text{EM} &=    H_\text{QED}  +   \sum_{i=1}^A     e A^\mu v_\mu \left( \frac{1 +  \tau^{(i)3}}{2} \right)  
   +   \dots,
\end{align}
where
$H_\text{QED}$ is the QED Hamiltonian describing interactions of electrons and photons. The last 
term is the LO 
nucleon coupling to ultrasoft photons, 
with the ellipsis representing suppressed terms such as 
magnetic moment and other recoil terms
(see, for example, Ref.~\cite{Bernard:1995dp}). 
The EW Hamiltonian is given by
\begin{align}
H_\text{EW} &= \sqrt{2} G_F V_{ud} \  \bar{e}_L \gamma_\mu \nu_L        \ {\mathcal J}_W^\mu,\notag   
\\
{\mathcal J}_W^\mu  
&= 
 \sum_{n=1}^A   \Big(   g_V \delta^{\mu 0} - g_A  \delta^{\mu i} \sigma^{(n)i}  \Big)  \tau^{(n)+}  
+    \left({\mathcal J}^{\text{2b}} \right)^\mu  +  \dots
\notag \\
&+
\delta^{\mu 0} \left( \mathcal V^0 + E_0  \mathcal V^0_E \right)
+ \delta^{\mu i} \mathcal V_i   
 + {p}_e^\mu \mathcal V_{m_e} + \dots 
\label{eq:Hpotential}
\end{align}
The first  two contributions to the weak nuclear current ${\mathcal J}_W^\mu$ represent the standard 1b and 2b terms, 
while the ellipsis refers to higher-order terms such as weak magnetism.
The remaining contributions, in the last line in Eq.~\eqref{eq:Hpotential}, represent the weak 2b 
currents  of order $\mathcal O(\alpha \epsilon_{\slashed{\pi}},  \alpha \epsilon_\chi)$, 
also called weak potentials in what follows. These are
induced by integrating out hard, soft, and potential photons, 
while the ellipsis denotes terms further suppressed in  $\epsilon_{\chi}$ and $\epsilon_{\slashed{\pi}}$.
Accordingly, 
the weak nuclear current ${\mathcal J}_W^\mu$ takes the general form 
\beq
{\mathcal J}_W^\mu = \sum_i   \ C_W^{(i)} (\mu) \, \big({\mathcal J}_W^{(i)} \big)^\mu
\label{eq:Hpotential2}
\eeq
in terms of scale-dependent effective couplings $C_W^{(i)} (\mu) = \{g_V (\mu), g_A (\mu), 
g^{N\!N}_{V1,V2} (\mu), \ldots\}$ that include 
EM effects due to hard, soft, and potential photons multiplying  one- and 
few-nucleon operators $\big({\mathcal J}_W^{(i)} \big)^\mu$. 
The matrix elements of these 
operators, dressed by ultrasoft photon exchanges according to Eq.~\eqref{eq:Hem},  
evaluated between initial and final nuclear states, 
eventually determine the RC to nuclear $\beta$ decays, see Sec.~\ref{sec:decay_rate}. 
In the next two sections, we describe the derivation of the operators $\big({\mathcal J}_W^{(i)} \big)^\mu$, 
their matrix elements, 
and their anomalous dimensions controlling the evolution of $C_W^{(i)} (\mu)$ for $\qe < \mu < k_F$.

\section{Potential contributions}
\label{sec:potentials}

\subsection{Electroweak potentials at \texorpdfstring{$\boldsymbol{\mathcal O(\alpha \epsilon_\chi)}$}{}
and \texorpdfstring{$\boldsymbol{\mathcal O(\alpha \epsilon_{\slashed{\pi}})}$}{}
}\label{sec:potentialsLO}

\begin{figure*}
    \centering
    \includegraphics[width=0.85\textwidth]{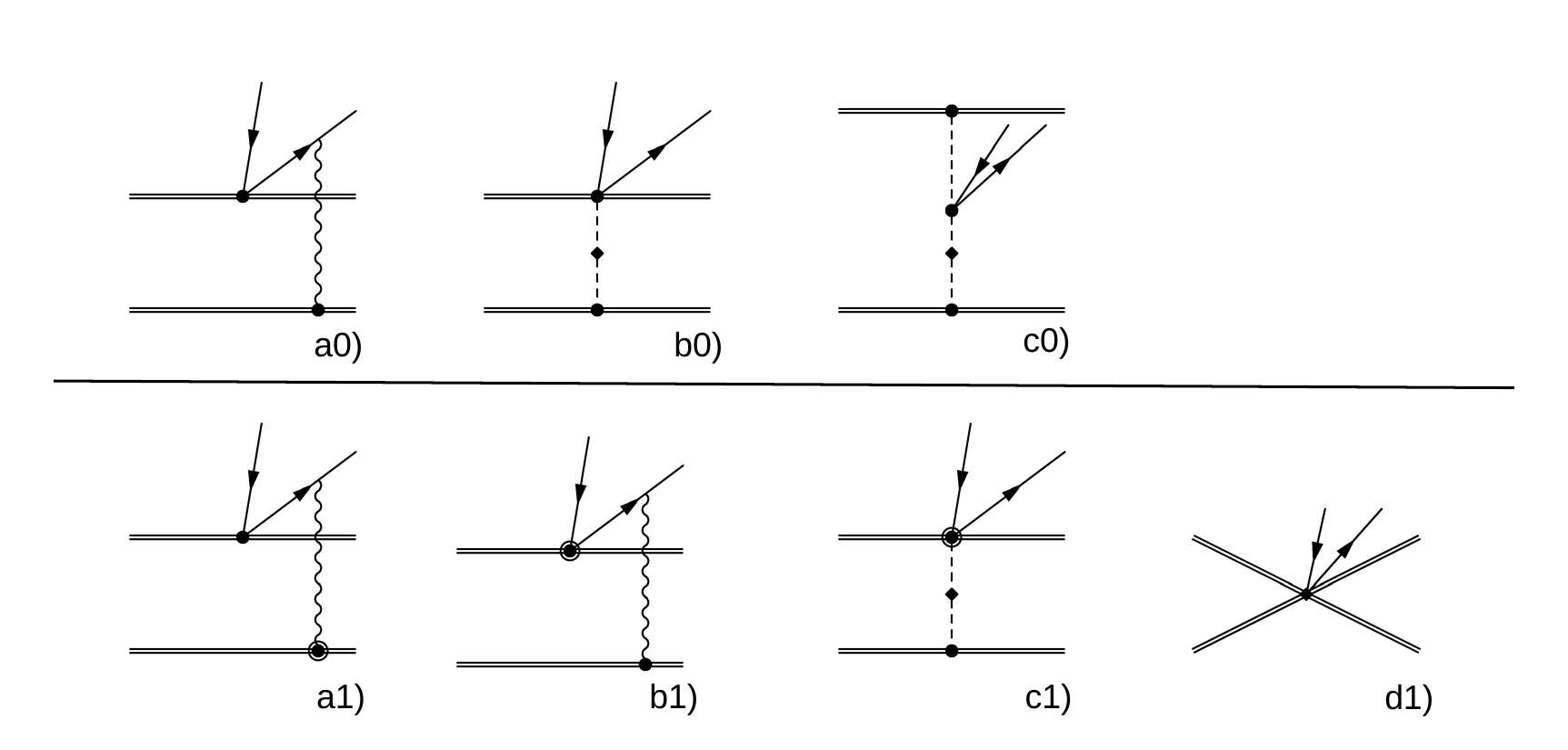}
    \caption{Lowest-order diagrams contributing to the EW potentials $\mathcal V^0_E$, $\mathcal V_{m_e}$, and $\mathcal V^{0}$. Single, double, and dashed lines denote leptons, nucleons, and pions, respectively. Dots and circled dots refer to interactions from the LO and NLO chiral Lagrangians, diamonds to isospin-breaking interactions. }
    \label{fig:potential}
\end{figure*}

Topologies such as diagram $(c)$ in Fig.~\ref{fig:diagrams}
receive contributions from the potential region, for which the  
transition operator reduces to a 2b current, or EW potential.
At lowest order, the potentials are calculated from tree-level diagrams, with the above assumptions on the scaling of the photon momentum.
The tree-level diagrams built
from the LO and NLO chiral Lagrangian are shown in the first and second row of Fig.~\ref{fig:potential}, respectively, with diagrams $a0)$--$c0)$ formally giving the leading contribution at $\mathcal O(\alpha)$. Considering diagram $a0)$ first, 
in the potential region the photon three-momentum $\qg$ is much larger than the external momenta, and we can thus expand the diagram in
powers of $|\pp_e|/|\qg|$.
Since the one-body LO vector and axial currents are momentum independent, because of the structure of the 
lepton propagator, the diagram is odd in photon three-momentum $\qg$, and thus its matrix element vanishes in $0^+$ states. 
Similarly, diagrams $b0)$ and $c0)$ are odd in $\qg$ and
we therefore find no correction at $\mathcal O(\alpha)$.

The first non-vanishing contribution 
to the EW potentials in Eq.~\eqref{eq:Hpotential} 
from the diagrams in the first row of Fig.~\ref{fig:potential}
is proportional to the electron or neutrino momenta
$p_e$ and $p_\nu$, and thus gives rise to corrections
scaling as $\mathcal O(\alpha E_e/\mpi ) = \mathcal O(\alpha \epsilon_\slashed{\pi} )$.
We find
\begin{align}\label{eq:FermiPot}
    \mathcal V^0_E &= \frac{1}{3}\left( \frac{1}{2} +  \frac{4 E_e}{E_0}\right) \mathcal V_E + \mathcal V^{\pi}_{E}, \notag\\
    \mathcal V_{m_e}  &= \frac{1}{2} \mathcal V_E + \mathcal V^{\pi}_{m_e},
\end{align}
see App.~\ref{app:energy} for details.
$\mathcal V_E$ is induced by photon exchange, with the result 
\begin{align}
    \mathcal V_E(\qq)  &=   g_V \, \sum_{j < k} e^2  \frac{1}{\qq^{4}}  \left( \tau^{+ (j)}  P^{(k)}_p  + P^{(j)}_p \tau^{+ (k)} \right). 
\end{align}    
Next, $\mathcal V_E^\pi$ is proportional to the pion mass splitting, and it has a more complicated structure
\begin{widetext}
\begin{align}\label{VEpi}
\mathcal V_{E}^\pi(\qq) & =    \frac{g_A^2 Z_\pi e^2}{3}\sum_{j<k}(\tau^{+(j)} \tau^{(k)}_3 + \tau^{(j)}_3 \tau^{+(k)})   \frac{1}{\left[\qq^{2} + \mpi^2\right]^2}
\bigg\{\bsigma^{(j)} \cdot  \bsigma^{(k)} \left(1-\frac{1}{3}\frac{\qq^{2}}{\qq^{2} + \mpi^2}
-\frac{2}{3}\frac{\qq^4}{(\qq^2+\mpi^2)^2} \right) \notag \\ 
& + \frac{2}{3}S^{(jk)} \left( \frac{1}{2}\frac{\qq^{2}}{\qq^{2} + \mpi^2} + 
\frac{\qq^4}{(\qq^2+\mpi^2)^2}\right) \bigg\},
\nonumber\\
\mathcal V_{m_e}^\pi(\qq) & =    -\frac{g_A^2 Z_\pi e^2}{3}\sum_{j<k}(\tau^{+(j)} \tau^{(k)}_3 + \tau^{(j)}_3 \tau^{+(k)})   \frac{1}{\left[\qq^{2} + \mpi^2\right]^2}
\bigg\{\bsigma^{(j)} \cdot  \bsigma^{(k)} \left(1-\frac{4}{3}\frac{\qq^{2}}{\qq^{2} + \mpi^2}
-\frac{2}{3}\frac{\qq^4}{(\qq^2+\mpi^2)^2} \right) \notag \\ 
& + \frac{2}{3}S^{(jk)} \left( 2\frac{\qq^{2}}{\qq^{2} + \mpi^2} + 
\frac{\qq^4}{(\qq^2+\mpi^2)^2}\right) \bigg\},
\end{align}
with 
\beq
P_{p,n}^{(j)}  = \frac{ \mathds{1}^{(j)}\pm \tau_3^{(j)}}{2},\qquad 
S^{(jk)} = \bsigma^{(j)} \cdot \bsigma^{(k)}  - \frac{3 \qq \cdot \bsigma^{(j)} \, \qq  \cdot \bsigma^{(k)}}{\qq^2}.
\eeq
\end{widetext}
The initial and final momenta of nucleon $j$ are labeled by
 $\pp_j$ and $\pp^\prime_j$, respectively, with $\qq = \pp_{j} - \pp^\prime_{j}
 = -(\pp_{k} - \pp^\prime_{k})
 $ and $\PP_j=\pp_j+\pp'_j$.
In momentum space 
these potentials scale as  $\mathcal O(e^2 \qe/k_F^4)$ and 
contribute to $\delta_\text{NS}$ to  $\mathcal O(\alpha \epsilon_{\slashed{\pi}})$ 
(recall that the LO diagram $(a)$ in Fig.~\ref{fig:diagrams} when evaluated between two nucleons scales as $\mathcal O(1/k_F^3)$). 
The EW potentials in coordinate space are given in App.~\ref{app:coordinate}.
The potentials induced by the pion mass splitting are given in agreement with our conventions for the isospin limit, defined by the mass of the neutral pion. If the isospin-symmetric calculation is performed for a different choice of the pion mass, all potentials depending on $Z_\pi$ need to be adapted accordingly.

The momentum dependence of the photon--nucleon interactions in the NLO chiral Lagrangian, given, for example, in Ref.~\cite{Bernard:1995dp},
allows one to build potentials that are independent of the lepton energy and momentum.
Focusing on spin/isospin structures that give non-vanishing contributions to $0^+ \rightarrow 0^+$ superallowed $\beta$ decays, 
we can write
\begin{equation}
    \mathcal V^0 = \mathcal V^\text{mag}_0 + \mathcal V^\text{rec}_0 + \mathcal V^\text{CT}_0.
\end{equation}
The magnetic potential is induced by diagram $a1)$,
while the recoil potential receives contributions from both photon exchange and the pion mass splitting, see diagrams~\ref{fig:potential} $a1)$, $b1)$, and $c1)$. We find
\begin{widetext}
\begin{align}\label{eq:MagPotential}
\mathcal V^\text{mag}_0(\qq) &= \sum_{j < k}  \frac{e^2}{3} \frac{g_A}{\mN}\frac{1}{\qq^{2}}  
\left( \bsigma^{(j)} \cdot \bsigma^{(k)} + \frac{1}{2} S^{(jk)}\right)  \Big[ (1+ \kappa_p)
\tau^{+ (j)}  P^{(k)}_p +  \kappa_n  \tau^{+ (j)}  P^{(k)}_n + (j\leftrightarrow k)\Big], \\
\mathcal V^\text{rec}_0(\qq, {\bf P}) &= \sum_{j < k}\bigg[  - i \frac{e^2 g_A}{4\mN} \frac{\tau^{+ (j)} P_p^{(k)}}{\qq^{4}}  
 (   (\PP_j - \PP_k) \times \qq  ) \cdot \bsigma^{(j)}
 -\frac{Z_\pi  e^2 g_A^2}{\mN}\frac{\tau^{+(j)} \tau^{(k)}_3}{(\qq^2+\mpi^2)^2}   \bsigma^{(j)} \cdot \qq \,\bsigma^{(k)} \cdot \PP_j
+ (j\leftrightarrow k)\bigg], 
\label{eq:RecPotential} 
\end{align}
\end{widetext}
where $\kappa_{p}=1.79$, $\kappa_n=-1.91$ are the proton and neutron anomalous magnetic moments.  The coordinate-space expression of Eqs.~\eqref{eq:MagPotential}
and~\eqref{eq:RecPotential} is given in Sec.~\ref{sec:light_nuclei} and
App.~\ref{app:coordinate}.
$\mathcal V_0^\text{mag}$ has a Coulombic scaling, $\simeq 1/\qq^2$, with an isospin-one/-two component proportional to $(1+\kappa_p) \pm \kappa_n$, respectively. 
In momentum space this class of  potentials scales as    $\mathcal O(e^2/(k_F^2 \Lambda_\chi))$  and 
contributes to $\delta_\text{NS}$ at  $\mathcal O(\alpha \epsilon_\chi)$. 

When applied to $^1S_0$ wave functions obtained at LO in  chiral EFT,
the Coulomb-like potential in Eq.~\eqref{eq:MagPotential}
gives rise to nuclear matrix elements that are logarithmically dependent on the ultraviolet (UV) cutoff used in the solution of the Lippmann--Schwinger or Schr\"odinger equation~\cite{Cirigliano:2018hja,Cirigliano:2019vdj}.
This signals sensitivity to UV physics, related to the exchange of hard photons with virtual momenta larger than $\Lambda_
\chi$, which can be absorbed by  the 2b short-range operators in Eq.~\eqref{eq:2nuc1}. To properly renormalize nuclear matrix elements, $g_{V1,V2}^{NN}$  need to scale as $\mathcal O(1/(F_\pi^2\Lambda_\chi))$. Their  contribution to the effective Hamiltonian is
\beq \label{V0CT}
\mathcal V_0^\text{CT} = e^2 \big(g^{N\!N}_{V1} O_1 + g^{N\!N}_{V2} O_2\big),
\eeq
where 
\beq
O_1 = \sum_{j\neq k} \tau^{+(j)}\mathds{1}^{(k)},\quad O_2 = \sum_{j<k} \big[\tau^{+(j)}\tau_3^{(k)}+(j\leftrightarrow k)\big].
\eeq

\begin{figure*}
    \centering
    \includegraphics[width=\textwidth]{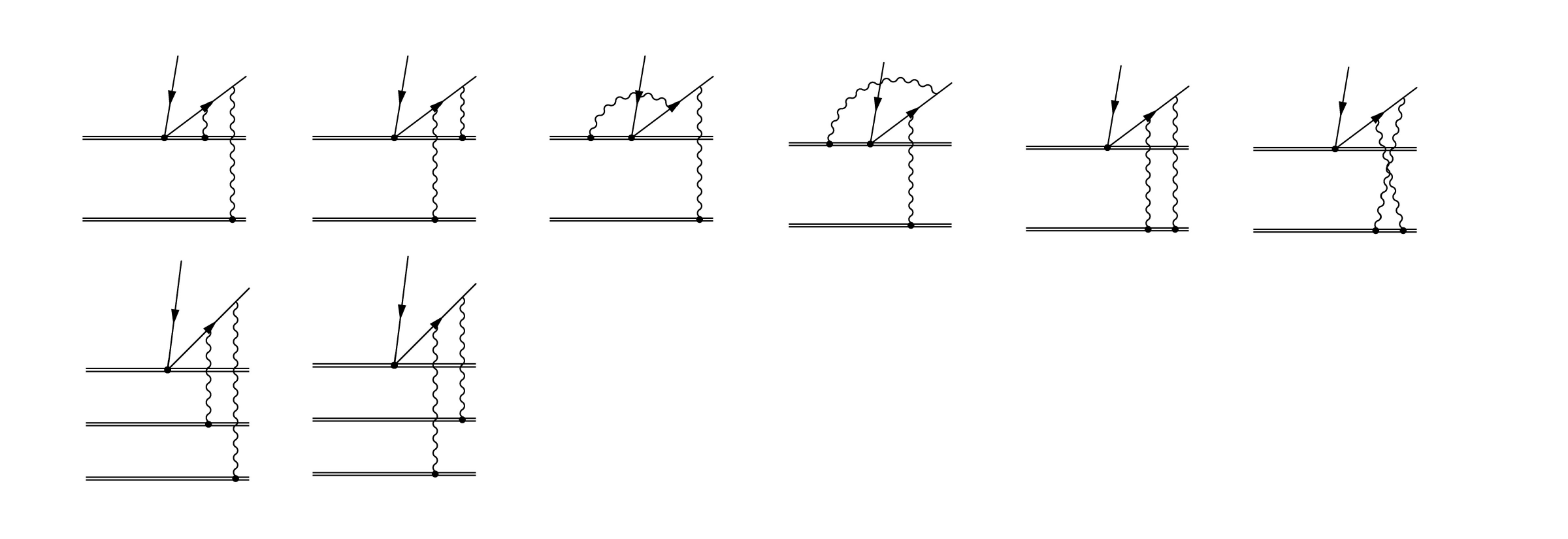}
    \caption{Diagrams contributing to the $\mathcal O(\alpha^2)$ 2b and 3b potentials. The diagrams in the top line  receive contributions from one soft and one potential photon. In the 3b diagrams, both photons follow potential scaling.   }
    \label{fig:alpha2}
\end{figure*}

Following essentially the same steps discussed in Refs.~\cite{Cirigliano:2018hja,Cirigliano:2019vdj} we can derive the cutoff dependence of $g^{N\!N}_{V1, V2}$.
First, we introduce the dimensionless couplings $\tilde{g}^{N\!N}_{V1, V2}$ as
\begin{equation}
    g^{N\!N}_{V1, V2} = \frac{1}{\mN}\left(\frac{\mN C_{^1S_0}}{4\pi} \right)^2 \tilde{g}^{N\!N}_{V1, V2},
\end{equation}
where $C_{^1S_0} = 3 C_T - C_S$ is the LO   $N\!N$ contact interaction in the $^1S_0$ channel.
At LO in chiral EFT,  the RG equations for $\tilde g^{NN}_{V1, V2}$ are the same in dimensional regularization and several cutoff schemes~\cite{Cirigliano:2019vdj} 
and are given by
\begin{align}
   \frac{d  \tilde{g}^{N\!N}_{V1}}{d \log \mu} = -g_A (1+\kappa_p + \kappa_n) = -1.12, \notag\\
   \frac{d  \tilde{g}^{N\!N}_{V2}}{d \log \mu} = - g_A (1+\kappa_p - \kappa_n) = -5.99,
\end{align}
where $\mu$ denotes the renormalization scale in the $\overline{\text{MS}}$ or power-divergence-subtraction schemes, or the UV cutoff scale. Beyond LO, the RG equations depend more explicitly on  the chosen scheme.

\subsection{The \texorpdfstring{$\boldsymbol{\alpha^2}$}{} potential}

At the precision of $\mathcal O(10^{-4})$ required for the analysis of superallowed $\beta$ decays, it is important to also consider subleading corrections in $\alpha$. We focus here on $\mathcal O(\alpha^2)$ corrections, which have an interplay 
with ultrasoft corrections that are enhanced by $Z^2 \log \mue/\mupi$ or $Z \log \mue/\mupi$. 
The diagrams in Fig.~\ref{fig:alpha2} generate 2b and 3b $\mathcal O(\alpha^2)$ potentials whose matrix elements are proportional to $Z$ and $Z^2$.
Subtracting the ultrasoft limit of the same diagrams, as discussed in App.~\ref{app:sub}, the diagrams in Fig.~\ref{fig:alpha2} induce $\mathcal O(\alpha^2)$ corrections to the potential $\mathcal V^0$.
These potentials can be captured by
\begin{equation}
    \mathcal V^0  \to \mathcal V^0 +C_\delta \mathcal V_\delta  + C_\delta^\text{3b} \mathcal V_\delta^\text{3b}  +C_+ \mathcal V_+ + C^\text{3b}_+ \mathcal V^\text{3b}_+.
\end{equation}
The diagrams in the first line of Fig.~\ref{fig:alpha2} induce the 2b potentials
\begin{align}
    \mathcal V_\delta(\qq) &= \sum_{j < k} (2\pi)^3 \delta^{(3)}(\qq) \Big(\tau^{+ (j)} P_p^{(k)} + \tau^{+ (k)} P_p^{(j)} \Big), \notag\\ 
    \mathcal V_+(\qq,\Lambda) &= \sum_{j < k} \frac{4 \pi^2}{\left[\qq^2 \right]_{+,\Lambda}^{\frac{3}{2}}} \Big(\tau^{+ (j)} P_p^{(k)} + \tau^{+ (k)} P_p^{(j)}\Big),
\end{align}    
where the $+$ distribution is defined as
\begin{align}\label{eq:plusdef}
   & \int \frac{d^3 q}{(2\pi)^3} \frac{1}{\left[\qq^2 \right]_{+,\Lambda}^{\frac{3}{2}}} f(\qq) \notag \\ &=
    \int \frac{d^3 q}{(2\pi)^3} \frac{1}{\left[\qq^2 \right]^{\frac{3}{2}}}  \left(f(\qq) - \theta(\Lambda e^{-\gamma_E +1} - |\qq|) f({\bf 0})\right). 
\end{align}
We calculate the diagrams 
in dimensional regularization,
with $d=4-2\epsilon$, and work  in the $\overline{\text{MS}}_\chi$ scheme defined in App.~\ref{app:sub}. In this scheme, the matching coefficients are given by
\begin{align}
    C_\delta &=  -g_V(\mu)\frac{\alpha^2}{2} \left(  \log \frac{\mu^2}{\Lambda^2} -\frac{13}{8}+ 2\gamma_E \right), \label{eq:cdelta} \\ 
    C_+ &=g_V(\mu)  \frac{\alpha^2}{2}\label{eq:cplus}. 
\end{align}
The + distribution depends on an arbitrary subtraction scale $\Lambda$, which, for convenience, we multiplied by the factor $\exp(-\gamma_E + 1)$.
When calculating matrix elements, the dependence on the subtraction scale $\Lambda$ cancels out between $C_\delta$ and $\mathcal V_+$.
It is instructive to also give the potentials in coordinate space
\begin{align}
  \mathcal  V_\delta(\rr) &= \sum_{j < k}  \Big(\tau^{+ (j)} P_p^{(k)} + \tau^{+ (k)} P_p^{(j)}\Big), \label{eq:Vdelta}\\
        \mathcal V_+(\rr,\Lambda) &=  -\sum_{j < k}  \log(r^2_{jk} \Lambda^2)  \Big(\tau^{+ (j)} P_p^{(k)} + \tau^{+ (k)} P_p^{(j)}\Big), \label{eq:Vplus}
\end{align}
where $r_{jk}=|\rr_j - \rr_k|$. 

The 3b potential is derived in App.~\ref{app:sub}.
For this discussion, the most important contribution has the form
\begin{align}\label{eq:Vdelta3b_mom}
   \mathcal V^\text{3b}_{\delta}  & = \sum_{i \neq j \neq k }  \tau^{+(i)}  P_p^{(j)} P_p^{(k)} (2\pi)^3 \delta^{(3)}(\qq_j) (2\pi)^3 \delta^{(3)}(\qq_k),  
\end{align}
with matching coefficient 
\begin{equation}
    C_{\delta}^\text{3b} = -g_V(\mu)\alpha^2 \left( \frac{1}{4}\log \frac{\mu^2}{ \Lambda^2} + \frac{\gamma_E}{2}  - \frac{3}{8}\right).
    \label{eq:cdelta3b}
\end{equation}
In coordinate space, ${\mathcal V}^\text{3b}_{\delta}$ assumes the simple form
\begin{equation}
 \mathcal  V^\text{3b}_{\delta}(\rr)  =  \, \sum_{i\neq j \neq k} \tau^{+(i)}  P_p^{(j)}  P_p^{(k)}.
\end{equation}
$\mathcal V_+^\text{3b}$
depends on the logarithm of the nucleon distances and, in coordinate space, it is given by 
\begin{align}
   C_+^\text{3b} \,    {\mathcal V}_+^\text{3b}(\rr, \Lambda) &=  - g_V(\mu)\frac{\alpha^2}{2}\notag\\
   &\times\sum_{i \neq j \neq k}  \log \bigg[\frac{\Lambda}{2} \Big(r_{ij} + r_{ik} + r_{jk} \Big) \bigg] \notag\\
   &\qquad\times\tau^{+(i)}  P_p^{(j)} P_p^{(k)}.
    \label{eq:V3plusMain}
\end{align}
The momentum-space expression is given in Eqs.~\eqref{eq:3ba} and~\eqref{eq:V3plus1}.
As for the 2b potential, the dependence on the subtraction scale $\Lambda$ cancels
between $\mathcal V^\text{3b}_{\delta}$
and $\mathcal V^\text{3b}_{+}$.
The matrix elements of $\mathcal V_\delta$ and $\mathcal V_{\delta}^\text{3b}$ are given in terms of Fermi matrix elements, and the sum over the additional nucleons induces factors of $Z$ and $Z^2$.
For $\beta^+$ decays, we have
\begin{align}
    \langle f | V_\delta | i \rangle &=  Z M^{(0)}_\text{F}, \notag\\
    \langle f | V_{\delta}^\text{3b} | i \rangle &=  Z (Z-1) M^{(0)}_\text{F},
\end{align}
while for $\beta^-$
\begin{align}
    \langle f | V_\delta | i \rangle &=  (Z-1) M^{(0)}_\text{F}, \notag\\
    \langle f | V_{\delta}^\text{3b} | i \rangle &=  (Z-2) (Z-1) M^{(0)}_\text{F},
\end{align}
where $Z$ is the charge of the final-state nucleus.

\section{Ultrasoft photons}
\label{sec:ultrasoft}

After integrating out the soft and potential photon modes we obtain a theory that features ultrasoft photons as dynamic degrees of freedom, augmented by the potentials collected in $H_\text{EW}$, see Eq.~\eqref{eq:Hpotential}, discussed in the previous section. The obtained potentials can be seen as the matching coefficients between the two theories. To minimize the logarithms that appear in these coefficients, it is natural to perform the matching at a scale $\mupi\simeq R^{-1}\simeq k_F$, as can be seen explicitly from the arguments of the logarithms in Eqs.~\eqref{eq:cdelta}--\eqref{eq:Vplus} and \eqref{eq:cdelta3b}--\eqref{eq:V3plusMain}. The remaining steps are then the evolution of the weak currents $\big({\mathcal J}_{W}^{(i)}\big)^\mu$ and their coefficients $C_W^{(i)}$ from $\mu\simeq \mupi$ to $\mu\simeq \mue$ and the computation of the matrix element at the low-energy scale. 

\subsection{Evolution to \texorpdfstring{$\boldsymbol{\mu\simeq\mue}$}{}}

The anomalous dimension  of $g_V$, which determines its RG equation, is known to ${\mathcal O}(\alpha^2)$ and equivalent to the case of neutron decay. New divergences appear when going beyond the 1b sector, which are sensitive to the charge of the external states and lead to enhancement factors of the charge of the final-state nucleus, $Z$. 
One finds that exchanges of ultrasoft photons between the electron and additional nucleon lines generate interactions proportional to factors of the conserved charge
\begin{equation}
\label{charge}
{\mathcal Q}\equiv \int_{\boldsymbol x} \bar N Q N(x),\qquad Q=\frac{\mathds{1}+\tau_3}{2},
\end{equation}
where $\int_{\boldsymbol{x}}=\int d^3 x$.
These contributions are divergent and require the inclusion of additional interactions that can be written as
\begin{align}\label{eq:Jusoft} 
H_\text{EW}&=\sqrt{2} G_F V_{ud}\bar e_L \gamma_\mu \nu_L\bigg[\sum_{n=0}^{\infty}c_W^{(i,n)}(\mu) {\mathcal Q}^n\bigg]\big({\mathcal J}_{W}^{(i)}\big)^\mu,\notag\\
\big({\mathcal J}_{W}^{(i)}\big)^\mu&=\{v^\mu \tau^+, v^\mu{\mathcal V}^0\,,  v^\mu E_0{\mathcal V}_E^0\,,p_e^\mu {\mathcal V}_{m_e} \,,v^\mu {\mathcal V}_+\,,v^\mu {\mathcal V}^\text{3b}_+\},
\end{align}
with the label $i$ running over the type of 1b, 2b, and 3b interactions, $i=\{ g_V, {\mathcal V}^0,\,{\mathcal V}_E^0,\, {\mathcal V}_{m_e},{\mathcal V}_+,{\mathcal V}^\text{3b}_+\}$. The appearance of the ${\mathcal Q}^n$ operators gives rise to factors of $Z^n$ when acting on the final state. 
The matching of the previous section mostly induces the interactions with $n=0$, while, for $i=g_V$, also the $n=1$ and $n=2$ terms are generated, corresponding to $ {\mathcal V}_\delta$ and $ {\mathcal V}_\delta^\text{3b}$.

As discussed in App.~\ref{app:usoftRGE}, after dressing the $c^{(i,n)}_W$ with additional ultrasoft photons exchanges 
one obtains divergences that are canceled by the counterterms of the $c^{(i,m>n)}_W$ interactions. 
These effects lead to an RG equation for the effective coupling, $C_\text{eff}^{(i)}(\mu) \equiv \sum_{n=0}^\infty c_W^{(i,n)}Z^n$, which is the combination that appears in the matrix element. 

Through ${\mathcal O} (\alpha^2 Z^2 \log \frac{k_F}{m_e} )$ and ${\mathcal O} (\alpha^2 Z\log \frac{k_F}{m_e} )$ we have
\begin{align}\label{eq:CeffRG}
&\frac{d C^{(i)}_\text{eff}(\mu)}{d\log \mu} = \gamma^{(i)}C_\text{eff}^{(i)}(\mu),\\
\gamma^{(g_V)}&=\frac{\alpha}{\pi}\tilde \gamma_0+\left(\frac{\alpha}{\pi}\right)^2\tilde \gamma_1  \notag\\
&\qquad+\left[\sqrt{1-\alpha^2 Z(Z\pm1)}-1\right],\notag\\
\gamma^{({\mathcal V}^0,\, {\mathcal V}_{m_e}\,,{\mathcal V}_+)}&=\left[\sqrt{1-\alpha^2 Z(Z\pm1)}-1\right]+{\mathcal O}(\alpha Z^0), 
\notag\\
\gamma^{({\mathcal V}^0_E)}&=\left[\sqrt{1-\alpha^2 Z^2}-1\right]+{\mathcal O}(\alpha^2 Z\,, \alpha Z^0),\notag
\end{align}
for $\beta^\pm$ decays. The quantities
\begin{equation}
\tilde \gamma_0 = -\frac{3}{4},\quad \tilde \gamma_1 = \frac{5 \tilde{n}}{24} + \frac{5}{32} - \frac{\pi^2}{6}, 
\end{equation}
with $\tilde n =1 $ for $\mu\leq \mpi$, are the one- and two-loop anomalous dimensions of $g_V$, while the terms in square brackets capture the effect from the $c_W^{(i,n>0)}$ coefficients. The matching of the previous section then gives the following boundary conditions at $\mu=\mupi$,
\begin{align}
C_\text{eff}^{(g_V)} &= g_V\left[c_W^{(g_V,0)}+Z c_W^{(g_V,1)}+ Z^2 c_W^{(g_V,2)}\right],\notag\\
c_W^{(g_V,0)}&=1+(-1\pm 1)\left(\frac{1}{2}C_\delta -C_\delta^\text{3b}\right),\notag\\
 c_W^{(g_V,1)}&= C_\delta - 2 C_\delta^\text{3b}  \pm C_\delta^\text{3b}, \qquad  c_W^{(g_V,2)}=C_\delta^\text{3b}, \notag\\
C_\text{eff}^{({\mathcal V}_+)} &=C_+, \qquad C_\text{eff}^{({\mathcal V}^\text{3b}_+)} =C_+^\text{3b}, \notag\\
C_\text{eff}^{({\mathcal V}^0,\, {\mathcal V}_{m_e}\,,{\mathcal V}_{E}^0)} &=1.
\label{eq:bc}
\end{align}

As can be seen from Eq.~\eqref{eq:CeffRG}, we do not control the $Z$-independent ${\mathcal O}(\alpha, \alpha^2)$ pieces for $i\in \{ {\mathcal V}^0,\, {\mathcal V}_{m_e}\,,{\mathcal V}_+\,,{\mathcal V}^\text{3b}_+\}$. In addition, although the $c_W^{(i,n>0)}$ coefficients affect most of the 1b and 2b interactions in the same way, this is not the case for the energy-dependent potential, $i={\mathcal V}^0_E$. Due to its different leptonic structure,\footnote{Its momentum dependence affects the loop integrals that determine the anomalous dimensions. One can show that, to ${\mathcal O}(\alpha^n Z^n)$, the effect reduces to the previous integrals multiplied by $E_e$, thanks to the $\delta$ functions of the internal photon energies discussed in App.~\ref{app:usoftRGE}. However, this is not guaranteed to hold at subleading powers in $Z$. These potentials include ultrasoft photon vertices through the covariant derivative, $\bar e_L v\cdot \overleftarrow D\gamma_0 \nu_L$, which we expect to affect the anomalous dimension at ${\mathcal O}(\alpha^2 Z)$.} we expect $\gamma^{({\mathcal V}^0_E)}$ to differ starting at ${\mathcal O} (\alpha^2 Z )$, as indicated in Eq.~\eqref{eq:CeffRG}. These uncontrolled anomalous dimensions only affect the potentials that appear at ${\mathcal O} (\alpha \epsilon_{\chi}  )$ or ${\mathcal O} (\alpha \epsilon_{\slashed \pi}  )$, so that their contributions are expected to appear beyond the order at which we work.\footnote{Note, however, that the leading terms can have a significant effect since $\alpha^2 Z^2\log \frac{\mpi}{m_e}\simeq 0.4$ for $Z=37$ in the case of the heaviest nuclei considered in Ref.~\cite{Hardy:2020qwl}.}
The above RG equation can be solved to give 
\begin{equation}\label{eq:usoftU}
C^{(i)}_\text{eff}(\mu) = U^{(i)}(\mu, \mupi) C_\text{eff}^{(i)}(\mupi),
\end{equation}
which allows us to evolve the Hamiltonian from the scale $\mupi\simeq k_F$ down to $\mue\simeq m_e$.
The explicit form of the kernel $U^{(g_V)}$ is given in Eq.~\eqref{eq:kernel} in App.~\ref{App:RGE}.

The RG equations in Eq.~\eqref{eq:CeffRG} capture large logarithms to order $(\alpha^2 Z^2L)^n $ for all effective couplings, to order $(\alpha^2 ZL)^n $ for $C_\text{eff}^{(i)}$ with $i\in \{g_V,{\mathcal V}_0,{\mathcal V}_{m_e},{\mathcal V}_+, {\mathcal V}^\text{3b}_+\}$, as well as terms of order $(\alpha L)^n$ and $(\alpha^2 L)^n$ for $C_\text{eff}^{(g_V)}$. 
 In the traditional approach, the first series is included in the standard Fermi function $F$, via logarithms of a fixed and somewhat arbitrary nuclear radius $R$, see Eq.~\eqref{eq:Fermistandard}. The first ($n=1$) term in the second series reproduces the logarithmic term in the $\alpha^2 Z$ correction first identified in Ref.~\cite{Jaus:1970tah}, and included in the $\delta_2$ correction to $\delta_R^\prime$ \cite{Sirlin:1986cc,Jaus:1986te}. Finally, the $(\alpha L)^n$ series is resummed in $\delta^\prime_R$~\cite{Hardy:2008gy}.
In principle,  additional contributions to the anomalous dimensions, at higher order in $\alpha$ or subleading in $Z$, are known as well~\cite{Borah:2024ghn}. To consistently include their effects, however, would require higher-order terms in the matrix element at $\mue$ and the matching at the scale $\mupi$. For example, including the ${\mathcal O}(\alpha^3)$ anomalous dimension would also require the matching and matrix element to ${\mathcal O}(\alpha^2)$. 
Exceptions are the ${\mathcal O}(\alpha^3Z^3)$ and ${\mathcal O}(\alpha^3Z^2)$ contributions to Eq.~\eqref{eq:CeffRG}. Due to the fact that there are no ${\mathcal O}(\alpha Z)$ terms in $\gamma^{(i)}$ we do not need knowledge of ${\mathcal O}(\alpha^2 Z)$ contributions to the matrix element in order to control all ${\mathcal O}(\alpha^3 Z^2 L)$ terms. The relevant anomalous dimensions have been computed in Refs.~\cite{Hill:2023acw,Borah:2024ghn} and add to Eq.~\eqref{eq:CeffRG} as
\begin{equation}
\delta \gamma^{(i)} = \frac{\alpha^3}{4\pi}Z^2\left(6-\frac{\pi^2}{3}\right),\qquad i\in\{g_V,{\mathcal V}_0,{\mathcal V}_{m_e},{\mathcal V}_+, {\mathcal V}^\text{3b}_+\},
\end{equation}
where a possible ${\mathcal O}(\alpha^3Z^3)$ term vanishes. Here $\delta \gamma^{(i)}$ captures terms at the same order as corrections that are usually included in $\delta_3$~\cite{Jaus:1972hua,Jaus:1986te}. We will refrain from including this anomalous dimension in the explicit example of $^{14}$O discussed in Sec.~\ref{sec:light_nuclei}, as its effects are ${\mathcal O}(10^{-5})$, and smaller than the uncertainty due to missing ${\mathcal O}(\alpha^2 Z)$ terms discussed below.

\subsection{The amplitude at \texorpdfstring{$\boldsymbol{\mu\simeq \mue}$}{}}

The final step is the calculation of the amplitude generated by the operators evaluated at $\mu\simeq \mue$,
\begin{align}\label{eq:MatrixElement}
-{\mathcal A} &=\langle f e\bar\nu |H_\text{EW}|i \rangle =   \sqrt{2} G_F V_{ud}\sum_i C^{(i)}_\text{eff}(\mu) {\mathcal M}^{(i)}(\mu),\notag\\
{\mathcal M}^{(i)}&=\langle f e\bar\nu| (\bar e_L\gamma_\mu \nu_L)\, \big({\mathcal J}_{W}^{(i)}\big)^\mu|i\rangle.
\end{align}
Here the ${\mathcal M}^{(i)}$ involve the matrix elements of the usual 1b operator and the subleading potentials in $\big({\mathcal J}_{W}^{(i)}\big)^\mu$, while the large logarithms and the effects of the $c_W^{(i,n)}$ are captured by $C_\text{eff}^{(i)}$. As the ${\mathcal M}^{(i)}$  do not involve large logarithms, one might expect an evaluation to  ${\mathcal O}(\alpha/(4\pi))$ to be adequate, as two-loop corrections scaling as ${\mathcal O}\left(\alpha^2/(4\pi)^2\right)$ would be below ${\mathcal O}(10^{-4})$. However, as is well known, certain loop contributions related to the Fermi function are enhanced with respect to this expectation by factors of $\pi^2$ and $Z$, which requires us to take into account certain classes of higher-loop diagrams.

\subsubsection{Ultrasoft loop contributions}

We can consider the loop expansion for each of the matrix elements, ${\mathcal M}^{(i)}={\mathcal M}^{(i)}_0+{\mathcal M}^{(i)}_1+\dots$, where ${\mathcal M}^{(i)}_n$ captures the effects of $n$-loop diagrams involving ultrasoft photons. To illustrate the structure of these contributions we first focus on the topologies of Fig.~\ref{fig:diagrams}(b,c) at one loop. The class of diagrams in which the photon connects the electron and nuclear lines leads to the following amplitude (in Feynman gauge),
\begin{align}\label{eq:Amp}
{\mathcal M }_1^{(i)} &=  \sum_n \int  \frac{d^4q}{(2\pi)^4} L_{\mu\nu}(q) \Bigg\{\frac{\langle f| \big({\mathcal J}^{(i)}_{W}\big)^\mu |n\rangle \langle n| {\mathcal J}^\nu |i\rangle}{E_i-E_n+q_0+i\epsilon} \notag\\
&\quad+\frac{\langle f| {\mathcal J}^\nu |n\rangle \langle n| \big({\mathcal J}_{W}^{(i)} \big)^\mu|i\rangle}{E_f-E_n-q_0+i\epsilon} \Bigg\},\notag\\
L_{\mu\nu}(q)&= i e^2\bar u(p_e) \gamma_\nu \frac{\slashed{p}_e+\slashed{q}+m_e}{(p_e+q)^2-m_e^2+i\epsilon}\gamma_\mu P_L v(p_\nu)\notag\\
&\qquad\times\frac{1}{q^2+i\epsilon},
\end{align}
where $\mathcal J^\nu$ is the EM current, while $|n\rangle$ and $E_n$ denote intermediate nuclear states and their energies.\footnote{The derivation of Eq.~\eqref{eq:Amp} requires few  assumptions. A very similar expression is able to capture the contributions from the soft and potential regions if one does not perform the ultrasoft expansion in $q/k_F$ and replaces the $\big({\mathcal J}_{W}^{(i)}\big)^\mu$ with the weak current in the theory with propagating soft pions and photons. In fact, one could derive an analogous expression at the quark level. The expression would involve the quark-level currents for ${\mathcal J}_\mu$ and ${\mathcal J}^\muW$, with the intermediate states running over all eigenstates of the QCD Hamiltonian, which would capture the $k\geq \Lambda_\chi$ region as well. } We can restrict the integration to the ultrasoft regime by expanding both the currents and energy denominators in $q/ k_F$. 
As we discuss in more detail in Sec.~\ref{sec:disp}, the only non-negligible effects then arise from the contributions of the LO EM current, $\mathcal J^\mu=\bar N Q v^\mu N$.
After expanding, the EM current effectively acts as a conserved charge, so that $\langle f| {\mathcal J}_\nu |n\rangle = v_\nu Z \delta_{nf} $. The factors in curly brackets in Eq.~\eqref{eq:Amp} then simplify 
\begin{align}\label{eq:Ausoft}
{\mathcal M }_1^{(i)}  &=\int  \frac{d^4q}{(2\pi)^4} \langle f| \big({\mathcal J}_{W}^{(i)}\big)^\mu |i\rangle  L_{\mu\nu}(q) v^\nu\notag\\
&\quad\times\bigg\{\frac{1}{-q_0+i\epsilon} -2\pi i\delta (q_0)(Z-1) \bigg\}.
\end{align}
Most of the $\big({\mathcal J}_{W}^{(i)}\big)^\mu$ are independent of the photon momentum, so that the matrix elements and the ultrasoft loop factorize. The exception is again the case of $i={\mathcal V}_E^0$, for which corrections to this factorization are expected to appear at ${\mathcal O}(\alpha/(4\pi))$. Since ${\mathcal V}_E^0$ itself contributes at ${\mathcal O}(\alpha \epsilon_{\slashed \pi})$, we neglect these corrections here.

The resulting loop integral in Eq.~\eqref{eq:Ausoft} is equivalent to what one would obtain in a theory with non-relativistic initial- and final-state nuclei as degrees of freedom, discussed in Refs.~\cite{Hill:2023acw,Hill:2023bfh}. 
Here the first term in curly brackets is identical to the contribution in the single-nucleon case. The remaining one-loop diagrams, as well as the contributions from real radiation graphs with an additional photon, $\simeq {\mathcal M}_{1\gamma}$, also reduce to the case of neutron decay and scale as $\alpha/(4\pi)$. This allows us identify their contributions to the squared amplitude with the same Sirlin function that appears in the 1b case, while enhanced terms, $\simeq \pi\alpha/\beta$ are collected in the Fermi function, $\bar F$, with $Z=1$. Finally, the $\delta$ function in the second term of Eq.~\eqref{eq:Ausoft} reduces the effects of the photon propagator to that of a static Coulomb potential. It therefore contributes to the Fermi function and effectively takes the ${\mathcal O}(\alpha)$ Fermi function, $\bar F(Z=1)$,  to $\bar F(Z)$. 

The one-loop and real-radiation corrections then combine into
\begin{equation}
|{\mathcal M}^{(i)}_0+{\mathcal M}^{(i)}_1|^2+|{\mathcal M}^{(i)}_{1\gamma}|^2d\Phi^{(\gamma)}\propto \bar F(\beta)\big[1+ \tilde \delta _{ R}'(\mu)\big],
\end{equation}
 where $\beta=|\pp_e|/E_e$ and $ \tilde \delta_{R}'$ captures the effect of the Sirlin function,
 \begin{align}\label{eq:drp}
\tilde      \delta_R' (E_e,\mu) &= \frac{\alpha \left( \mu \right)}{2\pi} \bigg[ \frac{3}{2} \log \frac{\mu^2}{m_e^2} + \frac{5}{4} + \hat g( E_e, E_0)   \bigg],\\
          \hat g( E_e, E_0) &= g( E_e, E_0) - \frac{3}{2} \log \frac{\mN^2}{m_e^2},
\label{eq:tdRprime}
 \end{align}
with $g( E_e, E_0)$ the Sirlin function of Ref.~\cite{Sirlin:1967zza},
\begin{align}
g(E_e,E_0) &= \frac{3}{2} \log \frac{\mN^2}{m_e^2} - \frac{3}{4} +  \frac{1 + \beta^2}{\beta} \, \log \frac{1+\beta}{1-\beta} \notag \\ &
+ \frac{1}{12 \beta} \left( \frac{\bar E}{E_e}\right)^2  \log \frac{1+\beta}{1-\beta}
\notag \\
&+4 \left[ \frac{1}{2 \beta} \log \frac{1+\beta}{1-\beta} - 1 \right] \left[ \log \frac{2 \bar E}{m_e}  - \frac{3}{2} + \frac{\bar E}{3 E_e} \right]
\notag \\
&-
\frac{1}{\beta} \left[4 \,  \text{Li}_2 \left( \frac{2 \beta}{1 + \beta} \right) + \log^2 \left( \frac{1+\beta}{1-\beta} \right)   \right],
\label{eq:gSirlin}
\end{align}
with $\bar E = E_0 - E_e$. 
To this order the Fermi function for $\beta^\pm$ decays is given by 
\begin{equation}
\bar F(\beta) =1\mp\frac{\pi \alpha Z}{\beta}.
\end{equation}

As mentioned above, we are interested in terms beyond the one-loop level that are enhanced by factors of $\pi^2$ or $Z$ compared to the naive expectation of $\alpha^2/(4\pi)^2$. This type of contribution arises from ladder diagrams in which ultrasoft photons are exchanged between the electron and the nucleus. Within the EFT framework, these diagrams reduce to the expressions one would obtain in a theory with non-relativistic initial (final) nuclei of charge $Z-1$ ($Z$). These graphs were computed to all orders in $\alpha^n Z^n$ in exactly this formulation in Refs.~\cite{Hill:2023acw,Hill:2023bfh}, which allows us to capture the enhanced terms. Combined with the non-enhanced ${\mathcal O}(\alpha/4\pi)$ terms in the Sirlin function, we obtain for the spin-summed squared amplitude
\begin{align}\label{eq:Ausoft2}
\sum_\text{spins}|{\mathcal A}|^2  &=4E_eE_\nu(1+a  {\boldsymbol{ \beta}}\cdot \hat{\pp}_\nu)
 \bar F(\beta,\mu)\big[1+\tilde \delta' _{R}(\mu)\big]\notag\\
&\times  |\langle f| \sum_i C_\text{eff}^{(i)}(\mu){\mathcal V}^{(i)} h_i(E_e)|i\rangle|^2,
\end{align}
where $h_{{\mathcal V}_E^0}= E_0$, $h_{{\mathcal V}_{m_e}}= m_e^2/E_e$, while $h_{i}= 1$ otherwise. The employed factorization between the Fermi and Sirlin functions 
holds 
up to corrections of ${\mathcal O}(\alpha^2 Z)$.
All potentials apart from ${\mathcal V}_{m_e}$  contribute equally to the electron--neutrino correlation coefficient $a$, so that 
\begin{equation}
a= \frac{|\langle f| \sum_{i\neq {\mathcal V}_{m_e}} C_\text{eff}^{(i)}(\mu){\mathcal V}^{(i)} h_i(E_e)|i\rangle|^2}{|\langle f| \sum_i C_\text{eff}^{(i)}(\mu){\mathcal V}^{(i)} h_i(E_e)|i\rangle|^2},
\end{equation}
to which we will come back in Sec.~\ref{sec:NME}.

After translating the results of Refs.~\cite{Hill:2023acw,Hill:2023bfh} from $\overline{\text{MS}}$ to $\overline{\text{MS}}_\chi$, the Fermi function to all orders in $Z^n\alpha^n$ is given by
 \begin{align}\label{eq:FF}
\bar F(\beta,\mu)  &= \frac{4\eta}{(1+\eta)^2}\frac{2(1+\eta)}{\Gamma(2\eta+1)^2}|\Gamma(\eta+iy)|^2 e^{\pi y}\notag\\
&\quad\times \left(\frac{2 |\pp_e|}{\mu}e^{1/2-\gamma_{E}}\right)^{2(\eta-1)},
\end{align}
with $\eta = \sqrt{1-\alpha^2 Z^2}$ and $y =\mp Z\alpha/\beta$, which differs from the traditionally employed Fermi function~\cite{Wilkinson:1982hu} by $4\eta/(1+\eta)^2\approx1-\alpha^4Z^4/16$.
It can be checked that the combination $|C_\text{eff}^{(i)}(\mu)|^2\bar F(\beta,\mu)\big[1+\tilde \delta '_{R}(\mu)\big]$ is independent of $\mu$ to ${\mathcal O}(\alpha)$, ${\mathcal O}(\alpha^2)$, and ${\mathcal O}(\alpha^n Z^n)$.\footnote{The $\mu$ dependence from the ${\mathcal O}(\alpha^n Z^{n-1})$ pieces of the anomalous dimensions should be canceled by terms of the same order in the Fermi function, which we currently do not control.}
To simplify the expression, the scale in the Fermi function is often chosen as 
$\mu = 1/R e^{1/2-\gamma_E}$, with a nuclear radius $R$, but to keep track of the scale dependence in a more transparent way we display the full expression~\eqref{eq:FF}. 

As we discuss in more detail below, Eq.~\eqref{eq:Ausoft2} contains all the needed ingredients to obtain the decay rate. In this form, the ultrasoft contributions, often referred to as outer corrections, are captured by $\bar F(\beta,\mu)\big[1+\tilde \delta '_{R}(\mu)\big]$, while the evolution between $\mu\simeq \mue$ and $\mu\gtrsim \mupi$ as well as physics from shorter distance scales is collected in the Wilson coefficients $C_\text{eff}^{(i)}$. Finally, the nuclear-structure dependence arises from the matrix elements of the ${\mathcal V}^{(i)}$, to which we will turn next.

\subsubsection{Nuclear matrix elements}\label{sec:NME}

In the EFT approach the required nuclear matrix elements of the EW potentials obtained in the previous section can be identified as contributions to $\delta_\text{NS}$,
\begin{align}
& |\langle f| \sum_i C_\text{eff}^{(i)}(\mu){\mathcal V}^{(i)} h_i(E_e)|i\rangle|^2 \notag\\
&\equiv  |C_\text{eff}^{(g_V)}(\mu)|^2|M_\text{F}|^2(1+\tilde \delta_\text{NS}),
\end{align} 
where $M_\text{F}$ is the Fermi matrix element and $\tilde \delta_\text{NS}$ is 
an electromagnetic correction that depends on the  nuclear structure. 

To account for isospin breaking in the nuclear states, 
the Fermi matrix element is traditionally written as   $M_\text{F}  = M_\text{F}^{(0)} ( 1 - \bar \delta_C/2)$, 
with  $M_\text{F} ^{(0)} =  \langle f^{(0)}| \tau^+|i^{(0)} \rangle$ computed 
in terms of the isospin-symmetric  nuclear states $|i^{(0)}\rangle$ and $|f^{(0)} \rangle$. 

For the nuclear structure correction, we find that
\begin{equation}
\tilde \delta_\text{NS}= \delta_\text{NS}^{(0)} (\mu)  +  {\delta^E_\text{NS}}(E_e,\mu)
\label{eq:deltaNS}
\end{equation}
receives $E_e$-independent contributions of  $\mathcal O(\alpha \epsilon_\chi)$, 
as well as $E_e$-dependent contributions of $\mathcal O(\alpha \epsilon_{\slashed{\pi}})$. 
To this order, $\delta_\text{NS}$ is entirely determined by matrix elements of appropriate potentials 
between the initial and final states
\beq
\tilde \delta_\text{NS} =\frac{2}{M_\text{F}}\sum_i \frac{C_\text{eff}^{(i)}(\mu)}{C_\text{eff}^{(g_V)}(\mu)}\langle f|{\mathcal V}^{(i)} h_i(E_e)|i\rangle.
\label{eq:dNS}
\eeq
As described above, the RG evolution is known to different orders for the different potentials. However, as the ${\mathcal V}^{(i)}$ contribute at ${\mathcal O }(\alpha \epsilon_{\chi,\slashed{\pi}})$, we neglect these differences and write  $C_\text{eff}^{(i)}(\mu)/C_\text{eff}^{(g_V)}(\mu)\simeq C_\text{eff}^{(i)}(\mupi)/g_V(\mupi)$. Similarly, $M_\text{F}$ in principle includes isospin-breaking corrections, but to the order we consider we can approximate the nuclear wave functions by those in the isospin limit, $i^{(0)}$ and $f^{(0)}$, which also allows the use of  $M_\text{F}\simeq M_\text{F}^{(0)} $.
With these simplifications one obtains
 \begin{align}
 \label{eq:dNS2}
 \delta_\text{NS}^{(0)}  &=  \frac{2\langle f^{(0)} | \mathcal V^0 +C_+\mathcal V_+  +C_+^\text{3b}{\mathcal V}^\text{3b}_+ | i^{(0)} \rangle}{g_V(\mupi)M_\text{F}^{(0)}}\notag\\
 &=  \frac{2}{g_V(\mupi)M_\text{F}^{(0)}}\langle f^{(0)} | \mathcal V^\text{mag}_0  + \mathcal V^\text{rec}_0  + \mathcal V^\text{CT}_0   \notag\\
 &\qquad+C_+\mathcal V_+  +C_+^\text{3b}{\mathcal V}^\text{3b}_+| i ^{(0)}\rangle, 
 \end{align}
and   
\begin{align}
{\delta^E_\text{NS}} &= \mp\frac{2}{g_V(\mupi)M_\text{F}^{(0)}} 
\langle i ^{(0)}| {\mathcal V}_E^0 E_0+ \frac{m_e^2}{E_e}{\mathcal V}_{m_e}| f ^{(0)}\rangle
\notag\\
&=\mp\frac{2}{g_V(\mupi)M_\text{F}^{(0)}} \bigg[
\langle f^{(0)} |  \mathcal V_E | i^{(0)}  \rangle \bigg(\frac{E_0+8E_e}{6} +  \frac{m_e^2}{2E_e} \bigg) \notag\\
&+ 
 \langle f^{(0)} |  \mathcal V^\pi_{E}  | i^{(0)}\rangle   E_0
 +\langle f^{(0)} |  \mathcal V^\pi_{m_e}  | i^{(0)}\rangle   \frac{m^2_e}{E_e} 
 \bigg],
\label{eq:dNSE}
\end{align}
while the neutrino--electron correlation simplifies to
\begin{equation}
a=1\pm\frac{2}{g_V(\mupi)M_\text{F}^{(0)}}\frac{m_e^2}{E_e}\langle f^{(0)} | \mathcal V_{m_e}   | i^{(0)} \rangle.
\end{equation}
The upper (lower) signs in the above equations refer to $\beta^+$ ($\beta^-$) decays.

Before combining these ingredients to form an expression for the decay rate in Sec.~\ref{sec:decay_rate}, we first discuss the ultrasoft effects due to subleading terms in the EM current, ${\mathcal J}_\mu$, and their connection to the dispersive approach.

\subsection{Comparison to the dispersive approach}
\label{sec:disp}

Although the dependence on the intermediate states in Eq.~\eqref{eq:Amp} disappeared in Eq.~\eqref{eq:Ausoft}, this is no longer the case when going beyond LO in $\mathcal J_\nu$. In particular, contributions from the magnetic moment allow the EM current to connect to excited states of the initial- or final-state nucleus. 
This leads to a sensitivity to the intermediate-state energies and requires knowledge of overlap factors of the form $\langle n| {\mathcal J}^\nu_\text{mag} |i\rangle\simeq \epsilon^{\nu \alpha\beta\gamma} v_\alpha q_\beta \langle n|S_\gamma |i\rangle$. 
These contributions capture similar effects to those discussed in Refs.~\cite{Seng:2022cnq,Gorchtein:2023naa}, in which the contributions due to low-lying nuclear states were studied and estimated to be sizable, due to an increased sensitivity to IR scales. In contrast, in the EFT we estimate
the impact of intermediate states on ${\mathcal A } _\text{usoft}$ as follows: first, the magnetic moment appears at ${\mathcal O}(q/\mN)$. Second, the only scales appearing in the integrand of Eq.~\eqref{eq:Amp} are $p_e\simeq  m_e$ or $E_{i,f}-E_n$, both of the order of $\qe$, implying that ${\mathcal A } _\text{usoft}^\text{mag}$ will scale as 
\beq
\label{usoft_mag}
{\mathcal O}\Big(\frac{\alpha}{\pi} \frac{\qe}{\mN}\Big)={\mathcal O}\left(\frac{\alpha}{\pi} \epsilon_\text{recoil}\right), 
\eeq
beyond the level of precision we need to consider.

To clarify the relation with the dispersive approach, we considered a toy model for $T_3(\nu,Q^2)$ that displays all the relevant features  expected from the magnetic ultrasoft contributions
\beq
\frac{iT_3^\text{toy}(\nu,Q^2)}{M\nu}=\frac{M}{\mN}\frac{g_A g_M}{s-\bar M^2+i\eps},
\eeq
where $s=M^2+\nu^2-\qq^2+2M\nu$, $M^2-\bar M^2=2M\Delta$, and $g_A$ ($g_M$) parameterizes the coupling to the EW (EM) current. In the dispersive approach, $T_3$ enters a master formula~\cite{Seng:2022cnq} similar to Eq.~\eqref{eq:boxbar}\footnote{For simplicity, we consider the limit $m_e=0$, which suffices to determine the relevant scales. See Ref.~\cite{Gorchtein:2023naa} for the general expression.}  
\begin{align}
\label{box_gW}
\Box_{\gamma W}&=-\frac{e^2}{M_\text{F}^{(0)}}\int\frac{\diff^4q}{(2\pi)^4} \frac{\mW^2}{Q^2+\mW^2} 
\notag \\
&\times \frac{T_3(\nu,Q^2)}{(p_e-q)^2Q^2}\frac{Q^2+M\nu\frac{p_e\cdot q}{p\cdot p_e}}{M\nu},
\end{align}
and for low-lying intermediate states with mass $\bar M$, corresponding to $\Delta >0$, it was found that $\Box_{\gamma W}$ becomes singular for $E_e\to 0$~\cite{Gorchtein:2023naa}, which would call into question the EFT prediction~\eqref{usoft_mag}. Evaluating the integral~\eqref{box_gW} by summing the three residues 
in the upper half plane we find
\beq
\label{toy_full}
\Box_{\gamma W}^{\text{toy}, \Delta}=\frac{3g_A g_M}{4M_\text{F}^{(0)}}\frac{\alpha}{\pi}\frac{\Delta}{\mN}\log\frac{2\Delta}{M}+\Order\big(\Delta^2\big),
\eeq
see App.~\ref{app:toy} for the individual residues. Accordingly, since $\Delta\simeq \qe$, the result does scale as expected in Eq.~\eqref{usoft_mag}.

However, we observe that the individual residues exhibit divergences for $E_e\to 0$, only the sum is again regular. Similarly, in the dispersive approach one finds that the so-called residue correction, required to be able to perform a Wick rotation in Eq.~\eqref{box_gW}, 
scales as
\beq
\label{toy_res}
\Box_{\gamma W}^{\text{toy, res}}=\frac{g_A g_M}{M_\text{F}^{(0)}}\sqrt{\frac{M}{\mN}}\frac{\alpha}{\pi}\sqrt{\frac{2\Delta}{\mN}}+\Order\big(\Delta^{3/2}\big),
\eeq
see App.~\ref{app:toy} for details. The result is finite for $E_e\to 0$, but it scales 
as $\Order(\alpha\sqrt{\eps_\text{recoil}})$ and could therefore be relevant numerically. This apparent mismatch is resolved because also the Wick-rotated integral scales with $\sqrt{\Delta}$ in such a way that the combined result indeed reproduces Eq.~\eqref{toy_full}
\beq
\Box_{\gamma W}^\text{toy}=\Box_{\gamma W}^\text{toy, Wick}-\Box_{\gamma W}^\text{toy, res}.
\eeq
While $\Box_{\gamma W}^\text{toy, res}$ itself could therefore indeed be enhanced compared to the ultrasoft scaling~\eqref{usoft_mag}, the EFT predicts that the most sizable contributions 
of size $\mathcal O(\alpha\sqrt{\eps_\text{recoil}})$ should cancel between residue and Wick-rotated contributions.

\section{Decay rate and factorization}
\label{sec:decay_rate}

\subsection{Grouping contributions according to EFT}

Putting together the various terms discussed in the previous sections, 
the  EFT-based master formula for the nuclear 
$0^+ \to 0^+$ decay rate takes the form
\begin{align}
\frac{d\Gamma}{dE_e d\Omega_e d\Omega_\nu} &= 
\frac{2   (G_F V_{ud})^2}{(2\pi)^5}  
W (E_e,\pp_e\,\pp_\nu)\,  
\tilde C (E_e) \, 
 \bar F (\beta, \mu) \notag\\ 
&\times \big[1 + \tilde \delta_R^\prime (E_e, \mu) \big] 
( 1-  \bar \delta_C) \, 
 \big[1+ \tilde \delta_\text{NS}  (E_e)  \big]\notag\\ 
&\times  \left[ C_\text{eff}^{(g_V)}(\mu) \right]^2, 
\label{eq:dGamma}
\end{align}
with 
\begin{align}
W (E_e,\pp_e\,\pp_\nu) &= w_0 (E_e)  \,
( 1+ a\, \boldsymbol{\beta} \cdot {\hat \pp}_\nu)\,\end{align}
and
\begin{align}
w_0 (E_e) 
&=   |\pp_e| E_e (E_0-E_e)^2.
\end{align}
The factor  $\tilde C (E_e)$   encodes corrections due to the 
nuclear EW form factor, nuclear recoil, 
atomic electron screening, and  atomic overlap~\cite{Hardy:2020qwl,Gorchtein:2023naa} not discussed in this work. 
We provide a detailed prescription on how to infer $\tilde C(E_e)$ from the standard calculation of the shape factor in Refs.~\cite{Hardy:2004id,Hardy:2008gy,Hayen:2017pwg} in App.~\ref{app:Cfactor}.  
The other correction  factors going from left to right represent the  effects of  photons of increasing virtuality: 
$\bar F$ and $\tilde \delta_R^\prime$ 
 arise from ultrasoft photons, 
$\bar \delta_C$  and $\tilde \delta_\text{NS}$  from soft, potential, 
and hard photons, and $C_\text{eff}^{(g_V)}$ encodes the effect of hard and soft photons through the running and matching from 
the EW scale  all the way down to $\mu \simeq \mue$. 
The key quantities contributing to the decay rate are:
\begin{enumerate}
\item  The Fermi function $\bar F$, given  in Eq.~\eqref{eq:FF}. 
\item The generalization of the traditional outer corrections, $\tilde \delta_R^\prime$, 
that can be read off from  Eqs.~\eqref{eq:drp}, \eqref{eq:tdRprime}, and \eqref{eq:gSirlin}. 
\item  The structure-dependent correction $\tilde \delta_\text{NS}$, which can be obtained from  Eqs.~\eqref{eq:deltaNS}--\eqref{eq:dNSE}, 
in terms of the effective couplings $C_\text{eff}^{(i)} (\mu \simeq \mupi)$ (with 
 $i=\{ g_V, {\mathcal V}^0,\,{\mathcal V}_E^0,\, {\mathcal V}_{m_e},{\mathcal V}_+,\,  {\mathcal V}^\text{3b}_+\}$) 
and  transition-dependent nuclear matrix elements.
The effective couplings at $\mu \simeq \mupi$ can be obtained from 
Eqs.~\eqref{eq:gVmu0},~\eqref{eq:cplus}, \eqref{eq:cdelta}, \eqref{eq:cdelta3b}, \eqref{eq:bc}. 
In Sec.~\ref{sec:light_nuclei}, we will provide 
the first ab-initio results for light nuclei, in particular  for the phenomenologically  relevant $^{14}\text{O}\to{}^{14}\text{N}$ decay. 

\item The effective vector coupling constant $C_\text{eff}^{(g_V)} (\mu \simeq \mue)$, 
which can be obtained  by solving the RG equations~\eqref{eq:CeffRG} with boundary condition  
at $\mu \simeq \mupi$ from  Eq.~\eqref{eq:gVmu0}. 

\end{enumerate}

We will give an explicit example of how these different ingredients can be combined for the case of $^{14}$O in Sec.~\ref{sec:example}.
As discussed in the previous section,  the dependence on the scale $\mu$ cancels among the various terms in Eq.~\eqref{eq:dGamma}, 
up to higher-order terms not included in our analysis.  Large logarithms appear in $C_\text{eff}^{(g_V)} (\mu \simeq \mue)$  
and are resummed using the RG equations.

Finally, upon integrating over the phase space we arrive at the final formula for the half-life:  
\begin{align}
    \frac{1}{t}  &=    \frac{G_F^2 |V_{ud}|^2  m_e^5}{\pi^3 \log 2}\,  
    \Big[C_\text{eff}^{(g_V)}(\mu)\Big]^2  \notag\\
    &\quad\times[1 + \bar \delta_R^\prime(\mu)]\,  (1 +  \bar \delta_\text{NS}) \,  ( 1 - \bar  \delta_C)  \,  \bar f(\mu),
    \label{eq:master2}
\end{align}
where
\begin{equation}\label{eq:PS}
 \bar f(\mu) =  \frac{1}{m_e^5} \int_{m_e}^{E_0} d E_e w_0(E_e)  
    \, \tilde C (E_e) \, 
 \bar F (\beta, \mu),
\end{equation}
and we defined the phase-space average 
\beq
\bar G(\mu) = \frac{\int_{m_e}^{E_0} d E_e w_0(E_e)  
    \, \tilde C (E_e) \, 
 \bar F (\beta, \mu) \,  
    \tilde G (E_e,\mu)  }{\int_{m_e}^{E_0} d E_e  w_0 (E_e)\, \tilde C (E_e) \, 
 \bar F (\beta, \mu)  },
\label{eq:PSaverage}
 \eeq 
 for $\tilde G(E_e,\mu)\in\big\{\tilde \delta_R'(E_e,\mu),\tilde \delta_\text{NS}(E_e)\big\}$.
 
At first sight, Eq.~\eqref{eq:master2} looks very similar to Eq.~\eqref{eq:master0}, but important differences arise in the details, most notably related to the separation of scales. For this reason, we next provide a discussion of how the above decay rate formula compares with the one commonly used in the literature.

\subsection{Comparison with the literature}
\label{sec:comparison}

\begin{table*}[t]
    \centering
\setlength{\tabcolsep}{8pt}
    \renewcommand{\arraystretch}{1.3}
    \begin{tabular}{lccl}\toprule
    & EFT & Traditional & Comments\\\colrule
{\bf    Vector coupling} & $C_\text{eff}^{(g_V)}$ & $\Delta_R^V,F,\delta_R'$ & 
\multirow{4}{10cm}{\noindent
    Contains the matching and RG evolution between $\muW$ and $\mue$, most of which is usually collected in $\Delta_R^V$. 
   Additionally, $C_\text{eff}^{(g_V)}$ resums terms $\simeq\alpha L$ as well as $\simeq \alpha^2 Z^2L $ and $\simeq\alpha^2Z L$, which are traditionally collected in the Sirlin function, the Fermi function, and $\delta_{2}$, respectively.
   }
    \\
    Eqs.~\eqref{eq:CeffRG},\eqref{eq:bc},\eqref{eq:gVmu0} & & &\\
    &&&\\
    &&&\\
{\bf    Outer correction} & $\tilde \delta_R^\prime$ & $\delta_R^\prime$ & \multirow{2}{10cm}{\noindent
$\delta_R'$ contains large logarithms in the Sirlin function and $\delta_{2}, \delta_{3}, 
\delta_{\alpha^2}$, while   
$\tilde \delta_R'$ does not, as they are captured by $C_\text{eff}^{(g_V)}$.  }
  \\Eqs.~\eqref{eq:drp}--\eqref{eq:gSirlin} & & &\\
{\bf Isospin breaking }& $\bar \delta_C$ & $\delta_C$ &
\multirow{3}{10cm}{\noindent In both approaches defined as the deviation of $\langle f|\tau^+|i\rangle$ from $\sqrt{2}$. In the EFT, computing $\bar \delta_C$ requires using chiral interactions consistent with those used to obtain $\tilde \delta_\text{NS}$.}
\\
&&&\\
&&&\\
{\bf Nuclear structure} & $ \delta_\text{NS}^{(0)}$ & $\delta_\text{NS,A},\delta_\text{NS,B}$ & 
\multirow{7}{10cm}{\noindent 
In the EFT, nuclear-structure dependence arises from the matrix elements of potentials. The parts of ${\mathcal V}_0^\text{mag}$ and ${\mathcal V}_0^\text{rec}$ induced by photon exchange correspond to the effects captured by $\delta_\text{NS,B}$. The (quenching) correction $\delta_\text{NS,A}$ does not appear in the EFT at the current order, however, the pion-induced parts of ${\mathcal V}_0^\text{rec}$ capture similar effects, as do 2b currents that renormalize $g_A$ and other higher-order diagrams. There is no analog of ${\mathcal V}_0^\text{CT}$ proportional to $g_{V1,V2}^{N\!N}$ nor the pion-exchange potentials in the traditional approach.
}  
\\Eq.~\eqref{eq:dNS2}&&&\\
&&&\\
&&&\\
&&&\\
&&&\\
&&&\\
 Eq.~\eqref{eq:dNSE}  & $ \delta_\text{NS}^{E}$ & $\delta_\text{NS,E},L_0,C_0$ & 
\multirow{2}{10cm}{\noindent 
$ \delta_\text{NS}^{E}$ gives contributions that scale as $\alpha Z R E_{0,e}$, which traditionally appear in the finite-size correction, $L_0$, and shape factor, $C_0$. 
}  \\
     &&& \\
{\bf Fermi function} & $\bar F$ & $F$ & 
\multirow{4}{10cm}{\noindent 
$\bar F$ is obtained diagrammatically, while $F$ is the solution of the Dirac equation. This results in differences at ${\mathcal O}(\alpha^4 Z^4)$. In addition, $F$ contains factors of $\alpha^2 Z^2 L$, while $\bar F$ does not include large logarithms, which, in the EFT approach,  are resummed in $C_\text{eff}^{(g_V)}$.
}
\\Eq.~\eqref{eq:FF}&&&\\
&&&\\
&&&\\
{\bf Other}  & $\tilde C$ & $L_0, C_0, U,S,r,R$ & 
\multirow{6}{10cm}{\noindent 
Several corrections are unchanged in the current EFT approach. These include atomic screening and overlap factors, $S$ and $r$, as well as recoil corrections, $R$. Similarly, for the corrections due to the finite size and charge distribution of the nucleus,  $L_0$, $C_0$, and $U$, 
we do not change terms that appear beyond ${\mathcal O}(\alpha)$, ${\mathcal O}(\alpha^2)$, or ${\mathcal O}(\alpha Z R E_{0,e})$. These effects are collected in $\tilde C$  following the traditional approach.
}\\App.~\ref{app:Cfactor}&&&\\
&&&\\
&&&\\
&&&\\
&&&\\
      \botrule
    \end{tabular}
    \caption{Comparison of the corrections in the EFT decomposition~\eqref{eq:dGamma} to the traditional form of the decay rate \cite{Hardy:2020qwl}, with the fourth column highlighting the main differences.}
    \label{tab:comparison}
\end{table*}

We have cast the EFT-based formula for the half-life,  Eq.~\eqref{eq:master2}, 
in a form that resembles the traditional master formula in Eq.~\eqref{eq:master0}, 
in order to facilitate the mapping between the two approaches. 
Comparison of the two formulae shows that 
$[C_\text{eff}^{(g_V)}(\mu)]^2 -1$ 
is related  to   $\Delta^V_R$ 
and that the quantities $\bar f$, $\bar \delta_R^\prime$, $\bar \delta_\text{NS}$, and $\bar \delta_C$ are related to the 
corresponding  unbarred quantities that appear in Eq.~\eqref{eq:master0}. However, we emphasize that these quantities do not coincide and can be quite different.
Foremost, these differences originate from the fact   that 
the traditional master formula does not fully exploit the separation of scales in the problem, 
while the EFT maximally does so. This has several implications, which we delineate in this subsection, summarized in Table~\ref{tab:comparison}. The main observations are as follows:
\begin{enumerate}
\item  The EFT clearly identifies corrections of size ${\mathcal O}(G_F \alpha \epsilon_\chi)$ that at the two-nucleon level appear as local interactions proportional to the LECs $g_{V1,V2}^{N\!N}$. 
These  are currently not  accounted for  in the traditional approach, where they appear implicitly, through the high-energy part of matrix elements 
of quark-level EW currents between nuclear states, the so-called nuclear $\gamma W$ box contribution. 

\item  The EFT power counting allows one to greatly simplify the calculation of nuclear-structure-dependent effects 
 ($\delta_\text{NS}$ versus  $\bar \delta_\text{NS}$), 
since the computation of $\bar \delta_\text{NS}$ in the EFT requires the matrix element of a 2b current   
between initial and final nuclear states, while the calculation of  $\delta_\text{NS}$ in the dispersive approach~\cite{Seng:2022cnq} 
requires a summation over intermediate nuclear states,
which can be very hard to accomplish in some ab-initio nuclear structure methods. The approach in Ref.~\cite{Towner:1992xm} is closer to ours, in that potentials are evaluated between initial and final states.

\item The EFT method allows one to sum large logarithms through the RG equations. 
For example,  already in the single-nucleon case, 
only in the EFT approach we can include the corrections to the vector amplitude 
to NLL accuracy, e.g., corrections to $g_V$ of order $\alpha^2 \log \frac{\mN}{m_e}$.  

\item Some effects that are present in both approaches end up being labeled differently.
For example, the large logarithms associated with the running of 
$C_\text{eff}^{(g_V)}(\mu)$, 
captured by $C_\text{eff}^{(g_V)} (\mu \simeq \mue)$  in the EFT, 
 in the traditional approach appear in multiple places, 
 such as $\Delta_R^V$,  $\delta_R^\prime$,
and in the Fermi function. The EFT labeling has the advantage that changes in the scale are properly taken into account via the RG evolution, while the decomposition at a fixed scale in the traditional approach requires an ultimately arbitrary choice.  
\end{enumerate}
We discuss two specific cases in more detail. First, subtleties arise when comparing 
$(1 + \Delta_R^V) (1 + \delta_R^\prime)$ to 
$[C_\text{eff}^{(g_V)}(\mu)]^2  (1 + \bar \delta_R^\prime)$.
In the standard approach, the large logarithm associated with the running of $g_V \to C_\text{eff}^{(g_V)}$ between $\mN$ and $\qe$ is 
taken into account in the outer corrections $\delta_R^\prime$. In fact,  the large logarithm of $\mN/m_e$  appears in the Sirlin function. 
Therefore, in the EFT approach the standard breakdown of RC corresponds to 
\begin{itemize}
\item [(i)] Evaluating the coupling  $g_V(\mu)$ at a scale $\mu \simeq \Lambda_\chi \simeq \mN$ 
and identifying 
$\Delta_R^V$ in the master formula~\eqref{eq:master0} with 
\begin{equation}
    \Delta^V_R = \big[ g_V( \mN) \big]^2 \left( 1 +  \frac{5 \alpha (\mN)}{8 \pi} \right) -1.
\end{equation}
Numerically, using the non-perturbative input for the single-nucleon matrix elements from 
Refs.~\cite{Seng:2018qru,Seng:2018yzq,Czarnecki:2019mwq,Shiells:2020fqp,Hayen:2020cxh,Seng:2020wjq}, 
we find $\Delta^V_R = 2.471(25)\%$~\cite{Cirigliano:2023fnz}.
\item[(ii)]  Shifting the large logarithm $\log \frac{\mN}{m_e}$  and the corresponding  LL RG evolution into the 
 Sirlin function $g (E_e, E_0)$ and hence  $\tilde \delta_\text{R}^\prime$. To LO this 
 simply amounts to replacing in Eq.~\eqref{eq:tdRprime}   $\tilde \delta_\text{R}^\prime  \to \alpha/(2 \pi) g (E_e, E_0) + \cdots$, where 
 the ellipsis represents  higher-order 
 corrections  of  $\mathcal O(Z \alpha^2)$ and 
 $\mathcal O(Z^2 \alpha^3)$, see Refs.~\cite{Jaus:1970tah,Jaus:1972hua,Jaus:1986te,Sirlin:1986cc,Sirlin:1986hpu}.
 \end{itemize}
Second, the relation between  
 $ f  (1 + \delta_\text{NS})$ and $ \bar f (1 + \bar \delta_\text{NS} )$ involves the following caveats.   
The traditional  Fermi function and $\bar F (\beta, \mu)$ differ in the logarithmic terms of  ${\mathcal O}(\alpha^2Z^2)$, which in the EFT are resummed in $C_\text{eff}^{(i)}$, so that one can only
identify   $\delta_\text{NS}$ in the master formula~\eqref{eq:master0} with 
the phase-space average  $\bar \delta_\text{NS}$, defined via Eq.~\eqref{eq:PSaverage}, up to ${\mathcal O}(\alpha^2)$ terms.

\section{Matrix elements in light nuclei}
\label{sec:light_nuclei}

Having derived the shape of RC corrections to superallowed $\beta$ decays, we now consider explicit transitions involving relatively light nuclei. We focus on three transitions: 
$^{6}\text{Li}(0^+) \rightarrow  {}^{6}\text{He}(0^+)$,
$^6\text{Be}(0^+)\rightarrow {}^6\text{Li}(0^+)$,
and 
$^{14}\text{O}(0^+)\rightarrow{}^{14}\text{N}(0^+)$.
The first two transitions do not happen in nature, 
but we can use them as a theoretical laboratory because the nuclear wave functions can be calculated to very high accuracy.
The decay of $^{14}$O is measured very accurately, 
with half-life of $t_{1/2} = 70619(11)\,\text{ms}$, 
branching fraction $\text{BR}= 99.446(13)\%$, and   $\mathcal Q_\text{EC} = 2831.543(76)\keV$~\cite{Hardy:2020qwl}, 
corresponding to a fractional uncertainties  below $1.6 \times 10^{-4}$.
$^{6}\text{Li}(0^+) \rightarrow {}^{6}\text{He}(0^+)$
is an example of 
a transition where the initial state has isospin $T_z=0$, 
while for the $^6$Be and $^{14}$O transitions 
the initial state has $T_z=-1$.\footnote{We adopt here the standard nomenclature in
the $0^+ \rightarrow 0^+$ literature, in which the proton has isospin $T_z = -1/2$
and the neutron $T_z=+1/2$, see, e.g., Ref.~\cite{Hardy:2020qwl}.
This is opposite to the more common nuclear and particle physics convention, in which the proton has $T_z = 1/2$.}

The nuclear wave functions are obtained using the variational Monte Carlo (VMC) method, described, for example, in Refs.~\cite{Carlson:2014vla,Lonardoni:2018nob}, with the next-to-next-to-leading-order (N$^2$LO) local chiral potential of Ref.~\cite{Gezerlis:2014zia}, and value of the cutoff $R_0 =1\fm$. We will perform a preliminary study here with just one set of wave functions, but a more comprehensive study should include wave functions obtained from different chiral potentials, cutoffs, and  variational methods. All isospin-breaking terms in the nuclear potential, including Coulomb, have been turned off. In this limit, the Fermi matrix element should be $M^{(0)}_\text{F} = \sqrt{2}$.
We obtain $M_\text{F}^{(0)}/\sqrt{2}=  1.0010(6) $ and $M_\text{F}^{(0)}/\sqrt{2} = 0.9990 (5)  $ for $A=6$ and $A=14$, respectively. The error corresponds to the statistical uncertainty of the VMC method.
In Sec.~\ref{sec:validation} we discuss the impact of including isospin breaking in the  nuclear potential and improving the nuclear wave function with the Auxiliary Field Diffusion Monte Carlo (AFDMC) method.

The potentials $\mathcal V_E$,
$\mathcal V^\pi_E$, $\mathcal V^\pi_{m_e}$, $\mathcal V_0^\text{mag}$, $\mathcal V_0^\text{CT}$,
and $\mathcal V_+$ are local potentials.
We will express a generic local EW potential as
a sum of a Fermi (F), Gamow--Teller (GT), and tensor (T) components
\begin{equation}
\label{V_O}
    \mathcal V_\mathcal O = \left(\frac{e^2}{4\pi} \right)^m \sum_{N = p,n} \left( \mathcal V^{\mathcal O} _{\text{F}, N}(\rr) + \mathcal V^{\mathcal O}_{\text{GT}, N}(\rr) + \mathcal V^{\mathcal O}_{\text{T}, N}(\rr) \right), 
\end{equation}
where we separated the contributions arising from couplings to neutrons and protons, and we have $m=1$ for the potentials in Sec.~\ref{sec:potentialsLO} and $m=2$ 
for $\mathcal V_+$. 

The F, GT, and T matrix components are defined as
\begin{align}
\label{V_O_F_GT_T}
    \mathcal V^{\mathcal O}_{\text{F}, N} &=  \sum_{j < k} h_\text{F}^{\mathcal O}(r_{jk}) \Big[ \tau^{+ (j)}  P^{(k)}_N  + (j \leftrightarrow k)  \Big], \\
    \mathcal V^{\mathcal O}_{\text{GT}, N} &=  \sum_{j < k} h_\text{GT}^{\mathcal O}(r_{jk}) \, \bsigma^{(j)} \cdot  \bsigma^{(k)} \Big[ \tau^{+ (j)}  P^{(k)}_N  + (j \leftrightarrow k)  \Big], \notag\\
    \mathcal V^{\mathcal O}_{\text{T}, N} &=  \sum_{j < k} h_\text{T}^{\mathcal O}(r_{jk}) S^{(ij)}({\bf \hat r}) \Big[ \tau^{+ (j)}  P^{(k)}_N  + (j \leftrightarrow k)  \Big],\notag
\end{align}
where $r_{jk}= |\rr_j - \rr_k|$ and \begin{equation}
S^{(ij)}(\hat \rr)=  3 \hat\rr \cdot \bsigma^{(i)} \, \hat\rr  \cdot \bsigma^{(j)} - \bsigma^{(i)} \cdot \bsigma^{(j)}.
\end{equation}
The radial functions $h$ for the $\mathcal O(\alpha)$
and $\mathcal O(\alpha^2)$
potentials are given in Apps.~\ref{app:coordinate} and~\ref{app:sub}.
Notice that all radial functions are defined to be dimensionless.
In the case of $\mathcal V_{E}$, $\mathcal V_{E}^\pi$, and $\mathcal V_{m_e}^{\pi}$ this is achieved by  introducing a factor of $R_A = 1.2 A^{1/3}\fm$.
We stress that $\bar{\delta}_\text{NS}$ does not depend on this choice.

Recoil corrections  induce non-local potentials, such as those given  in Eq.~\eqref{eq:RecPotential}. 
The evaluation of non-local potential is more time consuming. Since, as we will see, potentials induced by the pion mass splitting tend to yield smaller contributions to $\delta_\text{NS}$, 
we will focus in this study on the 
first term in Eq.~\eqref{eq:RecPotential}. This gives rise to a coupling of the spin and orbital angular momentum, which we denote by 
``spin-orbit'' (so) term. We write 
\begin{align}
\mathcal  V_\text{non-local} &= \frac{e^2}{4\pi} \mathcal V_\text{so}  \\ 
\mathcal V_\text{so}& = \sum_{j < k} h_\text{so}(r_{j k}) \Big[ \tau^{+ (j)} P_p^{(k)}  \LL_{jk} \cdot \boldsymbol{\sigma}^{(j)}  + (j \leftrightarrow k) \Big], \notag
\end{align}
where $\LL_{jk} = - i \rr_{jk}\times \left({\boldsymbol{\nabla}}_j - {\boldsymbol{\nabla}}_k\right)/2$
and the radial function is given in Eq.~\eqref{eq:radialso}.

In addition to the matrix elements, we 
will also show the 2b operator densities $C$. We define them through
\begin{align}
M^{\mathcal O}_{i, N} &=\int_0^\infty d r   \, C^{\mathcal O}_{i, N}(r) =    \langle f |  \mathcal V^{\mathcal O}_{i, N} | i \rangle,
\end{align}
where $i = \{\text{F} ,\text{GT}, \text{T}, \text{so}\}$.
With these definitions, the
$\mathcal O(\alpha \epsilon_\chi)$
and $\mathcal O(\alpha^2)$
corrections to $\delta_\text{NS}$ are given by
\begin{align}\label{deltaNS0}
    \delta^{(0)}_\text{NS}  &= \alpha \frac{2}{g_V(\mupi)M_\text{F}^{(0)}} \bigg[  \sum_{N = n,p} \big( M^\text{mag}_{\text{GT}, N} + M^\text{mag}_{\text{T}, N} +  M^\text{CT}_{\text{GT}, N} \big)  \notag \\ &   + M_\text{so}      \bigg] 
    + \alpha^2 \frac{2}{g_V(\mupi)M_\text{F}^{(0)}} M^+_{\text{F}, p}.
\end{align}
The evaluation of the CT matrix elements requires a choice for the numerical size of the LECs $g^{N\!N}_{V1}$ and $g^{N\!N}_{V2}$. We use RG-improved naive dimensional analysis, see the discussion surrounding Eq.~\eqref{V0CT}, for the linear combinations
\begin{equation}\label{eq:LECvalues}
 g_{V1}^{N\!N} \pm g_{V2}^{N\!N} = \frac{1}{\mN} \frac{1}{(2 F_\pi)^2},
\end{equation}
and we will treat their contributions as an uncertainty. 

\begin{figure*}[t]
\includegraphics[width=0.49\textwidth]{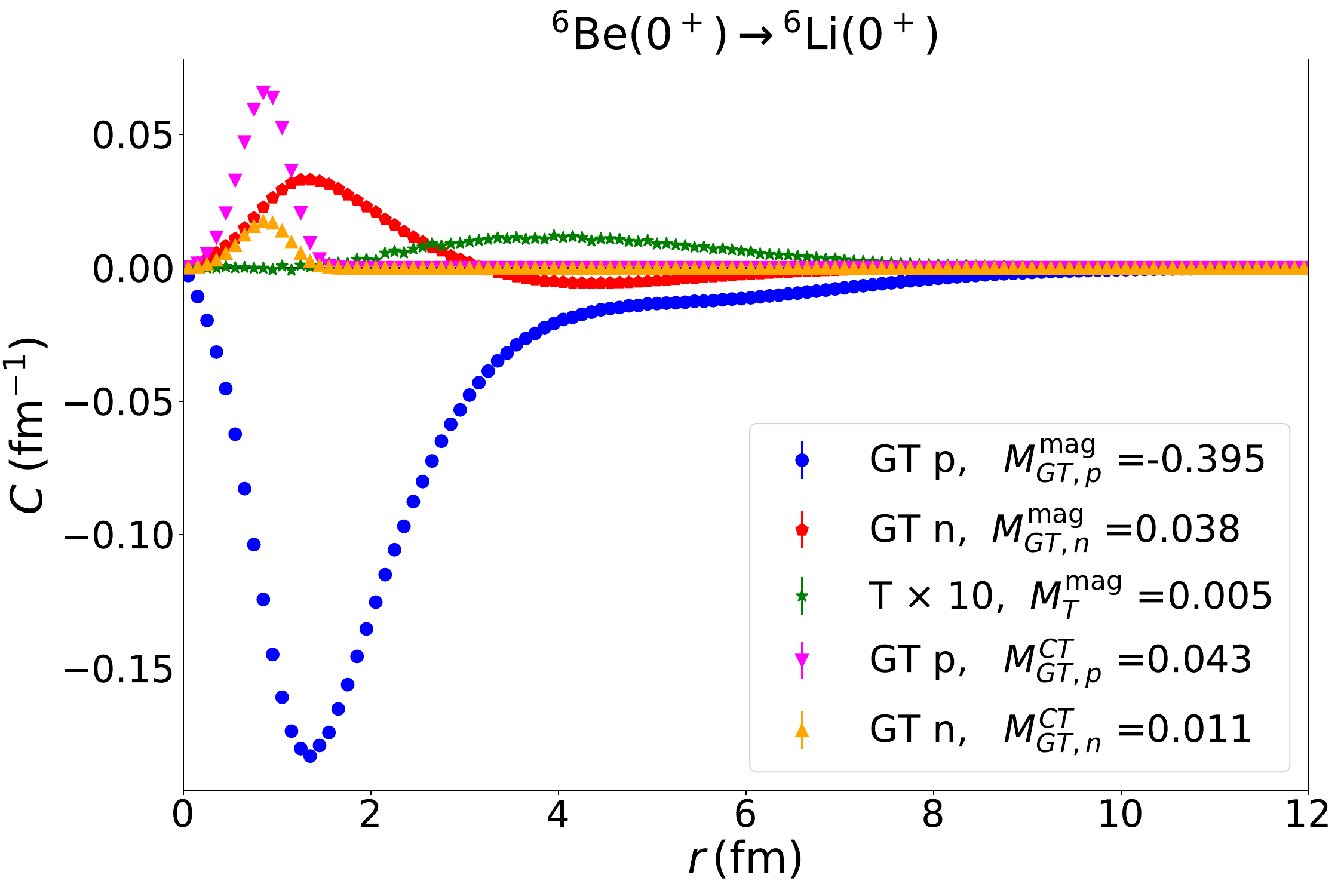}
\includegraphics[width=0.49\textwidth]{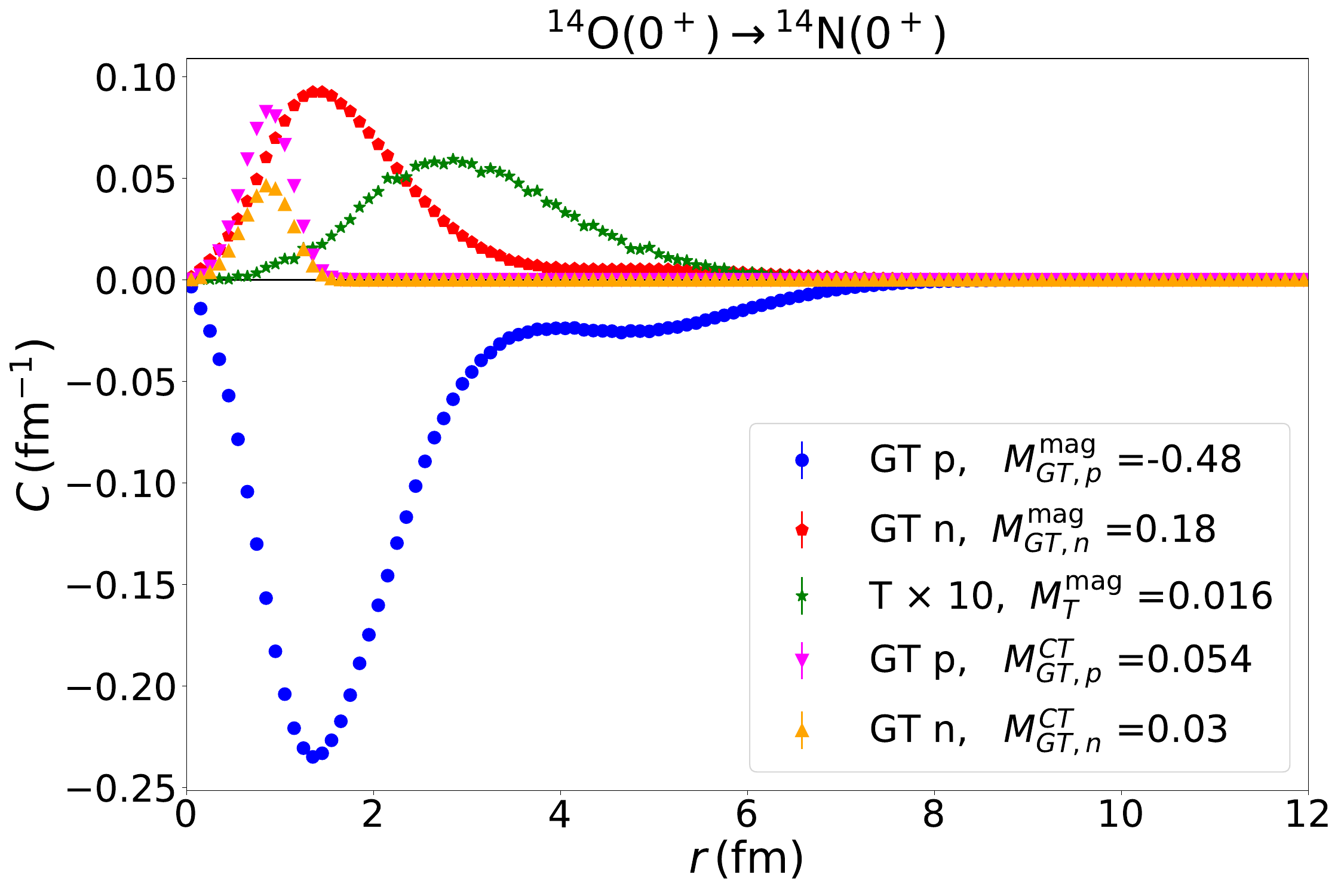}
\includegraphics[width=0.49\textwidth]{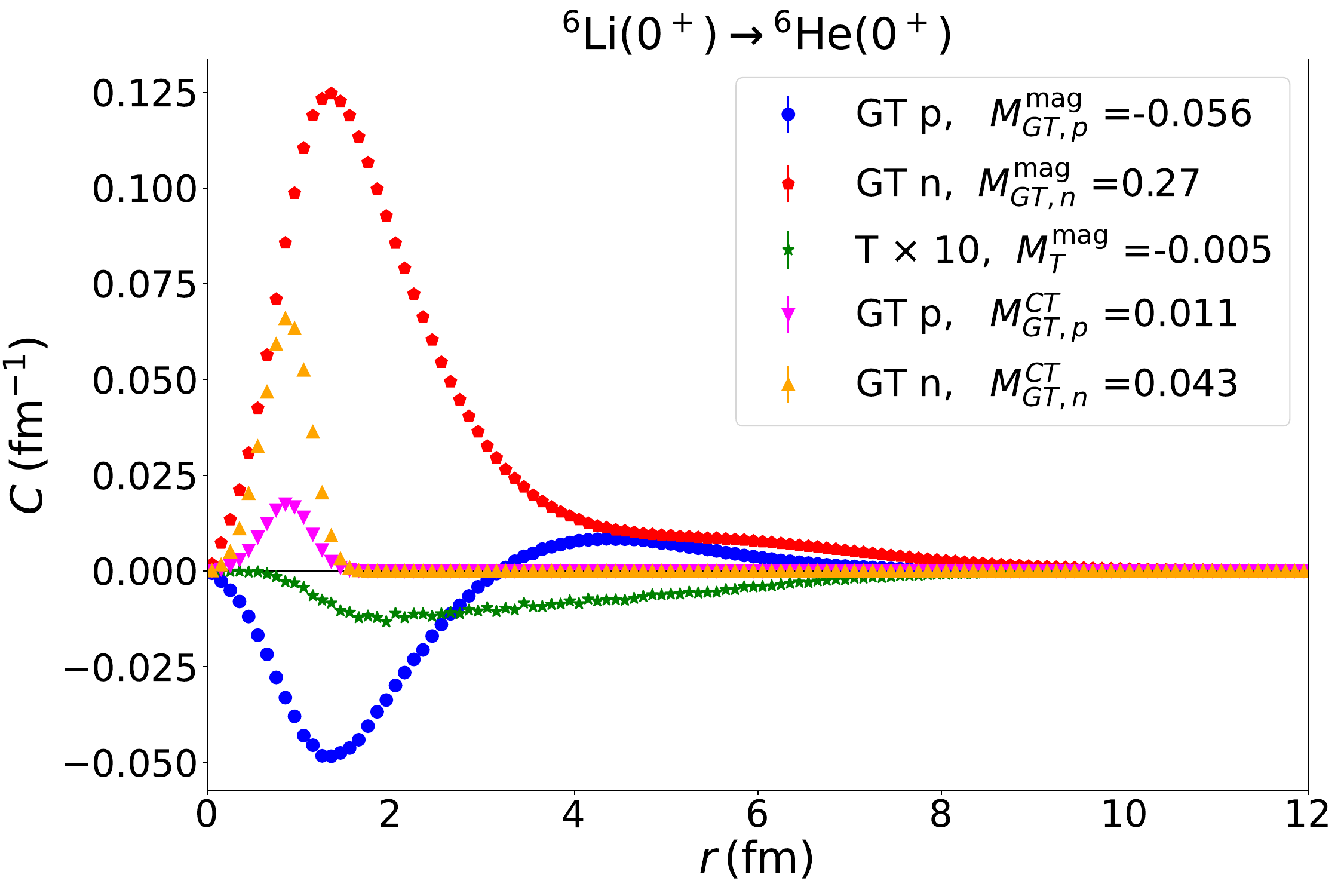}
\caption{
VMC magnetic and contact matrix elements for the 
$^6\text{Be} \rightarrow {}^{6}\text{Li}$,
$^{14}\text{O} \rightarrow {}^{14}\text{N}$,
and $^6\text{Li} \rightarrow {}^{6}\text{He}$
transitions. The tensor matrix element is multiplied by a factor of 10. Coulomb and isospin-breaking corrections in the nuclear potential have been turned off.
}\label{fig:local}
\end{figure*}

The energy-dependent $\mathcal O(\alpha \epsilon_{\slashed{\pi}})$ corrections are 
\begin{align}\label{deltaNSE}
\overline{\delta^E_\text{NS}} &= \alpha \frac{2}{g_V(\mupi)M_\text{F}^{(0)}} R_A E_0
\bigg[
\tilde{f}_E  M^E_{\text{F}, p} \notag \\ 
&  + 
\sum_{N=n,p} \left( M^{E\pi}_{\text{GT}, N}+ M^{E\pi}_{\text{T}, N}\right)\notag\\
&+
\tilde f_{m_e}^\pi
\sum_{N=n,p} \left( M^{m_e\pi}_{\text{GT}, N}+ M^{m_e\pi}_{\text{T}, N}\right)\bigg],
\end{align}
where the factors 
$\tilde f_E$
and $\tilde f_{m_e}^\pi$
arise from the phase-space average, see Eq.~\eqref{eq:PSaverage}, and are given by
\begin{align}
 \tilde{f}_E &= \frac{1}{E_0}   \left(\frac{4}{3}\langle E_e \rangle + \frac{1}{6} E_0 + \frac{1}{2}\left\langle \frac{m^2_e}{ E_e} \right\rangle \right),\notag\\
 \tilde{f}_{m_e}^\pi &=    \frac{1}{E_0}\left\langle \frac{m^2_e}{E_e} \right \rangle, 
\end{align}
with 
\beq
 \langle E_e^n \rangle = \frac{\int_{m_e}^{E_0} d E_e w_0(E_e)  
    \, \tilde C (E_e) \, 
 \bar F (\beta, \mu) \,  
    E_e^n }{\int_{m_e}^{E_0} d E_e  w_0 (E_e)\, \tilde C (E_e) \, 
 \bar F (\beta, \mu)  }~.
\label{eq:PSaverage2}
 \eeq 
For the $^{14}\text{O}\rightarrow{}^{14}\text{N}$ transition, 
the endpoint energy is
$E_0 = \mathcal Q_\text{EC} - m_e =  2320.544(76)\keV$, corresponding to $\tilde f_E = 0.95$, $\tilde f_{m_e}^\pi = 0.096$,
and $R_A E_0 = 3.4 \times 10^{-2}$.

\subsection{Monte Carlo methods}

In this work we use the VMC and the AFDMC techniques described in Refs.~\cite{Carlson:2014vla,Lonardoni:2018nob}. These methods have been extensively used to calculate diagonal matrix elements, i.e., observables of a given nuclear state. Here, for the first time, we extended those methods to calculate off-diagonal matrix elements.

We need to calculate matrix elements between different states, including
their normalization:
\begin{align} 
\langle M\rangle=\frac{\langle\Psi_f|O|\Psi_i\rangle}
{\sqrt{\langle\Psi_f|\Psi_f\rangle\langle\Psi_i|\Psi_i\rangle}}=
\frac{\langle\Psi_f|O|\Psi_i\rangle}
{\langle\Psi_i|\Psi_i\rangle}\sqrt{\frac{\langle\Psi_i|\Psi_i\rangle}{\langle\Psi_f|\Psi_f\rangle}}.
\end{align}

Let $\{W\}$ be a set of configurations (including the nucleons' positions and their spin and isospin amplitudes) that are obtained from VMC or AFDMC sampling, see Ref.~\cite{Lonardoni:2018nob} for details.
We can rewrite the above as:
\begin{align}
\frac{\langle\Psi_f|O|W\rangle\langle W|\Psi_i\rangle}
{\langle\Psi_i|W\rangle\langle W|\Psi_i\rangle}
\sqrt{\frac{\langle \Psi_i|W\rangle\langle W|\Psi_i\rangle}
{\langle \Psi_f|W\rangle\langle W|\Psi_f\rangle}}.
\end{align}

Within VMC, the configurations $\{W_i\}$ are sampled with probability $|\Psi_i|^2$.
The above can now be evaluated over the configurations as follows:
\begin{align}
\langle M\rangle=\sum_i\frac{\langle\Psi_f|O|W_i\rangle}{\langle\Psi_i|W_i\rangle}
\sqrt{\frac{1}{\sum_i\frac{|\langle\Psi_f|W_i\rangle|^2}{|\langle\Psi_i|W_i\rangle|^2}}}.
\label{eq:msum}
\end{align}

\begin{figure*}[t]
\includegraphics[width=0.49\linewidth]{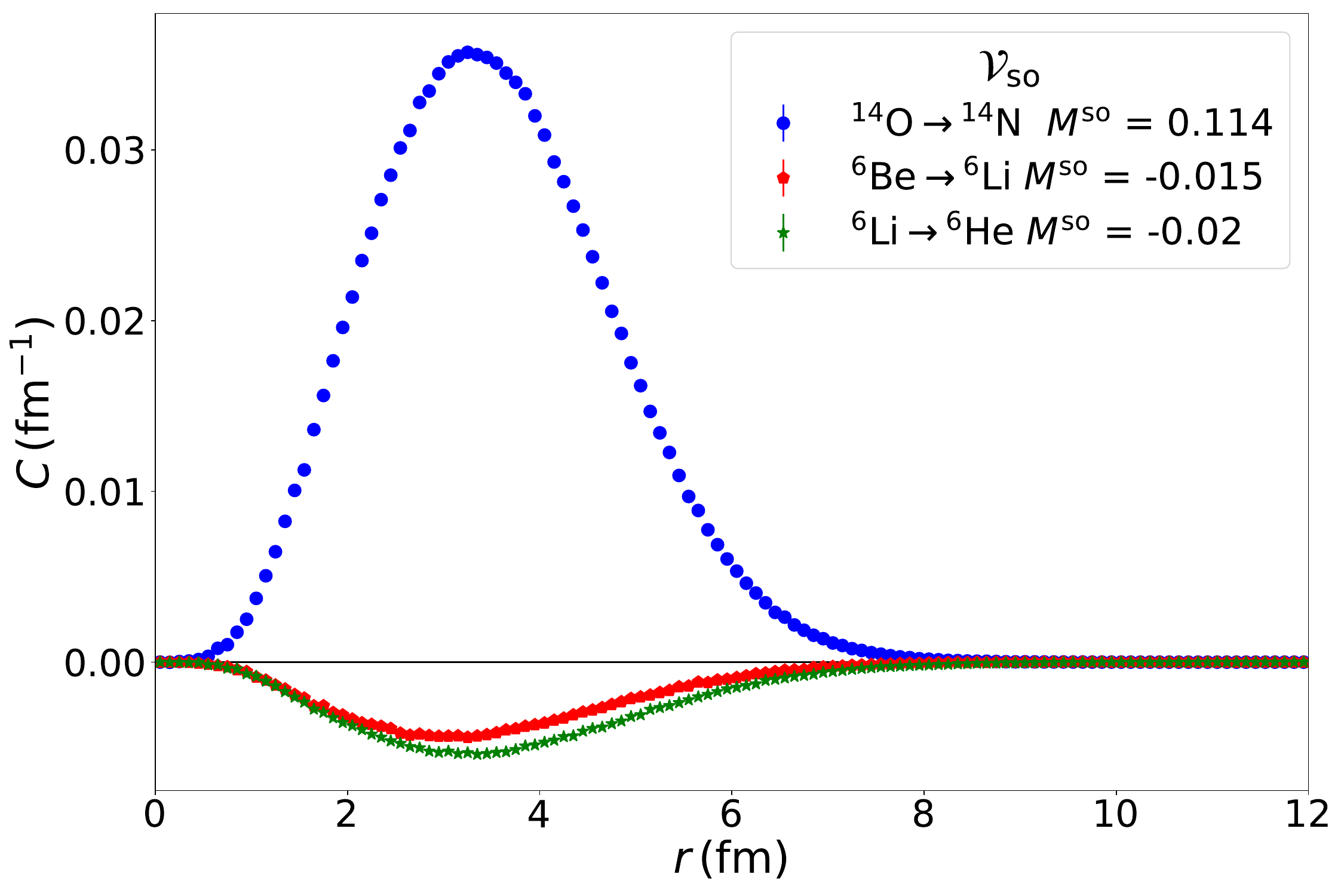}
\includegraphics[width=0.49\linewidth]{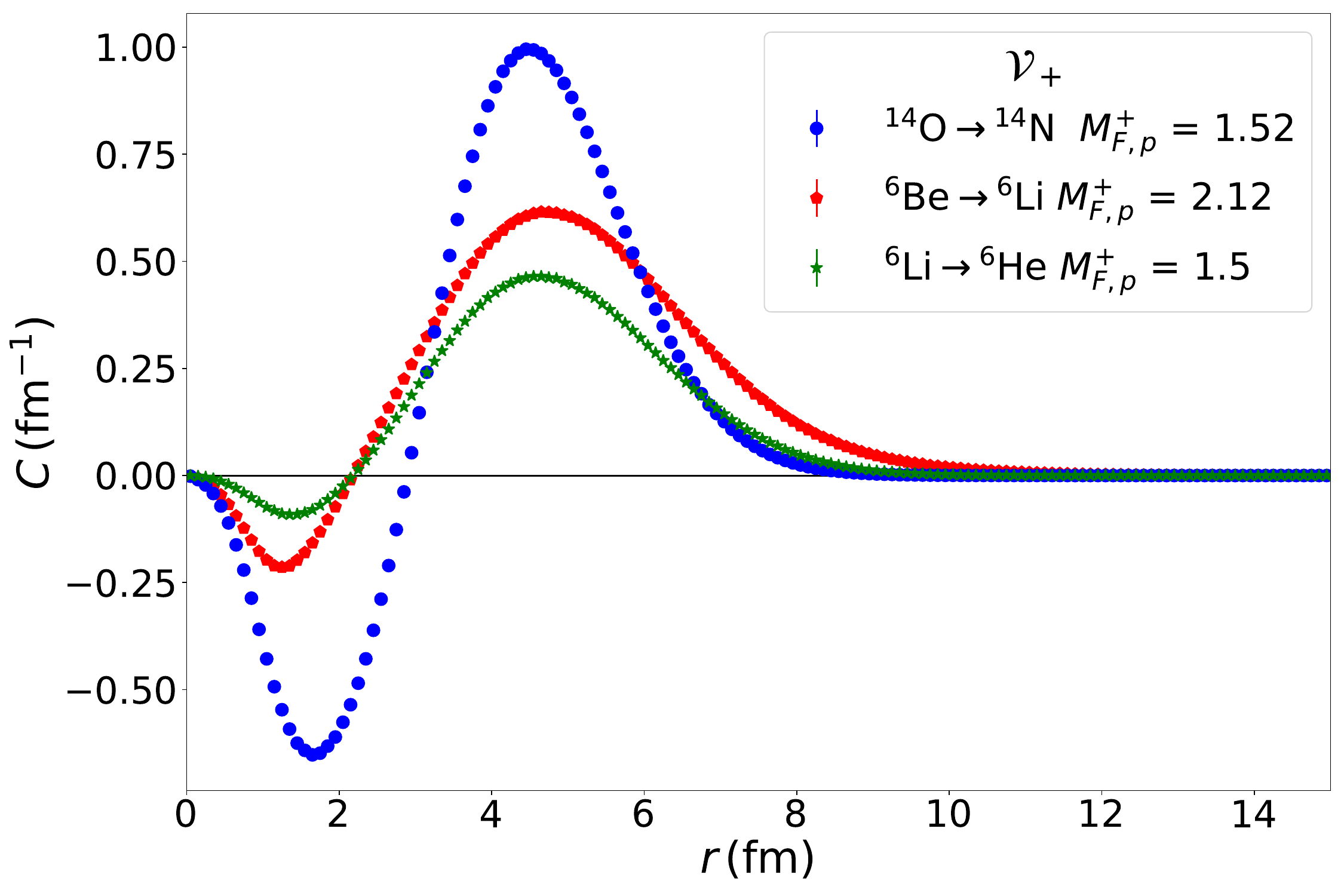}
\caption{
Left: spin-orbit density. Right: density
for the $\alpha^2$ potential, setting $\Lambda = \mu = R^{-1}_A$.}\label{fig:spinorbit}
\end{figure*}

Within AFDMC the matrix elements are obtained in a similar way, but the calculated observables are so-called ``mix,'' because the AFDMC propagation cannot be performed simultaneously for $\Psi_i$ and $\Psi_f$, but one of the two states is obtained within VMC. In practice, we perform three sets of calculations. $\langle M\rangle_v$ corresponds to the case with both initial and final wave functions obtained from VMC,
and $\langle M\rangle_{i(f)}$ to the ones in which the initial (final) wave function is obtained from AFDMC, respectively. 
The results referred to as AFDMC in this paper then amount to the extrapolation obtained by combining VMC and the mix calculations as:
\begin{align}
\langle M\rangle=\langle M\rangle_i+\langle M\rangle_f-\langle M\rangle_v ,
\end{align}
as described in Refs.~\cite{Pervin:2007sc,King:2020wmp}.

The AFDMC trial wave function we use takes the form:
\begin{align}
\label{eq:psi}
\langle RS|\Psi\rangle&=\langle RS|\prod_{i<j}f^1_{ij}\,\prod_{i<j<k}f^{3c}_{ijk}\\
&\times
\bigg[1+\sum_{i<j}\sum_{p=2}^6 f^p_{ij}\,\mathcal O_{ij}^p\, f_{ij}^{3p}
+\sum_{i<j<k}U_{ijk}\bigg]|\Phi\rangle_{J^\pi,T},\notag
\end{align}
where $|RS\rangle$ represents the sampled $3A$ spatial coordinates and the $4A$ spin/isospin amplitudes
for each nucleon, and the pair correlation functions $f^{p=1,6}_{ij}\equiv f^{p=1,6}(r_{ij})$ are obtained as the solution
of Schr\"odinger-like equations in the relative distance between two particles, as explained in Ref.~\cite{Carlson:2014vla}.

The term $|\Phi\rangle$ is taken as a shell-model-like wave function.
It consists of a sum of Slater determinants constructed using single-particle orbitals:
\begin{align}
\langle RS|\Phi\rangle_{J^\pi,T} = \sum_n c_n\Big[\sum \mathcal C_{J\!M}\,\mathcal D\big\{\phi_\alpha(r_i,s_i)\big\}_{J,M}\Big]_{J^\pi,T},
\end{align}
where $r_i$ are the spatial coordinates of the nucleons, and $s_i$ represents their spinor.
$J$ is the total angular momentum, $M$ its projection, $T$ the total isospin, and $\pi$ the parity.
The determinants $\mathcal D$ are coupled with Clebsch--Gordan coefficients $\mathcal C_{J\!M}$
in order to reproduce the experimental total angular momentum, total isospin, and parity $(J^\pi,T)$.
The $c_n$ are variational parameters multiplying different components having the same quantum numbers.
Each single-particle orbital $\phi_\alpha$ consists of a radial function multiplied by the spin/isospin trial states:
\begin{align}
\phi_\alpha(r_i,s_i)=\Phi_{nj}(r_i)\left[Y_{lm_l}(\hat{r}_i)\chi_\gamma(s_i)\right]_{j,m_j},
\label{eq:phi}
\end{align}
where the spherical harmonics $Y_{lm_l}(\hat{r}_i)$ are coupled to the spin state $\chi_\gamma(s_i)$
in order to have single-particle orbitals in the $j$ basis.
The radial parts $\Phi(r)$ are obtained from the bound-state solutions of the Woods--Saxon
wine-bottle potential:
\begin{align}
v(r)=V_s\left[\frac{1}{1+e^{(r-r_s)/a_s}}+\alpha_s\,e^{-(r/{\rho_s})^2}\right],
\end{align}
where the five parameters $V_s$, $r_s$, $a_s$, $\alpha_s$, and $\rho_s$ can be different for orbitals
belonging to different states, such as $1S_{1/2}$, $1P_{3/2}$, $1P_{1/2}$,\ldots, and they are
optimized in order to minimize the variational energy.
Details can be found in Ref.~\cite{Lonardoni:2018nob}.

It is important to note that the wave function essentially consists of three separate parts. The correlations, the shell-model components, and the single-particle orbitals.
If Coulomb interactions are neglected, it is possible to construct the wave function, for example, for $^6$Li by taking the $^6$He one and just flipping the isospin of one neutron. Likewise, the $^6$Be one can be obtained by taking $^6$He and flipping two neutrons into two protons. This will be what we call ``no-Coulomb.'' However, when  Coulomb interactions are included in the Hamiltonian, all the variational parameters mentioned earlier should be re-optimized in order to minimize the energy of the nucleus.

\subsection{Numerical results}

\begin{figure*}[t]
\includegraphics[width=0.49\linewidth]{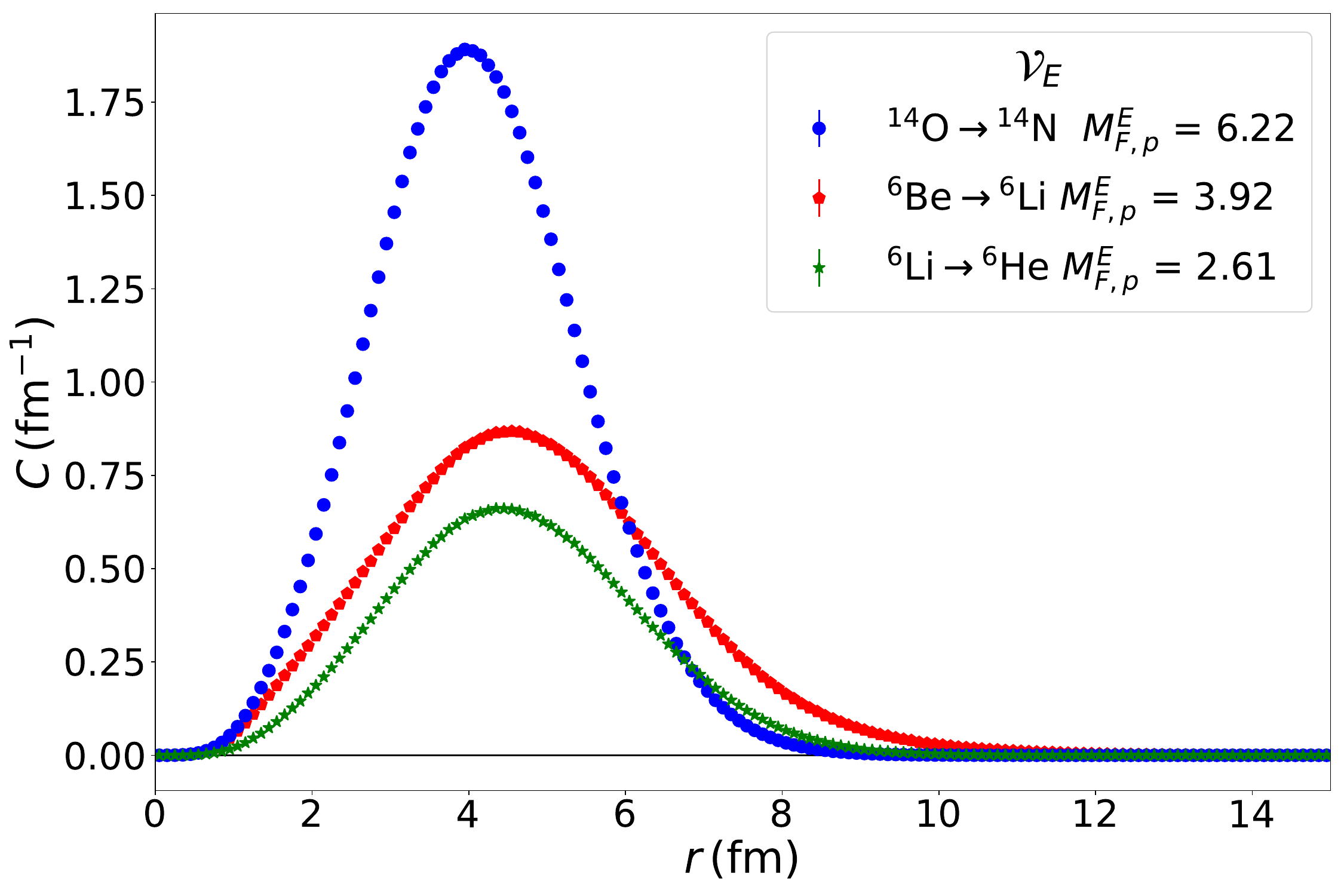}
\includegraphics[width=0.49\linewidth]{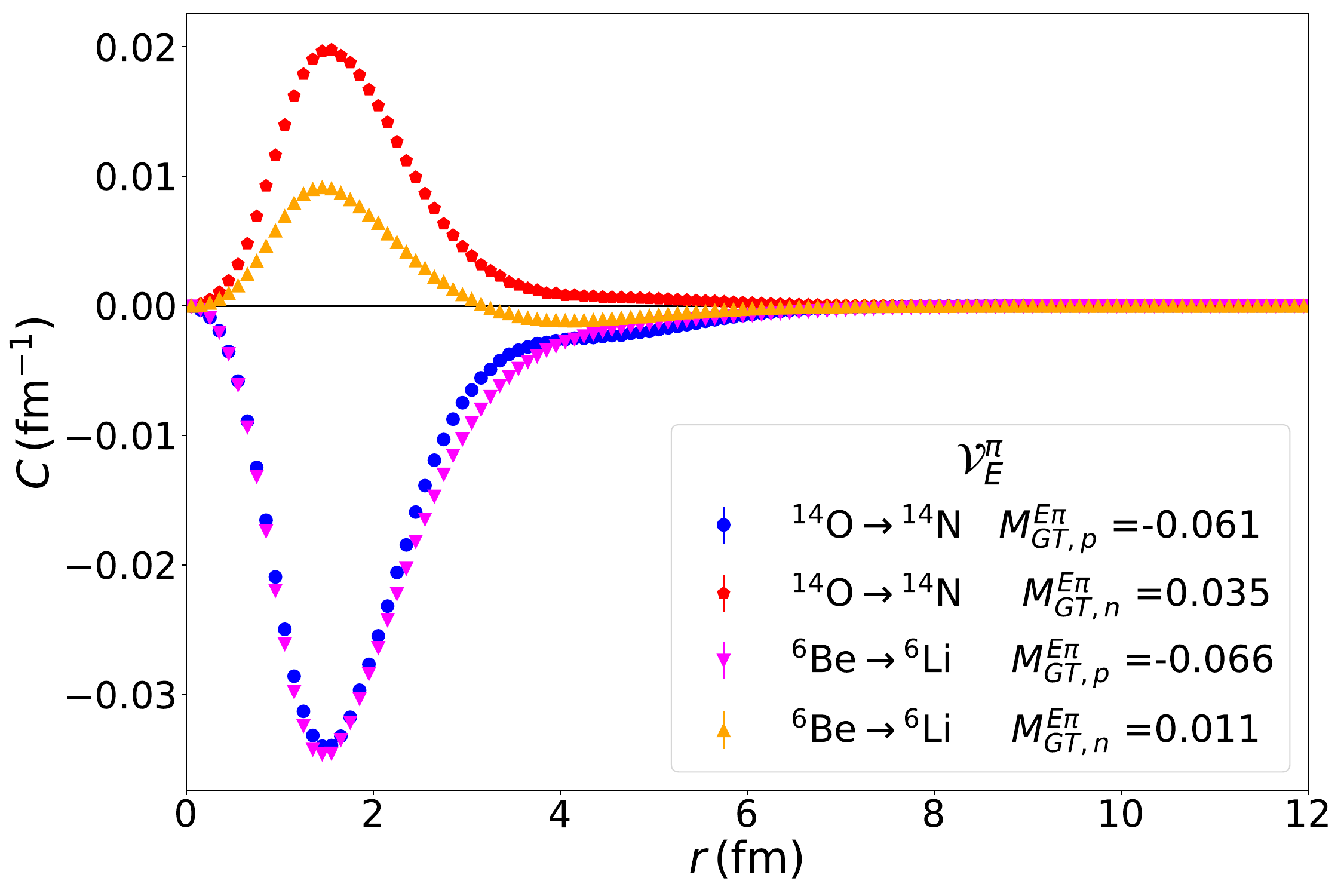}
\caption{Matrix elements of the energy-dependent potential $\mathcal V_E$ (left) and $\mathcal V^\pi_{E}$ (right).}\label{fig:energy}
\end{figure*}

In Fig.~\ref{fig:local}, we show the GT and T densities for $\mathcal V_0^\text{mag}$ and the GT densities\footnote{The Fermi and GT matrix elements for the short-distance operator are related through a Fierz relation: $M^{\text{CT}}_{\text{GT},N} = -3 M^{\text{CT}}_{\text{F},N}$.} for $\mathcal V_0^\text{CT}$ for the $A=6$ and $A=14$ transitions. The corresponding matrix elements are obtained by taking the integral of the densities, and the contributions to $\delta_{\text{NS}}^{(0)}$ are then obtained through Eq.~\eqref{deltaNS0} and are given in Table~\ref{tab:delta}.

The shape and relative importance of the different terms is very similar between the three transitions under consideration. The biggest difference is that for $T_z=-1$ transitions, the dominant contribution arises from the proton magnetic moment with a smaller component from the neutron magnetic moment. This behavior is opposite to that of the $T_z=0$ transition, where the neutron magnetic moment provides the biggest contribution. The tensor matrix element is very small for all three systems, which is also seen for neutrinoless double-$\beta$ decay. 

The short-distance densities always have the same sign, independent of the $T_z$ value. For our choice of LECs in Eq.~\eqref{eq:LECvalues}, the proton and neutron components add up and $\mathcal V_0^\text{CT}$ contributes at about 10\%  for ${}^6$Be, 25\% for ${}^{14}$O, and $20$\% for ${}^6$Li. We stress that this is just an estimate as it solely depends on the numerical values of the LECs in Eq.~\eqref{eq:LECvalues}. Replacing $2 F_\pi \rightarrow F_\pi$ would be as reasonable and would quadruple the short-distance effects leading to ${\mathcal O}(1)$ changes in $\delta^{(0)}_{\text{NS}}$. Clearly, the short-distance terms must be included in the analysis of RC corrections to superallowed $\beta$ decays.

In the left panel of  Fig.~\ref{fig:spinorbit}, we show the spin-orbit density. For $^{14}$O, it provides a $30$\% correction to the magnetic matrix elements. For the $A=6$ system, the spin-orbit contribution is smaller by a factor of five.
In the right panel of Fig.~\ref{fig:spinorbit} we show the matrix element of $\mathcal V_+$, setting the subtraction scale $\Lambda =\mu = R_A^{-1}$. While the peaks of the densities show a growing trend with $Z$, the matrix elements do not share this behavior. Numerically, the $\mathcal V_+$ contributions provide roughly $10\%$ corrections to the magnetic terms. 
We notice that $M^+_{\text{F}, p}$
can be set to zero for an appropriate choice of $\Lambda$, shifting this contribution to the matching coefficient $C_\text{eff}^{(g_V)}$. For $^{14}$O $\rightarrow$ $^{14}$N,  $M^+_{\text{F}, p}$
vanishes for $\Lambda^{-1} = 3.4\fm$, which, as we will see, is close to the scale set by the $^{14}$N charge radius.
 Since the matrix element of $\mathcal V_+$ is relatively small, we postpone the evaluation of the similar 3b potential $\mathcal V_+^\text{3b}$ to a future study.

\begin{table*}[t]
    \centering
    \renewcommand{\arraystretch}{1.3}
    \begin{tabular}{cccccccc}\toprule
$\delta^{(0)}_\text{NS}$         & $\mathcal V^\text{mag}_{\text{GT}, p}$ & $\mathcal V^\text{mag}_{\text{GT}, n}$
         & $\mathcal V^\text{mag}_\text{T}$ & $\mathcal V^\text{CT}_\text{GT,p}$& $\mathcal V^\text{CT}_\text{GT,n}$ & $\mathcal V_{\text{so}}$ & $\mathcal V_{+}$
\\
         \colrule 
      $^{6}$Be & $-4.07 \times 10^{-3}$& $0.40 \times 10^{-3}$ & $0.46 \times 10^{-4}$ & $4.44\times 10^{-4}$  & $1.17 \times 10^{-4}$    & $-1.57  \times 10^{-4}$ & $-1.60 \times 10^{-4} $           \\
      $^{14}$O    &  $-4.96 \times 10^{-3}$  & $ 1.86 \times 10^{-3}$ & $1.64 \times 10^{-4}$ & $5.61 \times 10^{-4}$ & $3.13 \times 10^{-4}$ & $+1.18 \times 10^{-3}$ & $-1.14 \times 10^{-4}$\\
      $^6$Li & $-0.58 \times 10^{-3}$ & $2.79 \times 10^{-3}$ & $-5.01 \times 10^{-5}$ & $1.12 \times 10^{-4}$ & $4.43 \times 10^{-4}$  & $-2.06 \times 10^{-4}$ & $-1.13 \times 10^{-4}$
      \\
      \colrule
$\overline{\delta^E_\text{NS}}$ & $\mathcal V^E_{\text{F}, p}$  & $\mathcal V^{E\pi}_{\text{GT}, p}$  & $\mathcal V^{E\pi}_{\text{GT}, n}$ & $\mathcal V^{m_e\pi}_{\text{GT}, p}$  & $\mathcal V^{m_e\pi}_{\text{GT}, n}$ & &\\ 
\colrule
 $^{14}$O  & $2.07 \times 10^{-3}$ & $-2.16 \times 10^{-5}$ & $1.22 \times 10^{-5} $ & $-7.65 \times 10^{-7}$ & $4.10 \times 10^{-7}$ & &\\
      \botrule
    \end{tabular}
    \caption{ 
 Contributions to $\delta_\text{NS}$
 from the EW potentials defined in Sec.~\ref{sec:potentials}.
 The top part of the table shows the energy-independent corrections induced 
 by the  GT and T components of $\mathcal V_0^\text{mag}$, by $\mathcal V_0^\text{CT}$, and by the spin-orbit term in $\mathcal V_0^\text{rec}$. The last column shows the  $\mathcal O(\alpha^2)$ correction from $\mathcal V_+$.
 The bottom part of the table shows the energy-dependent corrections from $\mathcal V_E$ and from the GT component of  $\mathcal V_E^\pi$. We neglect the T component of $\mathcal V_E^\pi$.}
    \label{tab:delta}
\end{table*}

We can now sum all contributions to $\delta^{(0)}_{\text{NS}}$ and compare to values obtained in the literature. Focusing on the $^{14}$O transition, we find
\begin{equation}
\delta^{(0)}_{\text{NS}}({}^{14}\text{O}) = -(1.76+0.11\pm 0.88)\times 10^{-3},
\label{eq:dNS0_14}
\end{equation}
where the first term encodes the magnetic and spin-orbit terms, the second is the $\mathcal O(\alpha^2)$ potential $\mathcal V_+$, and the  uncertainty is estimated from the short-distance contributions. Keeping in mind the caveats discussed in Sec.~\ref{sec:decay_rate}, it is still instructive to compare 
these contributions  to the results in Refs.~\cite{Towner:1992xm,Hardy:2020qwl}. $\delta^{(0)}_{\text{NS}}$ should correspond to $\delta_{\text{NS},B}$, which includes just the magnetic and spin-orbit terms, but not the short-distance effects nor $\mathcal V_+$. Reference~\cite{Hardy:2020qwl} quotes
\begin{equation}
\delta_{\text{NS},B}({}^{14}\text{O}) = -1.96(50)\times 10^{-3},
\end{equation}
whose central value is about 10\% larger than ours, if we neglect $\mathcal V_+$. This closeness is probably coincidental considering the rather different nuclear methods applied and the fact that the magnetic contributions depend on the applied regulator. That being said, this (qualitative) agreement is comforting. Numerically, the main difference lies in the short-distance contributions, which we have solely assigned to the overall uncertainty for now, leading to an error twice as large as in Ref.~\cite{Hardy:2020qwl}, but we stress again that this is based on Eq.~\eqref{eq:LECvalues}. 
 It will be crucial to pin down these contributions and we discuss strategies how to do so in Sec.~\ref{sec:LECs}.  

For the unphysical $^6\text{Be}\rightarrow{}^6\text{Li}$ and the $^6\text{Li}\rightarrow{}^{6}\text{He}$ transitions, we find
\begin{align}
    \delta^{(0)}_{\text{NS}}({}^{6}\text{Be}) = -(3.79 + 0.16 \pm 0.56)\times 10^{-3}, \notag\\
    \delta^{(0)}_{\text{NS}}({}^{6}\text{Li}) = +(1.95 - 0.11 \pm 0.56)\times 10^{-3}.
\end{align}
In addition to $\delta_{\text{NS},B}$, Ref.~\cite{Hardy:2020qwl} also includes the correction $\delta_{\text{NS},A}$, which in the EFT approach corresponds to diagrams further suppressed in the power counting, e.g., 3b corrections that lead to an apparent quenching of $g_A$. While their size seems to be roughly in line with the EFT expectation, this class of diagrams is largest among the omitted higher-order chiral corrections, and should be studied in future work, see Sec.~\ref{sec:conclusions}. In Ref.~\cite{Hardy:2020qwl}, $\delta_{\text{NS},A}$ is also estimated from quasi-elastic single-nucleon knockout processes, which in our approach would correspond to a weak axial and EM magnetic current acting on the same nucleon line, also entering at higher order in the power counting.

Next, we examine the energy-dependent potentials. 
In  Fig.~\ref{fig:energy} we show
the matrix element densities $C^E_{\text{F}, p}$ (left panel)
and $C^{E \pi}_{\text{GT}, N}$ (right panel),
corresponding to the potentials $\mathcal V_E$
and $\mathcal V_{E}^\pi$, for $A=6$ and $A=14$. We neglect the tensor potential.
In coordinate space, the radial function is $h^E_{\text{F},p}(r) = r/(2 R_A)$, so that $C^E_{\text{F},p}$ has significant support at large distances, $r \simeq (4\text{--}5)\fm$.
If one set $h^E_{\text{F},p}(r) = 1$, the  
integral of $C^E_{\text{F}, p}$ would simply count the protons in the final state. Even after restoring the $r$ dependence, we can see that the matrix element grows with $Z$ as there appear no nodes unlike in the $\mathcal V_+$ density.
The correction is sizable and gives rise to a contribution at the $10^{-3}$ level.

The matrix element $M^E_\text{F}$ is well approximated by replacing the radial function $h^E_{\text{F}, p}$with $\tilde h^{E}_{\text{F}, p}= R/(2 R_A)$ 
with $R = \sqrt{5/3} \sqrt{\langle r^2 \rangle}$, and $\sqrt{\langle r^2 \rangle}$ the charge radius of the daughter nucleus. For $^{14}$N, with $\sqrt{\langle r^2 \rangle}  = 2.558(7)\fm$ \cite{Angeli:2013epw}, we find
\begin{equation}
    \tilde{M}^E_{\text{F}, p} = \sqrt{2} \frac{Z R}{2 R_A}  = 5.56,
\end{equation}
which deviates from $M_E$ by 10\%.
For $^{14}$O, we can thus write the correction to $\delta_\text{NS}$ as    
\begin{align}\label{eq:dtENS}
    \overline{\delta^E_\text{NS}} &= \alpha Z R \left(\frac{4}{3}\langle E_e \rangle + \frac{1}{6} E_0   + \frac{1}{2}\left\langle \frac{m^2_e}{ E_e} \right\rangle \right) \notag \\ &
+ \alpha \frac{2}{M_\text{F}^{(0)}} R_A E_0 \tilde f_E \big(M^E_{\text{F}, p} - {\tilde M}^E_{\text{F}, p}\big),
\end{align}
where the term in the second line amounts to a correction of  $2.0 \times 10^{-4}$, significantly smaller than the first line.
The term in the first line has a dependence of $Z$ and $R$ that is similar to terms usually captured in the finite-nuclear-size corrections $L_0(Z,E_e)$
and in the shape correction $C(Z,E_e)$.
Using the analytic expressions from Refs.~\cite{Wilkinson:1993hx,Hayen:2017pwg}, one would find for the $\simeq \alpha R Z$ terms
\begin{align}
\label{L0C}
    &L_0(Z, E_e) C(Z, E_e) - 1 \notag\\
    &\supset \alpha Z R \bigg(\frac{48}{35} E_e + \frac{6}{35} E_0 + \frac{17}{35} \frac{m_e^2}{E_e} \bigg).
\end{align}
The numerical factors are very close to the ones in Eq.~\eqref{eq:dtENS}, indicating that the leading part of the EFT expression indeed captures similar physics. While the precise values of the coefficients in Refs.~\cite{Wilkinson:1993hx,Hayen:2017pwg} depend on the assumed charge distribution, see also Ref.~\cite{Behrens:1982}, the EFT allows one to systematically evaluate higher-order corrections. 
To avoid double counting with $\overline{\delta^E_\text{NS}}$, it is then necessary to subtract a set of $\mathcal O(\alpha Z R E_e)$ corrections to the shape factor. Our prescription is discussed in detail in App.~\ref{app:Cfactor}.

\begin{figure*}[t]
    \centering
    \includegraphics[width=0.49\textwidth]{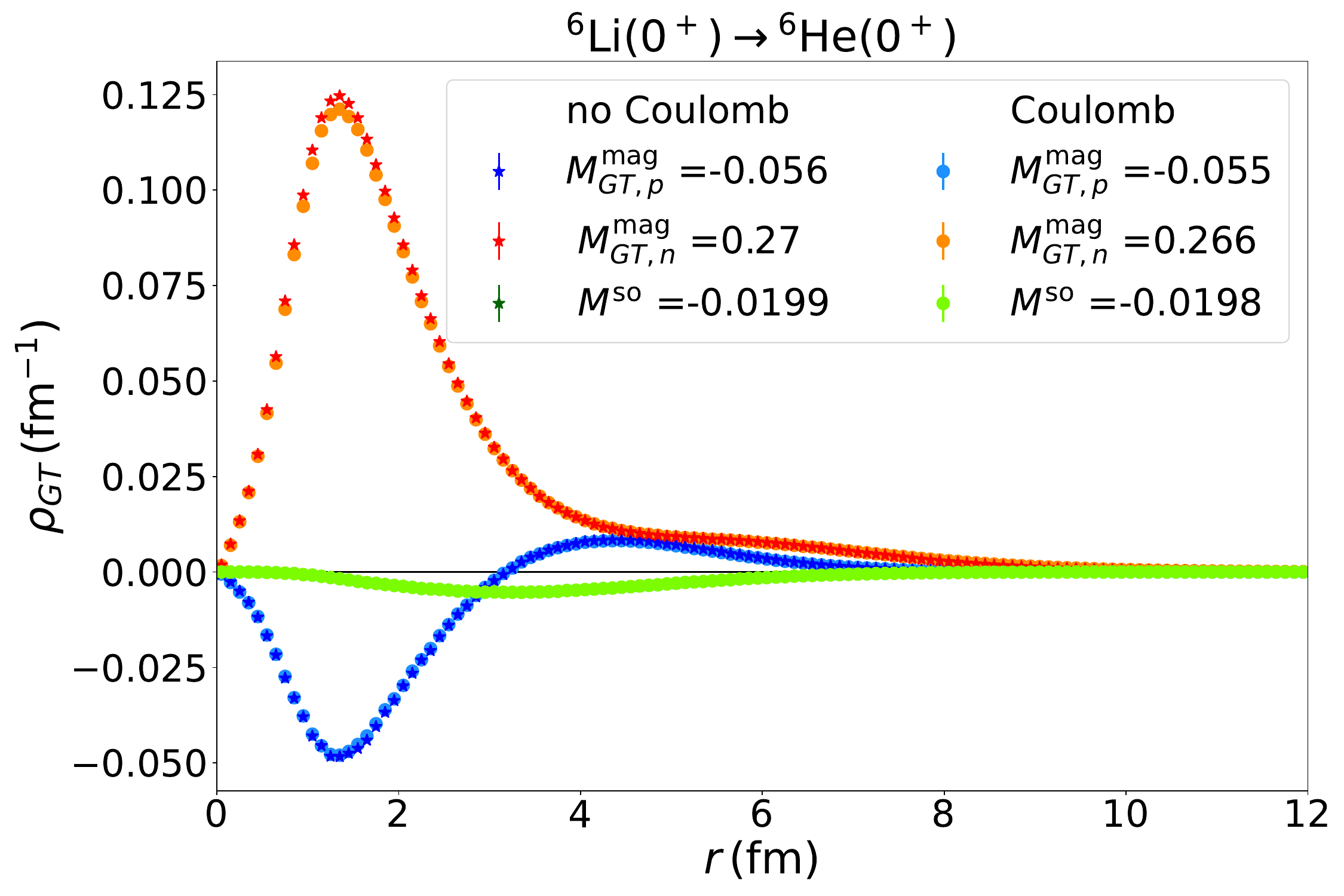}
        \includegraphics[width=0.49\textwidth]{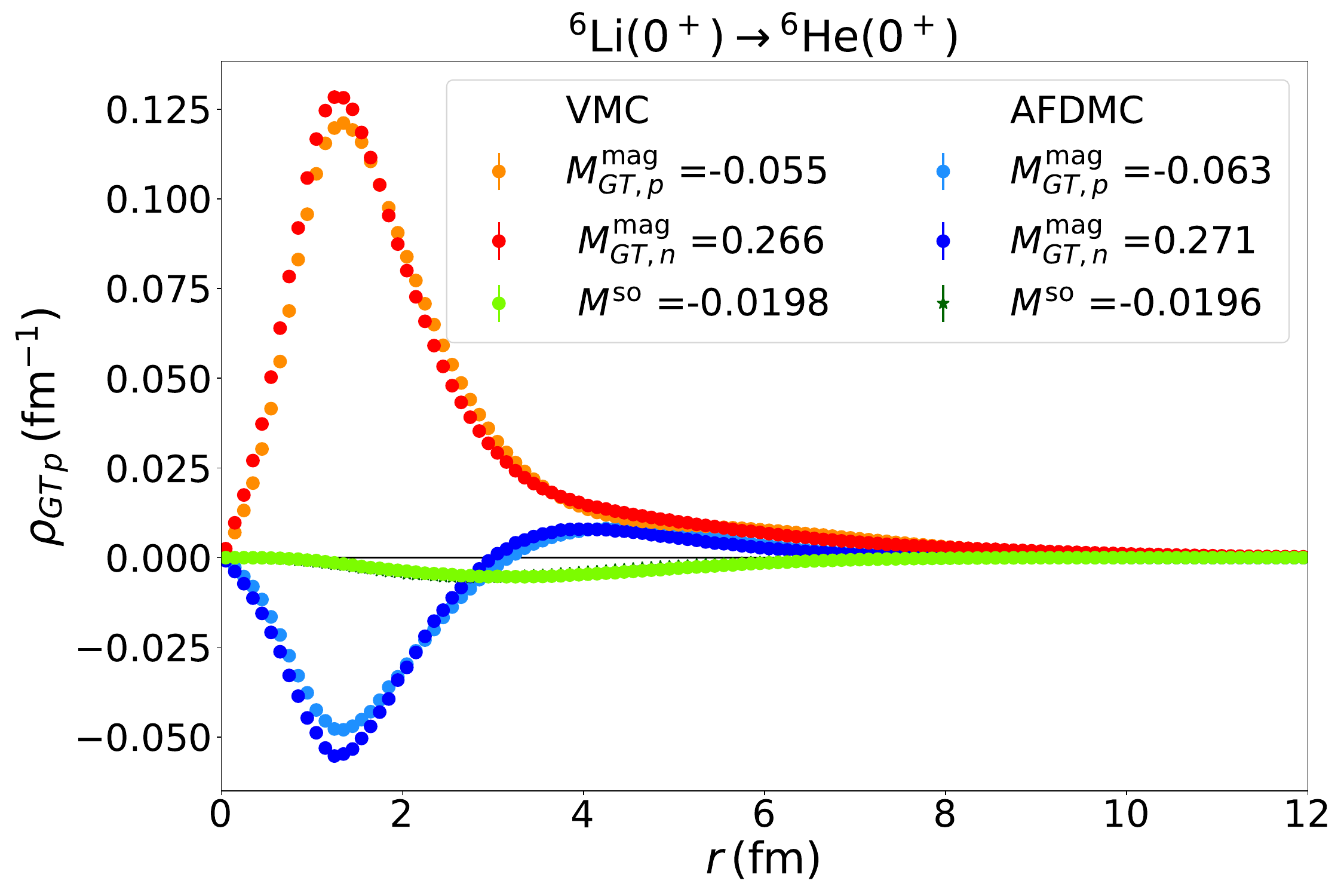}
    \caption{ Left: Impact on $\delta_\text{NS}$ matrix element from including the Coulomb potential in the nuclear Hamiltonian.
Right:  Comparison of VMC and AFDMC
matrix elements for $A=6$.}
    \label{fig:nuclear_checks}
\end{figure*}

In the right panel of Fig.~\ref{fig:energy}, we show the corrections induced by the pion mass splitting. Most of the support is in the region of $r \simeq \mpi^{-1}$, so that the overall size depends quite strongly on the behavior of the Fourier transforms~\eqref{VEpir} in this range. For instance, for ${\mathcal V}_{m_e}^\pi$ the GT radial wave function, proportional to $15 -21 \mpi r + \mpi^2 r^2$, has a zero crossing around $r \simeq 0.74 \mpi^{-1}$, suppressing the nuclear matrix element.\footnote{There is another zero crossing at $r \simeq 20 \mpi^{-1}$, but at these distances the radial functions are strongly suppressed due to the $e^{-\mpi r}$ dependence.} Accordingly, this contribution to 
 $\delta^E_ \text{NS}$ is at the $10^{-7}$ level and significantly smaller than anticipated from the power counting. 
In contrast, the GT wave function for ${\mathcal V}_{E}^\pi$ behaves as $12+12\mpi r-\mpi^2 r^2$, with zeros at $r\simeq -0.93\mpi^{-1}$ (and $r\simeq 13\mpi^{-1}$), which explains why this contribution does not suffer the same suppression due to an  accidental cancellation. Nevertheless, the total contribution is still small, at the $10^{-5}$ level, and an order of magnitude below our power-counting estimates. It remains to be seen whether this behavior persists in heavier nuclei.

Altogether we obtain 
\begin{equation}
\overline{\delta^{E}_{\text{NS}}}({}^{14}\text{O}) = 2.06(41)\times 10^{-3},
\end{equation}
where we assigned a $20\%$ uncertainty from higher-order chiral corrections.

\subsection{Validation of the Monte Carlo calculations}
\label{sec:validation}

A full analysis of the theoretical error on $\delta_\text{NS}$ requires using different nuclear Hamiltonians, cutoffs,
and many-body methods. 
We defer this important analysis to a future complete study. Here we validate the results discussed in the previous section in two ways.
First of all, we study the effect of restoring isospin-breaking components in the nuclear potential. In the left panel of Fig.~\ref{fig:nuclear_checks} we show the GT matrix elements $M^\text{mag}_{\text{GT}, N}$ for the $^6\text{Li}\rightarrow{}^6\text{He}$ transition, with and without turning on the Coulomb potential. We see that the effect of isospin-breaking interactions on the matrix elements is minimal. 
This gives us confidence that also the matrix element for $^{14}\text{O}\rightarrow{}^{14}\text{N}$ will be minimally affected by isospin breaking.
AFDMC uses the VMC wave functions as starting point, and, via an evolution in imaginary time, provides a more accurate description of the nuclear ground state~\cite{Carlson:2014vla}. Since AFDMC is computationally more demanding, especially for heavier nuclei, here we checked the impact of using AFDMC wave functions for the $^{6}$Li $\rightarrow$ $^6$He transition. The results are showed in the right panel of Fig.~\ref{fig:nuclear_checks}, where we compare the GT and spin-orbit matrix elements in VMC and AFDMC, including the Coulomb potential in both cases. 
Combining these results we find
\begin{equation}
    \delta^{(0)}_\text{NS} \Big|_\text{AFDMC}-   \delta^{(0)}_\text{NS} \Big|_\text{VMC} 
   = -2.9 \times 10^{-5}, 
\end{equation}
corresponding to a $1.5\%$ deviation. It will be important to confirm this behavior for $^{14}$O and larger nuclei.

\section{An explicit application: \texorpdfstring{$\boldsymbol{{}^{14}\text{O}\to{}^{14}\text{N}}$}{}}
\label{sec:example}

To illustrate how to use the EFT master formula in Eq.~\eqref{eq:master2} we now discuss in some detail the evaluation of the different ingredients for the $^{14}\text{O}\rightarrow{}^{14}\text{N}$ transition. For this particular decay we have an explicit computation of the nuclear-structure corrections in $\bar \delta_{\text{NS}}$, as provided in the previous section. 

We will now explicitly evaluate the various terms in Eq.~\eqref{eq:master2}. The experimental result for the lifetime is given in Ref.~\cite{Hardy:2020qwl} (using input for half-lives from Refs.~\cite{Alburger:1972zzc,Clark:1973rln,Azuelos:1974zz,Wilkinson:1978zz,Becker:1978,Gaelens:2001,Barker:2004tw,Burke:2006ze,Takau:2012,Laffoley:2013kma}, for branching fractions from Refs.~\cite{Kavanagh:1969nty,Wilson:1980zz,Hernandez:1981zz,Voytas:2015jlu}, and the electron-capture correction $P_\text{EC}=8.8\times 10^{-4}$),
\begin{equation}
    t = 71075(15) \times 10^{-3}\,\text{s},
\end{equation}
with an uncertainty of $0.02\%$, while the uncertainty in the prefactor~\cite{ParticleDataGroup:2022pth,MuLan:2012sih} 
\begin{equation}
    \frac{G_F^2 m_e^5}{\pi^3 \log 2} = 3.350722(3) \times 10^{-4}\,\text{s}^{-1}
\end{equation}
can be neglected.  

The next step involves $C^{(g_V)}_\text{eff}(\mu)$.
$C^{(g_V)}_\text{eff}(\mu)$ depends on the matching scale at which potentials and soft modes are integrated out, and on the low-energy scale $\mue$ at which we stop the RG evolution, see
Eq.~\eqref{eq:usoftU}. 
We evaluate 
$C^{g_V}_\text{eff}$ at three 
low-energy scales
\begin{equation}
\mue=\{E_0,\,2E_0,\,4E_0\},
\end{equation}
where for ${}^{14}$O \cite{Hardy:2020qwl,Valverde:2015bba,Ajzenberg-Selove:1991rsl,Koslowsky:1987zrb}
\begin{equation}
\qquad E_0 =2320.544(76)\keV.
\end{equation}
We will take
the spread in our final answer due to the variation of $\mue$ as an estimate of the uncertainty due to missing $\mathcal O(\alpha^2 Z)$ terms in the ultrasoft matrix element. 
For the matching scale, as discussed below Eq.~\eqref{eq:FF}, we set
\begin{equation}
    \mupi = R^{-1} \exp\left(\frac{1}{2} - \gamma_E\right) = 55.3 \MeV.
\end{equation}
To be consistent with the evaluation of $\delta_{\text{NS}}$ in the previous section we have to set $\Lambda = R_A^{-1}$.
From  Eq.~\eqref{eq:bc} we then obtain
\begin{equation}
    C^{(g_V)}_\text{eff}(\mupi) = 1.00060 \, g_V(\mupi) = 1.01721(12),
\end{equation}
and by solving the RG equations in Eqs.~\eqref{eq:CeffRG} and~\eqref{eq:usoftU} 
\begin{align}
    C_\text{eff}^{g_V} (\mue) &= \left\{1.01100, 1.00873, 1.00645 \right\} g_V(\mupi) \nonumber\\
    &= \left\{ 1.02778, 1.02547, 1.02315 \right\}.
 \end{align}
We obtained $g_V(\mupi)$ by evolving the value
at $\mu = M_{\pi^+}$ with the kernel given in Eq.~\eqref{eq:kernel2}
 \begin{equation}
    g_V(\mupi)= 1.01659(12).
\end{equation}
The error, which is dominated by the
non-perturbative contribution
$\overline{\Box}_\text{had}^V (\mu_0)$, is approximately scale independent.
We investigated the dependence on the matching scale by
varying it between $\mupi/2$
and $2\mupi$, and found a negligible change, of order $10^{-5}$.
We use the fine-structure constant in the $\overline{\text{MS}}_\chi$ scheme, defined in App.~\ref{App:RGE}, which gives
\begin{equation}
\alpha^{-1}(\mupi) = 136.145.
\end{equation}

The next term is $\bar{\delta}_R^\prime$, which is evaluated as in Eq.~\eqref{eq:drp} and then averaged through Eq.~\eqref{eq:PSaverage}. For ${}^{14}$O this procedure leads to
 \begin{equation}
\bar{\delta}_R^\prime(\mue)
= \left\{ -6.61712, -4.20287, -1.78341\right\} \times 10^{-3}.
\end{equation}
Before discussing the nuclear-structure corrections, we first address the phase-space factor in Eq.~\eqref{eq:PS}. This is rather complicated due to the factor $\tilde C(E_e)$, which includes various corrections not discussed in this work, while, at the same time, including corrections that do overlap with parts of our $\bar\delta_{\text{NS}}$ and therefore need to be separated. We present our procedure in detail in App.~\ref{app:Cfactor} and here give our numerical result 
\begin{equation}
    \bar f(\mue) = \{ 42.3632, 42.4318, 42.5009 \}.
\end{equation}
Combining the scale-dependent quantities then leads to 
\begin{equation}\label{combscale0}
   \left[C_\text{eff}^{(g_V)}\right]^2\bar f (1 + \bar\delta_R^\prime)= \left\{44.453, 44.433, 44.412 \right\}.
\end{equation}
 We see that the residual dependence on $\mue$ induces an uncertainty of the order $\simeq 2 \times 10^{-4}$.
We stress that this is much smaller than the scale variations of the individual pieces that make up Eq.~\eqref{combscale0}. The remaining scale dependence is dominated by the missing $\mathcal O(\alpha^2 Z)$ corrections in the amplitude at $\mu \simeq \mue$, and is thus expected within the EFT.
As key result, we obtain the estimate for the combination of phase-space, ``inner,'' and ``outer'' corrections
\begin{equation}\label{combscale}
   \left[C_\text{eff}^{(g_V)}\right]^2\bar f (1 + \bar\delta_R^\prime)= 44.433(11)_{g_V}(20)_{\mu},
\end{equation}
where we separated the uncertainty from $g_V(\mupi)$ and varying $\mue$. 

Finally, for the remaining corrections we use
$\bar \delta_C=\delta_C = 3.30(25) \times 10^{-3}$~\cite{Hardy:2020qwl} and 
\begin{align}
\label{eq:dNS0}
\bar \delta_\text{NS}&= \overline{\delta_\text{NS}^{(0)}}   +  \overline{{\delta^E_\text{NS}}},\\
\overline{\delta^{(0)}_{\text{NS}}} &= -1.87(88)\times 10^{-3},\qquad 
\overline{\delta^{E}_{\text{NS}}} = 2.06 (41)\times 10^{-3},\notag
\end{align}
as obtained in the previous section. 
Inserting everything into Eq.~\eqref{eq:master2}, we can extract the CKM element
 \begin{align}
    V_{ud} &= 0.97364(10)_\text{exp}(12)_{g_V}(22)_{\mu}(12)_{\delta_C}(43)_{g_V^{N\!N}}(20)_{\delta^E_{\text{NS}}}\notag\\&=0.97364(56)_{\text{total}},
    \label{Vud14O}
\end{align}   
with a total uncertainty of $0.06\%$. This uncertainty is dominated by the unknown LECs appearing in $\overline{\delta_{\text{NS}}^{(0)}}$, and would reduce to $\Delta V_{ud}=3.6\times 10^{-4}$ if this error could be eliminated. Among the remaining uncertainties, the experimental one is still subleading, at the level of $\Delta V_{ud}=1.0\times 10^{-4}$. 

It is instructive to compare our results with other determinations in the literature. Equation~\eqref{Vud14O} is consistent with the determination from neutron decay~\cite{Cirigliano:2023fnz} 
\beq
V_{ud}^\text{neutron}= 0.97402(42),
\eeq
where we quoted the variant based on Refs.~\cite{UCNt:2021pcg,Markisch:2018ndu} for lifetime and asymmetry, respectively, while PDG averages including the scale factor almost double the (experiment-dominated) uncertainty. Equation~\eqref{Vud14O} is also consistent with the global survey from Ref.~\cite{Hardy:2020qwl}
\begin{equation}
\label{Vud:HT}
    V_{ud}^\text{\cite{Hardy:2020qwl}}= 0.97373(31),
\end{equation}
but for a more detailed comparison we concentrate on the $^{14}\text{O}\to{}^{14}\text{N}$ transition alone. In this case, Ref.~\cite{Hardy:2020qwl} quotes the different error components of the $\mathcal F t$ value as
\beq
\label{FT14O}
\mathcal Ft=3070.2(0.8)_\text{exp}(2.0)_{\delta_\text{NS}}(0.8)_{\delta_C}[2.3]_\text{total}\,\text{s},
\eeq
where we added all uncertainties in quadrature (the experimental error being derived from the $ft$ value), resulting in a slightly larger error than quoted in Ref.~\cite{Hardy:2020qwl}, $\mathcal Ft=3070.2(1.9)\,\text{s}$.\footnote{On the other hand, the uncertainty in Eq.~\eqref{Vud14OTH} is slightly underestimated, because $\delta_R'$ is only included in the error analysis of the global fit in Ref.~\cite{Hardy:2020qwl}, so that the overall comparison should be realistic.}  From the breakdown in Eq.~\eqref{FT14O}, one obtains
\beq
\label{Vud14OTH}
V_{ud}^\text{\cite{Hardy:2020qwl}}\big[{}^{14}\text{O}\big]=0.97405(13)_\text{exp}(9)_{\Delta_R^V}(12)_{\delta_C}(31)_{\delta_\text{NS}}[37]_\text{total},
\eeq
where the total error is close to the full analysis~\eqref{Vud:HT} because it is dominated by the systematic uncertainty in $\delta_\text{NS}$. 
In our analysis~\eqref{Vud14O}, we find a lower central value, albeit consistent within uncertainties. The experimental error is close, as is the uncertainty propagated from the single-nucleon hadronic matrix elements, contained in $g_V$ and $\Delta_R^V$, respectively. The uncertainties on $\delta_C$ are identical by construction, so that the main difference originates from the effects represented by $\delta_\text{NS}$ and $\bar\delta_\text{NS}$, see the discussion in Sec.~\ref{sec:comparison}. Here, the EFT allows one to separate uncertainties related to RG corrections, labeled by $\mu$ in Eq.~\eqref{Vud14O}, from the genuine uncertainties of the matrix elements, and therein higher-order corrections from LECs. 

In particular, from this breakdown there is a clear path towards establishing $V_{ud}$ at a similar level as quoted in Eq.~\eqref{Vud:HT} once the LECs are determined following the strategies outlined in the subsequent section. In view of the error analysis presented here for $^{14}\text{O}\to{}^{14}\text{N}$, a few light transitions together with the corresponding nuclear-structure calculations should suffice to obtain a competitive determination of $V_{ud}$, including a robust estimate of the nuclear-structure uncertainties.

\section{Determination of the low-energy constants}
\label{sec:LECs}

A key finding of our EFT analysis of superallowed $\beta$ decays is that at the required level of precision contributions from contact terms have to be included, renormalizing  ${\mathcal O}(G_F\alpha\eps_\chi)$ potential-photon corrections. The two associated LECs $g^{N\!N}_{V1,V2}$, see Eq.~\eqref{eq:2nuc1}, encode effects of hard photons that are not predicted by symmetry arguments, and their values thus have to be determined by other means. This situation is similar to neutrinoless double-$\beta$ decay~\cite{Cirigliano:2018hja,Cirigliano:2019vdj}, ultimately tracing back to potential matrix elements evaluated with chiral $^1S_0$ wave functions~\cite{PavonValderrama:2014zeq}, but the crucial difference is that for superallowed $\beta$ decays a purely data-driven strategy to determine the LECs is possible. 

That is, while a reduction of the number of LECs using large-$N_c$ arguments might be possible~\cite{Richardson:2021xiu} and independent theoretical determinations using lattice QCD or a Cottingham-like approach~\cite{Cirigliano:2020dmx,Cirigliano:2021qko} could be envisioned, the contact terms can also be determined from a global fit to measured superallowed transitions, with $V_{ud}$ and $g_{V1,V2}^{N\!N}$ as degrees of freedom. Given that there are ${\mathcal O}(10)$ precisely measured decays, such a simultaneous extraction is feasible if the matrix elements for systems with  different $A$ are not degenerate and if their errors can be quantified in a reliable manner. The latter should be possible for a wide range of targets with modern ab-initio nuclear-structure techniques. The results we presented in Sec.~\ref{sec:light_nuclei} make it appear   unlikely that degeneracies in the $A$ dependence of the matrix elements occur. 

As further refinements of such a data-driven strategy to extract the LECs simultaneously with $V_{ud}$ one may consider decays with initial  
$m_I=-1$ or $m_I=0$ separately. The LECs appear in the linear combinations
$g_{V1}^{N\!N} \langle f||O_1||i\rangle \mp \sqrt{3/5} g_{V2}^{N\!N} \langle f||O_2||i\rangle$, respectively, where 
$\langle f||O_{1,2}||i\rangle$ are the reduced matrix elements of the operators from Eq.~\eqref{eq:2nuc1}. 
If the reduced matrix elements were (approximately) proportional for different isotopes, 
the combined analysis of systems with the same $m_I$ would be (primarily) sensitive to a single unknown together with $V_{ud}$, and the comparison of determinations from $m_I=-1$ and $m_I=0$ decays could then be taken as a consistency check of the $V_{ud}$ determination from superallowed $\beta$ decays. It remains to be seen how the reduced matrix elements behave, empirically, for the nuclei in question.

\section{Conclusions and outlook} 
\label{sec:conclusions}

In this work we provided the details of a comprehensive EFT analysis of superallowed $\beta$ decays~\cite{Cirigliano:2024rfk}, spanning scales that range from EW physics down to nuclear transitions. In particular, 
we identified the set of contributions that needs to be included at ${\mathcal O}(10^{-4})$ precision relevant for a competitive determination of $V_{ud}$, see classes~\ref{ultrasoft_modes}--\ref{alpha2_corrections} in Sec.~\ref{sec:PC}, finding that the nuclear-structure-dependent terms usually represented by 
$\delta_\text{NS}$ can be expressed as potential
matrix elements of EW transition operators evaluated between initial and final nuclear wave functions. 
We provided a detailed  account of these potential corrections in Sec.~\ref{sec:potentials}. Among the identified corrections are terms such as the magnetic and recoil potentials already present in the literature~\cite{Towner:1992xm},
while others have not been considered in the past. Most importantly, the EFT allows for a systematic evaluation of all contributions, including effects from pion exchange, and predicts that renormalization requires the consideration of short-range operators at the same order. 

We also provided a detailed account of ultrasoft modes, see Sec.~\ref{sec:ultrasoft}, as well as a careful consideration of scale dependence and RG corrections. The understanding of ultrasoft photons is critical to be able to map our results onto the traditional decomposition of the decay rate and justify factorization assumptions, see Sec.~\ref{sec:decay_rate}. This mapping,  together with the appropriate caveats, is summarized in Table~\ref{tab:comparison}.
Moreover, ultrasoft contributions play a prominent role in the comparison to a dispersive approach 
for $\delta_\text{NS}$~\cite{Gorchtein:2018fxl,Seng:2022cnq}, and we demonstrated how the EFT scaling applies in the presence of low-lying states, such as 
the $3^+$ and $1^+$ levels of $^{10}$B  in the $^{10}\text{C}\to{}^{10}\text{B}$ transition. We showed that individual terms can display an enhancement by $\sqrt{\eps_\text{recoil}}$, but the total effect should comport with the EFT expectation.  Finally, we confirmed the EFT power counting with VMC calculations of $^{6}\text{Be}\to{}^{6}\text{Li}$,
$^{6}\text{Li}\to{}^{6}\text{He}$,
and $^{14}\text{O}\to{}^{14}\text{N}$ transitions, see Sec.~\ref{sec:light_nuclei}, and outlined a data-driven strategy to determine the coefficients of the ${\mathcal O}(G_F\alpha\epsilon_\chi)$ contact operators, see Sec.~\ref{sec:LECs}.  

Combined with advances in ab-initio nuclear-structure calculations to evaluate the nuclear matrix elements identified in this work with quantified uncertainties, our framework should allow one to systematically address the dominant uncertainty in $V_{ud}$ as determined from superallowed $\beta$ decays. To this end, we addressed all contributions expected to be relevant at $\mathcal O(10^{-4})$, but there are several subleading effects whose role should be investigated in future work:
\begin{enumerate}
\item 2b and 3b $\mathcal O(\alpha \epsilon_\chi^2)$ corrections: the largest class of omitted diagrams identified in Sec.~\ref{sec:EFT}, Fig.~\ref{fig:diagrams}(f)  and (h), includes corrections that amount to a modification of the axial-vector coupling via 2b currents, contributing to the apparent quenching of $g_A$ in $\beta$ decays. In the literature, see, e.g., Ref.~\cite{Hardy:2020qwl}, similar corrections are included in an ad-hoc quenching of $g_A$ and other shell-model parameters. Numerically, the hierarchy of these corrections does appear to comply with the power-counting expectation, but a dedicated ab-initio evaluation would clearly be desirable. 
\item Shape corrections in the phase-space factor: the standard evaluation of the phase-space factor in Eq.~\eqref{eq:master0} involves  corrections related to the EW form factor and nuclear recoil. Both effects are, in principle, present in the EFT, and therefore care has to be taken to not double count the same effects at different places in the calculation, see App.~\ref{app:Cfactor}. In this work, we presented the decomposition of the decay rate in the EFT, leaving a dedicated study of the phase-space average to future work.   
 \item Subleading terms in the Fermi function $\mathcal O(\alpha^2 Z)$: the calculation from Refs.~\cite{Hill:2023acw,Hill:2023bfh} captures the leading effects $\alpha^nZ^n$ in $\overline{\text{MS}}$, converted to the 
 the $\overline{\text{MS}}_\chi$ scheme~\cite{Gasser:1983yg}
in Eq.~\eqref{eq:FF},
but neglecting terms $\mathcal O(\alpha^n Z^{n-1})$ with $n\geq 2$. Corrections of size $\mathcal O(\alpha^2 Z)$ could potentially be relevant for large $Z$. In analogy to neutron decay~\cite{Cirigliano:2023fnz}, one could consider a matching to a non-relativistic theory to capture the leading terms in an expansion in $\beta$, but since this expansion will be less accurate than for neutron decay, a dedicated calculation appears to be necessary.    
\end{enumerate}
In addition, while our focus has been on $\delta_\text{NS}$, we stress that $\delta_C$ and $\delta_\text{NS}$ should be calculated in the same ab-initio framework, e.g., to ensure that the isospin-breaking corrections contained in $\delta_C$ are consistent with the definition of the isospin limit in $\delta_\text{NS}$ (in this work, we used the mass of the neutral pion).

With these caveats in mind, our results as summarized in the master formula for superallowed $\beta$ decays in Eqs.~\eqref{eq:dGamma} and \eqref{eq:master2} pave the way for a modern EFT reinterpretation of the experimental program, see Sec.~\ref{sec:example} for an illustrative application to the $^{14}\text{O}\to {^{14}\text{N}}$ transition, and enables ab-initio nuclear many-body computations of nuclear-structure-dependent corrections. Our findings can be used to perform state-of-the-art extractions of $V_{ud}$ from nuclear processes with controlled uncertainty quantification and to use precision $\beta$-decay experiments to search for physics beyond the Standard Model.

\begin{acknowledgments}
We thank Mikhail Gorchtein, Vaisakh Plakkot,  Chien-Yeah Seng,
and Oleksandr Tomalak
for valuable discussions. Financial support by the Dutch Research Council (NWO) in the form of a VIDI grant, the U.S.\ DOE (Grant No.\
DE-FG02-00ER41132), Los Alamos
National Laboratory's Laboratory Directed Research and Development program under projects
20210190ER and 20210041DR, and the  SNSF (Project No.\ PCEFP2\_181117) is gratefully acknowledged.  Los Alamos National Laboratory is operated by Triad National Security, LLC,
for the National Nuclear Security Administration of U.S.\ Department of Energy (Contract No.\
89233218CNA000001). 
We acknowledge support from the DOE Topical Collaboration ``Nuclear Theory for New Physics,'' award No.\ DE-SC0023663. The work of S.G.\ is also supported by the Office of Advanced Scientific Computing Research, Scientific Discovery through Advanced Computing (SciDAC) NUCLEI program
and  by the Network for Neutrinos, Nuclear Astrophysics, and Symmetries (N3AS).
This research used resources provided by the Los Alamos National Laboratory Institutional Computing Program, which is supported by the U.S.\ Department of Energy National Nuclear Security Administration under Contract No.\ 89233218CNA000001.
M.H.\ and E.M.\ thank the Institute for Nuclear Theory at the University of Washington for its kind hospitality and support during the program ``New physics searches at the precision frontier (INT-23-1b),'' when this project was initiated. 
\end{acknowledgments}

\newpage
\appendix

\begin{widetext}

\section{Energy-dependent potentials}
\label{app:energy}

The potentials ${\mathcal V}_E$ and ${\mathcal V}_E^\pi$ are obtained from diagrams a0)--c0) in Fig.~\ref{fig:potential}. The energy dependence of these potentials results from expanding in small $\pp_{e,\nu}/\qg$, which leads to terms $\simeq \bar e_L \boldsymbol \gamma \cdot \pp_{e,\nu}\nu_L$ that can be rewritten in terms of $E_0$ and $m_e$ using the equations of motion of the leptons. In addition, however, one obtains contributions proportional to $q_0$, where $q=(p'_1-p_1)/2-(p'_2-p_2)/2$ is the difference of the relative momenta of the nucleons. It is not immediately clear how to deal with such terms, which are also encountered, for example, in the construction of the 2b contributions to weak and EM currents~\cite{Serot:1978vj} and the one-pion-exchange potential~\cite{Adam:1993zz}. In principle, one could use the equations of motion of the nucleons to relate $q_0$ to the kinetic energies of the nucleons, however, this implicitly relies on the use of a field redefinition~\cite{Furnstahl:2000we}. As we discuss below, the required field redefinition generates a shift in the potential.

Focusing on the somewhat simpler example of the ${\mathcal V}_E$ potential, the terms proportional to $q_0$ lead to a term in the Lagrangian of the form:
\begin{equation}\label{eq:VElag}
{\mathcal L}_{q_0} = -\sqrt{2}G_F V_{ud}\bar e_L\slashed{v} \nu_L\int_{\rr}\bigg[\left(\bar N Q N\right)(x+\rr/2) \bigg(i\frac{\partial}{\partial x_0}-i\frac{\overleftarrow\partial}{\partial x_0}\bigg)V(\rr) \left(\bar N\tau^+ N\right)(x-\rr/2)\bigg],
\end{equation}
where $Q$ is defined in Eq.~\eqref{charge} and  $V(\rr)\simeq \alpha |\rr|$ is the ${\mathcal V}_E$ potential in coordinate space, while the derivatives in brackets give rise to a factor of $2q_0$ if we take $p_1$ ($p_2$) to be the momenta of the nucleons that couple to the EM (weak) currents. Here the $x$ coordinate has a timelike component, while only the spacelike components of $r$ appear, so that all the fields are evaluated at equal times. When writing the potential as a term in the Lagrangian, the nucleon fields should in principle appear in a different ordering, namely, $\simeq \int_{\rr}\bar N(x+\rr/2) \bar N(x-\rr/2) V(\rr) N(x-\rr/2) N(x+\rr/2)$. However, in this case, the difference is proportional to $V(0)=0$. 

Before discussing the consequences of the abovementioned field transformation in more detail, we discuss another way of rewriting Eq.~\eqref{eq:VElag}, which will lead to the same conclusions. In particular, we can use the fact that the time evolution of an operator is determined by its commutator with the Hamiltonian. Using integration by parts to get rid of the derivative on the weak current, together with  $i\partial_0 O = [O,H]$, we obtain
\begin{align}\label{eq:VElag3}
{\mathcal L}_{q_0} &= 2\sqrt{2}G_F V_{ud}\bar e_L\slashed{v} \nu_L\int_{\rr}\big[\bar NQ N, H\big](x+\rr/2)V(\rr) \left(\bar N\tau^+ N\right)(x-\rr/2)\nonumber\\
&\quad+\sqrt{2}G_F V_{ud}iv\cdot \partial\left(\bar e_L\slashed{v} \nu_L\right)\int_{\rr}\big(\bar N Q N\big)(x+\rr/2)V(\rr) \left(\bar N\tau^+ N\right)(x-\rr/2),
\end{align}
where the second term is proportional to $E_0$, since $iv\cdot \partial\left(\bar e_L\slashed{v} \nu_L\right)\to -E_0\bar e_L\slashed{v} \nu_L$ when the lepton fields act on the external state, while the first term becomes $\simeq [H, J^0_\text{EM}]\simeq \partial\cdot J_\text{EM}=0$ up to the 2b part of the EM current. This 2b part would lead to 3b terms in Eq.~\eqref{eq:VElag3} and can be neglected. 
Comparing Eq.~\eqref{eq:VElag} with Eq.~\eqref{eq:VElag3} then implies the following replacement rule for $q_0$,
\begin{equation}\label{eq:q0rule}
q_0\to  E_0/2,
\end{equation}
so that the $q_0$ terms contribute to the part of the potential $\simeq E_0$ in Eq.~\eqref{eq:FermiPot}.

A very similar argument holds for the ${\mathcal V}_E^\pi$ potential although we can no longer use $V(0)=0$, since ${\mathcal V}_E^\pi(\rr)$  does not vanish as $|\rr|\to0$. In this case, it is simpler to use the ordering of the fields corresponding to a genuine potential instead of rewriting everything in terms of currents as in Eq.~\eqref{eq:VElag}. Doing so leads to a slightly modified form of Eq.~\eqref{eq:VElag3},  
 \begin{align}\label{eq:VElag4}
{\mathcal L}_{q_0} &= 2\sqrt{2}G_F V_{ud}\bar e_L\slashed{v} \nu_L\int_{\rr}\bar N_i(x-\rr/2)\big[\bar NQ N(x+\rr/2), H\big]V(\rr) \left(\tau^+ N (x-\rr/2)\right)_{i}\nonumber\\
&\quad+\sqrt{2}G_F V_{ud}iv\cdot \partial\left(\bar e_L\slashed{v} \nu_L\right)\int_{\rr}\big(\bar N Q N\big)(x+\rr/2)V(\rr) \left(\bar N\tau^+ N\right)(x-\rr/2),
\end{align}
where $i$ is an isospin index. This form is equivalent to Eq.~\eqref{eq:VElag3} when $V(0)=0$, but differs otherwise. With this change, the rest of the argument goes through unchanged and the same replacement rule of Eq.~\eqref{eq:q0rule} applies.

\subsection{Field redefinitions}

Alternatively to using $i\partial_0 O = [O,H]$, the ${\mathcal L}_{q_0}$ term can be removed through a transformation of the form,
\begin{equation}\label{eq:fieldredef}
N(x)\to N(x)+\delta N(x)=N(x)+\xi \,\int_{\yy} V(\yy) \left[\bar N Q_2 N(x+\yy)\right] \,  Q_1 N(x) ,
\end{equation}
where $Q_{1,2}$ are (isospin) operators. Similar transformations, which allow one to alter terms in the potential, have been considered, e.g., in Refs.~\cite{Furnstahl:2000we,Cirigliano:2021qko}. The shift resulting from the kinetic part of the Lagrangian, $\bar N i v\cdot D \delta N+\text{h.c.}$, leads to a new term of the same form as Eq.~\eqref{eq:VElag}. In other words, the transformation can remove ${\mathcal L}_{q_0}$ from the Lagrangian, for some choice of the operators $Q_{1,2}$ and $\xi$. 

To evaluate the complete shift in the Lagrangian we can first use integration by parts to write
\begin{align}\label{eq:Lq0}
{\mathcal L}_{q_0} &= -2\sqrt{2}G_F V_{ud}\bar e_L\slashed{v} \nu_L\int_{\rr}\bigg[\big(\bar NQ N\big)(x+\rr/2) \bigg(-i\frac{\overleftarrow\partial}{\partial x_0}\bigg)V(\rr) \left(\bar N\tau^+ N\right)(x-\rr/2)\bigg]\nonumber\\
&\quad+\sqrt{2}G_F V_{ud}iv\cdot \partial\left(\bar e_L\slashed{v} \nu_L\right)\int_{\rr}\big(\bar NQ N\big)(x+\rr/2)V(\rr) \left(\bar N\tau^+ N\right)(x-\rr/2).
\end{align}
Using the field redefinition in Eq.~\eqref{eq:fieldredef} with
\begin{equation}\label{eq:fieldredef2}
Q_1=Q=\frac{\mathds{1}+\tau_3}{2},\qquad Q_2 = i\sqrt{2} G_F V_{ud}\tau^+\bar e_L\slashed{v}\nu_L +\text{h.c.}, \qquad \xi=2i,
\end{equation}
then removes the first line in Eq.~\eqref{eq:Lq0} due to a shift in the kinetic term, $\simeq\delta ( \bar N v\cdot D N)$. 

The remaining terms in the Lagrangian are also transformed, for which it is useful to write  $\delta N$ as a commutator
\begin{equation}\label{eq:fieldredef3}
\delta N(x)=- 2i \int_{\yy,\zz}V(\yy)\left( \bar N Q_2 N(z+\yy)\right)\, \left[\bar N Q_1 N(z), N(x)\right] ,
\end{equation}
where $z_0=x_0$ so that all the appearing fields are again evaluated at equal times.
This allows us to write the total shift in the Lagrangian as
\begin{align}\label{eq:Lagshift}
\delta {\mathcal L}(x) &=-2i \int_{\yy,\zz} \left[\bar N Q_1 N(z), {\mathcal L}(x)\right] V(\yy) \bar N Q_2 N(z+\yy)\nonumber\\
&=-2i \int_{\yy,\zz} \left[\bar N Q_1 N(z), i\bar N v\cdot D N(x)\right] V(\yy) \bar N Q_2 N(z+\yy)\nonumber\\
&\quad-2i \int_{\yy,\zz} \left[\bar N Q_1 N(\zz), {\mathcal L}(x)-i\bar N v\cdot D N(x)\right] V(\yy) \bar N Q_2 N(z+\yy),
\end{align}
here the shift from the kinetic term in the second line removes the first line in Eq.~\eqref{eq:Lq0}, while the term in square brackets in the last line corresponds to the Hamiltonian density. The term $\simeq q_0$ together with Eq.~\eqref{eq:Lagshift} contributes to the action as follows,
\begin{align}
\int d^4x\left[{\mathcal L}_{q_0}(x)+\delta {\mathcal L}(x)\right] &=\int d^4x \bigg[-2i \int_{\yy,\zz} \left[\bar N Q_1 N(x), H\right] V(\yy) \bar N Q_2 N(x+\yy)\nonumber\\
&\quad+\sqrt{2}G_F V_{ud}iv\cdot \partial\left(\bar e_L\slashed{v} \nu_L\right)\int_{\rr}\big(\bar N Q N\big)(x+\rr/2)V(\rr) \left(\bar N\tau^+ N\right)(x-\rr/2)
\bigg],
\end{align}
where we relabeled $\xx$ and $\zz$ and used the fact that $H = \int_{\xx} \left[{\mathcal L}(x)-i\bar N v\cdot D N(x)\right]$. After plugging in the expressions for $Q_{1,2}$, this reproduces the Lagrangian in Eq.~\eqref{eq:VElag3}.

\subsection{Comparison to relativistic corrections in the traditional approach}

Instead of integrating by parts first to obtain Eq.~\eqref{eq:VElag3}, we could have used $i\partial_0 O = [O,H]$ directly, leading to
\begin{align}\label{eq:VElag2}
{\mathcal L}_{q_0} &= -\sqrt{2}G_F V_{ud}\bar e_L\slashed{v} \nu_L\bigg[-\int_{\rr}\big[\bar N Q N, H\big](x+\rr/2)V(\rr) \left(\bar N\tau^+ N\right)(x-\rr/2)\nonumber\\
&\quad+\int_{\rr}\big(\bar NQ N\big)(x+\rr/2)V(\rr) \left[\bar N\tau^+ N,H\right](x-\rr/2)\bigg].
\end{align}
Again neglecting $ [H, J^0_\text{EM}]\simeq \partial\cdot J_\text{EM}=0$  leaves the commutator with the weak current. This way of rewriting Eq.~\eqref{eq:VElag} leads to a contribution that looks similar to the relativistic terms discussed, e.g., in Ref.~\cite{Hayen:2017pwg}, around Eq.~(130). In particular, the nucleon mass splitting and the Coulomb potential in $H$ give rise to terms of the form
\begin{align}
{\mathcal L} _{q_0}&= \sqrt{2}G_F V_{ud}\bar e_L\slashed{v} \nu_L\int_{\rr}\big(\bar NQ N\big)(x+\rr/2)V(\rr) \bigg[(m_p-m_n)\bar N\tau^+ N(x-\rr/2)\nonumber\\
&\quad+\int_{\rr'} \big(\bar N QN\big)(x-\rr/2)V_C(\rr')(\bar N \tau^+ N)(x-\rr/2-\rr')\bigg].
\end{align}
Assuming that the matrix element of the potentials are roughly given by their value at $|\rr|=R_A$, i.e., that $\langle V(R_A)-V(\rr)\rangle $ is small, allows us to evaluate the appearing currents as conserved charges, $\int_{\xx}\big(\bar NQ N\big)(x)={\mathcal Q}\to Z$.
This then gives a contribution proportional to
\begin{equation}\label{eq:VELeendert}
 \alpha Z R \bigg[m_p-m_n+\frac{\alpha Z}{R}\bigg],
\end{equation}
while the relativistic corrections discussed in Ref.~\cite{Hayen:2017pwg} take the same form up to a factor $6/5$ in front of the Coulomb contribution. This factor depends on the assumed charge distribution of the nucleus, see the analog discussion around Eq.~\eqref{L0C}. Nevertheless, the  
form of Eq.~\eqref{eq:VELeendert} is qualitatively similar, implying that the two approaches are capturing the same physical effects.

\section{\texorpdfstring{$\boldsymbol{\mathcal O(\alpha \epsilon_\chi)}$}{} and \texorpdfstring{$\boldsymbol{\mathcal O(\alpha \epsilon_\slashed{\pi})}$}{} potentials in coordinate space}
\label{app:coordinate}

For the numerical implementation, it is convenient to also provide the matrix elements in coordinate space, see Eqs.~\eqref{V_O} and~\eqref{V_O_F_GT_T}.
The radial functions needed for the energy-dependent corrections $\mathcal V_{E}$ and $\mathcal V_{E}^\pi$ are 
\begin{align}\label{VEpir}
    h^E_{\text{F}, p}(r)  &= - \frac{r}{2 R_A},\notag\\
    h^{E \pi}_{\text{GT}, p}(r)  &= - h^{E \pi}_{\text{GT}, n}(r) = \frac{g^2_A Z_\pi}{3 }
     \frac{e^{- \mpi r}}{72 \mpi R_A}  \big(12 + 12 \mpi r - \mpi^2 r^2 \big), \notag\\
     h^{m_e \pi}_{\text{GT}, p}(r)  &= - h^{m_e \pi}_{\text{GT}, n}(r) = \frac{g^2_A Z_\pi}{3 }
     \frac{e^{- \mpi r}}{72 \mpi R_A}  \big(15 - 21 \mpi r + \mpi^2 r^2 \big), \notag\\
     h^{E\pi}_{\text{T}, p}(r) &=-h^{E\pi}_{\text{T}, n}(r) = \frac{g^2_A Z_\pi}{3}\frac{e^{- \mpi r}}{72 \mpi R_A }  \big( 9 \mpi r - \mpi^2 r^2\big),\notag\\
     h^{m_e\pi}_{\text{T}, p}(r) &=-h^{m_e\pi}_{\text{T}, n}(r) = -\frac{g^2_A Z_\pi}{3}\frac{e^{- \mpi r}}{72 \mpi R_A }  \big( 18 \mpi r - \mpi^2 r^2\big).
\end{align}
$\mathcal V_E$ 
only has a Fermi-component coupling to protons, while the pion-mass-splitting contributions only induce GT and T components.
The factor of $R_A$ was introduced to make the radial functions, and thus the matrix elements, dimensionless.
The magnetic contribution induces both a Gamow--Teller and tensor component, with radial functions given by
\begin{align}
    h^\text{mag}_{\text{GT}, p}(r) 
    &= 4 h^\text{mag}_{T, p}(r)
    = \frac{g_A}{3 \mN}   \frac{1+\kappa_p}{r}, \notag\\ 
    h^\text{mag}_{\text{GT}, n}(r) &=4 h^\text{mag}_{\text{T}, n}(r) 
     = \frac{g_A}{3 \mN}   \frac{\kappa_n}{r}.
\end{align}
The recoil terms in Eq.~\eqref{eq:RecPotential}
are non-local, and their Fourier transform is given by
\begin{align}
 \mathcal  V^\text{rec}_0(\rr) &= \frac{e^2}{4\pi} \frac{g_A}{2\mN} \sum_{j < k}\bigg[ - \frac{\tau^{+ (j)} P_p^{(k)}}  {r_{jk}}
 \LL_{jk} \cdot \boldsymbol{\sigma}^{(j)} 
 - Z_\pi g_A \tau^{+(j)} \tau^{(k)}_3  \frac{e^{- \mpi r}}{r}
\left(2 \bsigma^{(j)} \cdot \rr_{j k} \,\bsigma^{(k)} \cdot {\boldsymbol{\nabla}} + \bsigma^{(j)} \cdot \bsigma^{(k)} \right)  
+ (j\leftrightarrow k)\bigg],
\end{align}
so that the spin-orbit radial function is given by
\begin{equation}\label{eq:radialso}
    h_\text{so}(r) =  -\frac{g_A}{2\mN} \frac{1}{r}.
\end{equation}

Finally, the short-range potential is given by
\begin{equation}\label{eq:deltacoor}
    \mathcal V^\text{CT}_0(\rr) = e^2 \sum_{j < k }  \left( g_{V1}^{N\!N}\tau^{+ (j)} + g_{V2}^{N\!N}\tau^{+ (j)} \tau_3^{(k)} \right) \delta^{(3)}(\rr_{jk}).
\end{equation}
To compare with the magnetic potential, it is convenient to perform a Fierz transformation on 
$O_1$ and $O_2$ and write them as
\begin{align}
    N^\dagger \tau^+ N \, N^\dagger N  \rightarrow - \frac{1}{3} N^\dagger \boldsymbol{\sigma} \tau^+ N \cdot N^\dagger \boldsymbol{\sigma} N, \qquad 
     N^\dagger \tau^+ N \, N^\dagger \tau^3 N  \rightarrow - \frac{1}{3} N^\dagger \boldsymbol{\sigma} \tau^+ N \cdot N^\dagger \boldsymbol{\sigma} \tau^3 N. 
\end{align}
The $\delta$ function in Eq.~\eqref{eq:deltacoor} can be regularized in various ways. 
In this work, we use nuclear wave functions obtained with the local N$^2$LO chiral potential of Ref.~\cite{Gezerlis:2014zia},
in which the $\delta$ function is replaced by
\begin{equation}
    \delta^{(3)}(\rr) \rightarrow \delta_{R_0}(r) =\frac{1}{\pi \Gamma\left(\frac{3}{4}\right) R_0^3} \exp\left(-\frac{r^4}{R^4_0}\right).
\end{equation}
We can thus express the short-range potential via the radial functions
\begin{align}
    h^\text{CT}_{\text{GT}, p}(r) &= - \frac{4\pi}{3} (g^{N\!N}_{V1} + g^{N\!N}_{V2}) \delta_{R_0}(r), \notag\\  h^\text{CT}_{\text{GT}, n}(r) &= - \frac{4\pi}{3} (g^{N\!N}_{V1} - g^{N\!N}_{V2}) \delta_{R_0}(r).
\end{align}

\section{Subtraction of the ultrasoft region 
in the derivation of the \texorpdfstring{$\boldsymbol{\mathcal O(\alpha \epspi)}$}{}
and \texorpdfstring{$\boldsymbol{\mathcal O(\alpha^2)}$}{} potentials
}
\label{app:sub}

The calculation of the photon-exchange potentials at $\mathcal O(\alpha \epspi)$
and $\mathcal O(\alpha^2)$
requires some care as the
ultrasoft and potential modes can overlap when $\qq \rightarrow 0$. To properly define the EW potentials we then need to subtract the ultrasoft region. We discuss here how we perform these subtractions.

\subsection{\texorpdfstring{$\boldsymbol{\mathcal O(\alpha)}$}{}  two-body potential}

We start by discussing $\mathcal V_E$, which, in momentum space, behaves like $1/\qq^4$ and it is thus sensitive to IR contributions.
In a 2b calculation, the momentum-space matrix element of $\mathcal V_E$ 
  could be written as
\begin{equation}\label{eq:q4}
    \langle f |{\mathcal V}_E | i \rangle \equiv 
\int \frac{d^3 q_1}{(2\pi)^3} \frac{1}{\qq_1^4} \int \frac{d^3 q_2}{(2\pi)^3} \Big[  \psi^*(\qq_2)  \psi(\qq_1 + \qq_2) -  \psi^*(\qq_2)  \psi(\qq_2)  \Big],
\end{equation}
where $\psi$ is a 2b wave function. In many-body calculation $\psi$ would correspond to the many-body wave function after integrating over all relative momenta but one. The second term in Eq.~\eqref{eq:q4}
corresponds to subtracting
the ultrasoft limit $|\qq_1| \ll |\qq_2| $,
so that the matrix element of $\mathcal V_E$ is well defined in the IR.
Going to coordinate space, this expression becomes 
\begin{equation}
    \langle f | {\mathcal V}_E | i \rangle = \int d^3 r \, \psi^*(r) \left[ \int \frac{d^3 q}{(2\pi)^3} \frac{1}{\qq^4} \left(e^{i \qq \cdot \rr} - 1\right)\right] \psi(r)=
  -\frac{1}{4\pi} \int d^3 r \, \psi^*(r) \frac{r}{2}  \psi(r) ,
\end{equation}
coinciding with the result one would obtain by taking the Fourier transform of $1/\qq^4$
in dimensional regularization.
Equation~\eqref{eq:q4} is reminiscent of the ``zero-bin'' subtraction devised in Ref.~\cite{Manohar:2006nz}, which is needed in order to avoid double counting due to the different photon modes with overlapping (IR) momentum regions.
We will use the same idea for the more complicated $\mathcal O(\alpha^2)$ potentials.

\subsection{\texorpdfstring{$\boldsymbol{\mathcal O(\alpha^2)}$}{} two-body diagrams}

The diagrams in the first row of Fig.~\ref{fig:alpha2} lead to the amplitude
\begin{equation}\label{eq:a2mom}
    \mathcal A =  \sum_{i < j} g_V \frac{e^4}{(4\pi)^2} (\bar\mu^2)^{2\epsilon}\frac{2 \pi^2}{\left[\qq^2 \right]^{\frac{3}{2}+  \epsilon}} \bigg[1 + \epsilon\Big(\frac{3}{4} - \gamma_E + \log (16\pi)\Big) \bigg]  \bar u(p_e) \gamma^0 P_L v(p_\nu) \, \, \left(\tau^{+(i)}  P_p^{(j)} +\tau^{+(j)}  P_p^{(i)}\right).
\end{equation}
Here we work in dimensional regularization, with $d=4-2\epsilon$ dimensions and in
the $\overline{\text{MS}}_\chi$ scheme~\cite{Gasser:1983yg},
which subtracts the combination 
\begin{equation}
    \frac{1}{\epsilon} - \gamma_E + \log (4\pi) + 1,
\end{equation}
including 
an additional finite piece compared to the standard $\overline{\text{MS}}$ scheme.
At $\mathcal O(\alpha^2)$, we implement this scheme by introducing the scale
\begin{equation}\label{eq:muchi}
    \bar \mu^2 = \mu^2 \frac{e^{\gamma_E-1}}{4\pi}.
\end{equation}
To interpret Eq.~\eqref{eq:a2mom}
as a potential, and obtain the matching coefficients in Eqs.~\eqref{eq:cdelta}
 and~\eqref{eq:cplus}, 
we follow a strategy very similar to Eq.~\eqref{eq:q4}. We consider the amplitude $\mathcal A$ to be applied to a test function $\varphi(\qq)$ (which stands here for the product of nuclear wave functions)
and we subtract the value 
$\varphi({\bf 0})$, which corresponds to the regime in which the photon momentum becomes ultrasoft. Schematically, we have to consider matrix elements of the form
\begin{equation}
\int \frac{d^{d-1} q}{(2\pi)^{d-1}}   (\bar \mu^2)^{2\epsilon}\frac{2 \pi^2}{\left[\qq^2 \right]^{\frac{3}{2}+  \epsilon}} \left(\varphi(\qq)  - \varphi({\bf 0}) \right),
\end{equation}
which we can rewrite as
\begin{equation}
\int \frac{d^{d-1} q}{(2\pi)^3}   (\bar\mu^2)^{2\epsilon}\frac{2 \pi^2}{\left[\qq^2 \right]^{\frac{3}{2}+  \epsilon}} \left(\varphi(\qq)  - \theta(\Lambda e^{-\gamma_E + 1} - | \qq|) \varphi({\bf 0}) \right) - 
\int \frac{d^{d-1} q}{(2\pi)^{d-1}}   (\bar\mu^2)^{2\epsilon}\frac{2 \pi^2}{\left[\qq^2 \right]^{\frac{3}{2}+  \epsilon}} \theta(|\qq|- \Lambda e^{-\gamma_E + 1}) \varphi({\bf 0}).
\end{equation}
The first term is IR and UV finite, and we can simply drop the dimensional regulator $\epsilon$ and obtain the plus distribution in Eq.~\eqref{eq:plusdef}. 
The second term is 
equivalent to the application of a potential that is a $\delta$ function in momentum space, with coefficient 
\begin{equation}
    - 
\int \frac{d^{d-1} q}{(2\pi)^3}   (\bar\mu^2)^{2\epsilon}\frac{2 \pi^2}{\left[\qq^2 \right]^{\frac{3}{2}+  \epsilon}} \theta(|\qq|- \Lambda e^{-\gamma_E + 1})  = -  \frac{1}{4\epsilon}  - \frac{1}{2} \log \frac{\mu^2}{\Lambda^2} + 1 - \frac{5}{4} \gamma_E  + \log (16\pi). 
\end{equation}
When combined with the $\mathcal O(\epsilon)$ contribution coming from the loop, and subtracting the divergence in the $\overline{\text{MS}}_{\chi}$ scheme, we obtain
\begin{equation}
    C_\delta = - g_V \frac{\alpha^2}{2} \bigg[  \log \frac{\mu^2}{\Lambda^2} - \frac{13}{8} + 2 \gamma_E  \bigg].
\end{equation}

\subsection{Three-body diagrams}

For the 3b diagrams we encounter amplitudes of the form
\begin{align}\label{eq:A3b}
    \mathcal A_\text{3b} &= g_V \frac{e^4}{2} \sum_{i \neq j\neq k } \bar u(p_e) \gamma^0 P_L v(p_\nu) \, \, \tau^{+(i)}  P_p^{(j)} P_p^{(k)} \left(   \frac{1}{\qq_i^2} \frac{1}{\qq_j^2} \frac{1}{\qq_k^2} + \frac{1}{\left[\qq^2_k\right]^2} \Big( \frac{1}{\qq_j^2} - \frac{1}{\qq_i^2}\Big) \right),
\end{align}
with the momenta satisfying $\qq_i + \qq_j + \qq_k ={\bf 0}$.
This potential acts on 
functions of $\qq_j$ and 
$\qq_k$. As in the 2b case, the amplitudes receive contributions from the regions in which 
$\qq_j$ and $\qq_k$ are ultrasoft, 
$\qq_j, \qq_k \rightarrow {\bf 0}$, which need to be subtracted to obtain a 3b potential. We focus here on the first term of Eq.~\eqref{eq:A3b}, which leads
to logarithmic divergences.
We can thus write
\begin{align}\label{eq:A3b2}
&    (\bar\mu^2)^{2\epsilon}  \int \frac{d^{d-1} q_j }{(2\pi)^{d-1}}
    \int \frac{d^{d-1} q_k }{(2\pi)^{d-1}}
    \frac{1}{\qq_j^2} \frac{1}{\qq_k^2} \frac{1}{(\qq_j + \qq_k)^2} \left[ \varphi(\qq_j,\qq_k) - \varphi({\bf 0},{\bf 0})    \right] \notag\\
& = (\bar\mu^2)^{2\epsilon}  \int \frac{d^{d-1} q_j }{(2\pi)^{d-1}}
    \int \frac{d^{d-1} q_k }{(2\pi)^{d-1}}
    \frac{1}{\qq_j^2} \frac{1}{\qq_k^2} \frac{1}{(\qq_j + \qq_k)^2} \Big[ \varphi(\qq_j,\qq_k) -   \theta(\tilde \Lambda - |\qq_j|)\theta(\tilde \Lambda - |\qq_k|) \varphi({\bf 0},{\bf 0})
    \Big]  \notag \\
& \quad -   (\bar\mu^2)^{2\epsilon}  \int \frac{d^{d-1} q_j }{(2\pi)^{d-1}}
    \int \frac{d^{d-1} q_k }{(2\pi)^{d-1}}
\frac{1}{\qq_j^2} \frac{1}{\qq_k^2} \frac{1}{(\qq_j + \qq_k)^2}
    \Big[  
    \theta(\tilde \Lambda - |\qq_j|)\theta(-\tilde \Lambda + |\qq_k|) + \theta(-\tilde \Lambda + |\qq_j|)\theta(\tilde \Lambda - |\qq_k|)  \notag \\ 
    &\quad\quad + \theta(-\tilde \Lambda + |\qq_j|)\theta(-\tilde \Lambda + |\qq_k|)  
    \Big]\varphi({\bf 0},{\bf 0}),
\end{align}
where $\varphi$ is a test function, $\tilde \Lambda = \Lambda e^{-\gamma_E +1}$,
and $\bar \mu$ is defined in Eq.~\eqref{eq:muchi}.

The first term is now IR finite. We can 
set $d=4$, and this term is represented by the distribution
\begin{equation}
 \left[ \frac{1}{\qq_j^2} \frac{1}{\qq_k^2} \frac{1}{(\qq_j + \qq_k)^2} \right]_{+, \Lambda} 
\end{equation}
defined as
\begin{align}
&\int \frac{d^3 q_j}{(2\pi)^3}
\int \frac{d^3 q_k}{(2\pi)^3}  \left[ \frac{1}{\qq_j^2} \frac{1}{\qq_k^2} \frac{1}{(\qq_j + \qq_k)^2} \right]_{+, \Lambda}  \varphi(\qq_j, \qq_k) \notag \\ & \equiv
\int \frac{d^3 q_j}{(2\pi)^3}
\int \frac{d^3 q_k}{(2\pi)^3}  \frac{1}{\qq_j^2} \frac{1}{\qq_k^2} \frac{1}{(\qq_j + \qq_k)^2} \Big[ \varphi(\qq_j,\qq_k) -   \theta(\tilde \Lambda - |\qq_j|)\theta(\tilde \Lambda - |\qq_k|) \varphi({\bf 0},{\bf 0})
    \Big]. 
\end{align}
The integrals in the second and third line of Eq.~\eqref{eq:A3b2}
can be performed, giving
\begin{equation}
    - (2\pi)^3 \delta^{(3)}( \qq_j)
    (2\pi)^3 \delta^{(3)}(\qq_j)
    \frac{1}{(4\pi)^2}
\left[ \frac{1}{4} \left( \frac{1}{\epsilon} + 2  \log \frac{\mu^2}{\tilde\Lambda^2}  \right) + \frac{7 \zeta(3)}{ 2\pi^2} + \frac{1}{4} \right].
\end{equation}
The 3b diagrams thus lead to the following correction to $\mathcal V_0$
\begin{equation}
     \mathcal V^{0}\big|_\text{3b}
    = \tilde{C}_\delta^\text{3b}\, \mathcal V_{\delta}^\text{3b} + C^\text{3b}_+ \,\tilde{\mathcal V}^\text{3b}_{+}, 
\end{equation}
with $\mathcal V_{\delta}^\text{3b}$ defined in Eq.~\eqref{eq:Vdelta3b_mom},
and  $\mathcal V_+^\text{3b}$
\begin{equation}\label{eq:3ba}
   \tilde{\mathcal V}_+^\text{3b}(\qq) =
    \sum_{i \neq j\neq k } (4\pi)^2  \left[ \frac{1}{\qq_j^2} \frac{1}{\qq_k^2} \frac{1}{(\qq_j + \qq_k)^2} \right]_{+, \Lambda} \tau^{+(i)}  P_p^{(j)} P_p^{(k)}.
\end{equation}
The matching coefficients are given by
\begin{equation}
   \tilde{C}_\delta^\text{3b} = - g_V \alpha^2 \left( \frac{1}{4}\log \frac{\mu^2}{ \Lambda^2} + \frac{\gamma_E}{2} + \frac{7 \zeta(3)}{4\pi^2}   - \frac{3}{8}\right), \qquad  C_{+}^\text{3b} =  g_V \frac{\alpha^2}{2}.
\end{equation}
In coordinate space, we can obtain the Fourier transform of $\mathcal V^\text{3b}_+$ in the limit of large $\Lambda$
\begin{equation}
    \tilde{\mathcal V}_+^\text{3b}(\rr) = - \sum_{i \neq j\neq k } \left( \log  \bigg[\frac{\Lambda}{2} \Big(r_{ij} + r_{ik} + r_{jk} \Big) \bigg]  - \frac{7 \zeta(3)}{2 \pi^2} \right) \tau^{+(i)}  P_p^{(j)} P_p^{(k)}.
\end{equation}
Notice that the term proportional to $\zeta(3)$ cancels between $\tilde C_{\delta}^\text{3b}$ and $\tilde{\mathcal V}_+^\text{3b}$.
It is thus convenient to define a coefficient 
\begin{equation}
    C_\delta^\text{3b} = \tilde C_\delta^\text{3b} + g_V \alpha^2 \frac{7 \zeta(3)}{4 \pi^2} = - g_V \alpha^2 \left( \frac{1}{4}\log \frac{\mu^2}{ \Lambda^2} + \frac{\gamma_E}{2}  - \frac{3}{8}\right), 
\end{equation}
and a potential
\begin{align}
    {\mathcal V}_+^\text{3b}(\qq) &= \tilde{\mathcal V}_+^\text{3b}(\qq) -  \frac{7 \zeta(3)}{2 \pi^2} \mathcal{V}^\text{3b}_\delta(\qq), \label{eq:V3plus1} \\
    {\mathcal V}_+^\text{3b}(\rr) &=  - \sum_{i \neq j\neq k }  \log  \bigg[\frac{\Lambda}{2} \Big(r_{ij} + r_{ik} + r_{jk} \Big) \bigg]\tau^{+(i)}  P_p^{(j)} P_p^{(k)}
    \label{eq:V3plus2}.
\end{align}
so that terms scaling as $\mathcal O(\alpha^2 Z (Z-1))$ are fully captured by $C_\delta^\text{3b}$, while $\mathcal V^\text{3b}_+$ is a purely logarithmic correction.
In coordinate space,
the 3b potentials read
\begin{equation}
    {C}_\delta^\text{3b}\, \mathcal V_{\delta}^\text{3b} + C^\text{3b}_+ \,\mathcal V^\text{3b}_{+} = 
    - g_V \frac{\alpha^2}{2}\sum_{i \neq j\neq k } \left( \log  \bigg[\frac{\mu}{2} \Big(r_{ij} + r_{ik} + r_{jk} \Big)\bigg] + \gamma_E - \frac{3}{4}  \right) \tau^{+(i)}  P_p^{(j)} P_p^{(k)}.
\end{equation}

\section{Renormalization group equations below \texorpdfstring{$\boldsymbol{\mu=\mupi}$}{}}
\label{app:usoftRGE}

As discussed in Sec.~\ref{sec:decay}, integrating out potential and soft photons leads to an effective theory containing ultrasoft photons as propagating degrees of freedom, supplemented by static potentials. Compared to the theory above $\mu=\mupi$, additional divergences arise that depend on the charge of the nucleus, instead of the nucleon charges. This can be seen from the effective action generated by the exchange of $n$ photons between a single electron line and up to $n$ nucleon lines
\begin{align}
 S_\text{eff}^{(n)}&\supset e^{2n} \int_{q_1\dots q_n}  \int_{y,x_1\dots x_n}\bar e_L(0)  \slashed{v} \frac{\slashed p+\slashed q_1}{(p+q_1)^2}\dots \slashed{v} \frac{\slashed p+\slashed q_1+\dots +\slashed q_n}{(p+q_1+\dots +q_n)^2}  \gamma_\mu \nu_L(y) \frac{e^{-i  q_1\cdot (x_1-y)}}{q_1^2}\dots \frac{e^{-i q_n\cdot (x_n-y)}}{q_n^2}\notag\\
 &\times T\left[ j_0(x_1)\dots  j_0(x_n) j_W^\mu(y)\right],
\end{align}
where $\int _q\equiv \int \frac{d^dq}{(2\pi)^d}$ for momenta and $\int _x\equiv \int d^d x$ for positions, while $j_W^\mu  = -\sqrt{2} G_F V_{ud}g_V\bar N v^\mu\tau^+ N$ and $j_\mu = \bar N v_\mu Q N$ are the EW and EM currents, with $Q$ defined in Eq.~\eqref{charge}, and $p$ and $q_i$ are the momenta of the electron and $i^\text{th}$ photon, respectively. In principle, there are additional contributions that correspond to diagrams in which one of the photons connects to the same electron/nucleon line. Using the symmetry arguments discussed in Ref.~\cite{Borah:2024ghn}, one can show that such terms first contribute at ${\mathcal O}(\alpha^2 Z^0)$, or when going beyond ${\mathcal O}(\alpha^2)$. Here we focus on the terms $\simeq (\alpha^2 Z^2)^n$ and $(\alpha^2 Z)$. Since the exchanged photons are ultrasoft, small ratios of $q_i/k_F$ should be expanded. In particular, we have  $|\xx_i-\yy|\simeq 1/k_F$, so that the exponentials become $e^{-i q_i\cdot (x_i-y)}\simeq e^{-i q_i^0 (x_i^0-y^0)}$. The only $\xx_i$ dependence then appears in the EM currents, which, after integration, lead to time-independent conserved charges, ${\mathcal Q}\equiv \int_{\xx_i}j_0(x_i) $, allowing us to write
\begin{equation}
\label{eq:Seff2}
 S_\text{eff}^{(n)}\supset e^{2n} \int_{q_1\dots q_n}  \int_{y,x^0_1\dots x^0_n}\bar e_L  \slashed{v} \frac{\slashed p+\slashed q_1}{(p+q_1)^2}\dots \slashed{v} \frac{\slashed p+\slashed q_1+\dots +\slashed q_n}{(p+q_1+\dots +q_n)^2}  \gamma_\mu \nu_L \frac{e^{-i  q^0_1 (x^0_1-y^0)}}{q_1^2}\dots \frac{e^{-i q^0_1 (x^0_n-y^0)}}{q_n^2}
\, T\left[ {\mathcal Q}^n j_W^\mu(y)\right].
\end{equation}
Once the factors of ${\mathcal Q}$ in the time-ordered product act on states, they give rise to factors of the charge of the initial- or final-state nucleus, depending on whether they appear before or after $j_W$. In the following we organize these terms by powers of $Z$.

\subsection{Contributions of \texorpdfstring{$\boldsymbol{{\mathcal O}(\alpha^nZ^n)}$}{}}

Focusing on the terms $\simeq Z^n$, with $Z$ the charge of the daughter nucleus, we can move all factors of ${\mathcal Q}$ to the left of $j_W^\mu$ using $j_W^\mu {\mathcal Q} = {\mathcal Q}j_W^\mu +[j_W^\mu,{\mathcal Q}]$. Neglecting the commutator contributions, each term in the time-ordered product gives rise to ${\mathcal Q}^n j_W^\mu$, multiplied by Heaviside functions, $\theta(\pm (x_i^0-y^0))$. As each time ordering, or combination of Heaviside functions, comes with the same coefficient they simply add up to one, leading to
\begin{align}
 S_\text{eff}^{(n)}&\supset e^{2n} \int_{q_1\dots q_n}  \int_{y}\bar e_L  \slashed{v} \frac{\slashed p+\slashed q_1}{(p+q_1)^2}\dots \slashed{v} \frac{\slashed p+\slashed q_1+\dots +\slashed q_n}{(p+q_1+\dots +q_n)^2}  \gamma_\mu \nu_L \frac{2\pi \delta(q_1^0)}{q_1^2}\dots \frac{2\pi \delta(q_n^0)}{q_n^2}
\notag\\&\times {\mathcal Q}^n j_W^\mu(y)+ {\mathcal O} \left({\mathcal Q}^{n-1}\left[ j_W^\mu,\, {\mathcal Q}\right]\right)+\dots,
\end{align}
where the terms involving one or more commutators lead to fewer factors of ${\mathcal Q}$ and are subleading in $Z$.

The remaining integrals over $\qq_i$ lead to divergences for even values of $n$. The result is proportional to the original operator structure, but comes with additional factors of $\mathcal Q$, namely, $\simeq \bar e_L\gamma_\mu\nu_L \, {\mathcal Q}^{n} j_W$. In contrast, terms at odd $n$ give structures that involve the electron momentum, $\simeq \pp\cdot \boldsymbol{\gamma}/|\pp|$. The integrals can be done iteratively by noticing that each combination of two photon exchanges leads to integrals of the same form. The required integrals are very similar to those discussed in Ref.~\cite{Hill:2023bfh} and given by
\begin{equation}
\label{eq:usoftInt}
\int_{\qq}\frac{\slashed{v}\boldsymbol{\gamma}\cdot(\pp+\qq)}{\left[(\pp+\qq)^2\right]^{1+a}}\frac{1}{\qq^2}\equiv  J_1(a)\frac{\slashed v \boldsymbol{\gamma} \cdot \pp}{\left[ \pp^2\right]^{\frac{5-d+2a}{2}}},\qquad \int_{\qq_1,\qq_2}\frac{\boldsymbol{\gamma}\cdot(\pp+\qq_1)}{(\pp+\qq_1)^2} \frac{\boldsymbol{\gamma} \cdot(\pp+\qq_1+\qq_2)}{\left[(\pp+\qq_1+\qq_2)^2\right]^{1+a}}\frac{1}{\qq_1^2}\frac{1}{\qq_2^2}\equiv - J_2(a)\frac{1}{\left[ \pp^2\right]^{4-d+a}},
\end{equation}
with
\begin{equation}
 J_1(a) = \frac{\Gamma\left(\frac{d-3}{2}\right)
 \Gamma\left(\frac{d-1-2a}{2}\right) \Gamma\left(\frac{5-d+2a}{2}\right)}
 {(4\pi)^{\frac{d-1}{2}}\Gamma(1+a)\Gamma(d-2-a)},
 \qquad  J_2(a) = \frac{\Gamma\left(\frac{d-3}{2}\right)^2
 \Gamma\left(\frac{d-1-2a}{2}\right) \Gamma\left(4-d-a\right) \Gamma\left(d-3-a\right)}
 {(4\pi)^{d-1}\Gamma(1+a)\Gamma(d-2-a)\Gamma\left(\frac{3d-9-2a}{2}\right)}.
\end{equation}
The effective action induced by an even or odd number of photon exchanges then becomes
\begin{align}
\label{eq:Seff}
 S_\text{eff}^{(n)}&\supset e^{2n}  {\mathcal Q}^n  \int_y j^\muW(y) \bar e_L\gamma_\mu\nu_L \frac{1}{\left[ \pp^2\right]^{n\epsilon}} \prod_{k=0}^{n/2-1} J_2(2k \epsilon),\qquad (n\,\, \text{even}  ),\notag\\
 S_\text{eff}^{(n)}&\supset -e^{2n}  {\mathcal Q}^n \int_yj^\muW(y)  \frac{\bar e_L\slashed v \boldsymbol{\gamma} \cdot \pp \gamma_\mu \nu_L}{\left[ \pp^2\right]^{1/2+n\epsilon}} J_1((n-1) \epsilon)\prod_{k=0}^{\frac{n-3}{2}} J_2(2k \epsilon),\qquad (n\,\, \text{odd}  ).
 \end{align}
The divergences contained in these expression need to be absorbed by counterterms and require the presence of additional terms in the Lagrangian, namely,
\begin{equation}
\label{eq:LagUsoft}
{\mathcal L} \supset \left[\sum_{n=0}^\infty c_n {\mathcal Q}^n\right]\, \bar e_L\gamma_\mu \nu_L \, j_W^\mu,
 \end{equation}
 where $c_0=1$ gives rise to the 1b EW current, while $c_{n\neq0}$ absorb the divergences induced by terms with $n>0$ in Eq.~\eqref{eq:Seff}.  Combining the amplitudes generated by $S_\text{eff}^{(n)}$ with those from $c_n$ and demanding the sum to be finite, we can determine the counterterms contained in $\cc=(c_0,c_1\dots, c_n)^T$. This procedure requires including terms in which the $c_n$ operators are dressed with additional photon exchanges, whose contributions are described by integrals of the same form as in Eq.~\eqref{eq:usoftInt}.
 The bare couplings $c_i^{(b)}$ are renormalized  by the renormalization constants $Z_{ij}=1+\sum_{n=1} Z^{(n)}_{ij}/\epsilon^n$ according to $c_i^{(b)} = Z_{ij} c_j$. 
After obtaining the counterterms, $Z_{ij}$, one finds that the amplitudes involving $\pp$-dependent structures, resulting from odd numbers of photons in Eq.~\eqref{eq:Seff}, are rendered finite. 
 
 With the above definitions, the anomalous dimensions are given by $\gamma_{ij} = 2\frac{d}{d\log \alpha}Z^{(1)}_{ij}$, for which one finds
 \begin{equation}
 \label{eq:usoftRG}
 \frac{d}{d\log\mu}\cc=\gamma \cc,\qquad
\gamma=-  
\left(\begin{array}{*{20}c}
0&&&&\dots&&&&&&0\\
0&0\\
\frac{\alpha^2}{2} & 0 & 0\\
0 & \frac{\alpha^2}{2} &0 & 0\\
 \frac{\alpha^4}{8} &0 & \frac{\alpha^2}{2} &0 & 0&&&&&&\vdots\\
0&  \frac{\alpha^4}{8} &0 & \frac{\alpha^2}{2} &0 & 0\\
\frac{\alpha^6}{16}&0&  \frac{\alpha^4}{8} &0 & \frac{\alpha^2}{2} &0 & 0\\
0& \frac{\alpha^6}{16}&0&  \frac{\alpha^4}{8} &0 & \frac{\alpha^2}{2} &0 & 0\\
\frac{5\alpha^8}{128}&0& \frac{\alpha^6}{16}&0&  \frac{\alpha^4}{8} &0 & \frac{\alpha^2}{2} &0 & 0&\\
\vdots &&&&&&&&&\ddots
\end{array}\right).
 \end{equation}
These RG equations imply that each $c_i$ contributes to $c_{i+2}$, with an anomalous dimensions $-\alpha^2/2$, which determines the LL contributions $\simeq \left[\alpha^2 Z^2 \log \mu\right]^n$. NLL $\simeq \left[\alpha^4 Z^4 \log \mu\right]^n$ arise from the contributions of $c_i$ to $c_{i+4}$, proportional to $-\alpha^4/8$. It turns out that this sequence of (sub)leading anomalous dimensions sums up to a square root~\cite{Hill:2023acw,Hill:2023bfh}. This can be seen by noting that the matrix elements of the Lagrangian in Eq.\ \eqref{eq:LagUsoft}  are proportional to an effective coupling of the form $C(\mu)=\sum_n c_n(\mu) Z^{n} $, which indeed follows the RG equation,
 \begin{equation}
  \frac{d}{d\log\mu} C=\gamma'  C,\qquad \gamma' =\sqrt{1-\alpha^2 Z^2}-1.
  \end{equation}

\subsection{Contributions of \texorpdfstring{$\boldsymbol{{\mathcal O}(\alpha^nZ^{n-1})}$}{}}

The anomalous dimensions are not just an expansion in $(\alpha^2Z^2 )^n$, but also involve contributions with fewer powers of $Z$. To capture the first of these, we can go back to Eq.~\eqref{eq:Seff2} and, instead of neglecting all commutator terms, focus on the terms that involve a single commutator. Such contributions are proportional to $\simeq \alpha^n {\mathcal Q}^{n-1}\to \alpha^n Z^{n-1}$, giving the first subleading terms in $Z$. Following similar steps as before, most of the integrals over $x_i^0$ again lead to $\delta$ functions, $\simeq \delta (q_i^0)$. However, as we are interested in terms with one commutator, say $[j_W,j_0(x_i)]$, the $i^\text{th}$ coordinate only contributes with one Heaviside function, $\theta(-(x_i^0-y^0))$ (since $j_0(x_i)$ does not need to be moved through $j_W$ for the other time-ordering). This leads to the following terms
\begin{equation}
\label{eq:SeffSubleading}
 S_\text{eff}^{(n)}\supset e^{2n} \int_{q_1\dots q_n}  \int_{y}\bar e_L  \slashed{v} \frac{\slashed p+\slashed q_1}{(p+q_1)^2}\dots \slashed{v} \frac{\slashed p+\slashed q_1+\dots +\slashed q_n}{(p+q_1+\dots +q_n)^2}  \gamma_\mu \nu_L  \sum_{l=1}^n\left[\prod_{k\neq l}^n\frac{2\pi \delta(q_k^0)}{q_k^2}\right]\frac{1}{q_l^2}\frac{i}{v\cdot q_l+i\epsilon}\, {\mathcal Q}^{n-1} [j_W^\mu(y),{\mathcal Q}].
\end{equation}
The factor of $\left[v\cdot q_l+i\epsilon\right]^{-1}$ can be written as the sum of a symmetric and antisymmetric term in $v\cdot q_l$. The antisymmetric piece requires an odd number of $q_l^0$ factors in the numerator of the electron line in order to contribute. The denominators of the electron propagators are even as all other $q_i^0$ are set to zero through $\delta$ functions, and we can take $p^0$ to vanish as  well, since $p$ only serves as an IR regulator. One can show that these terms therefore become proportional to the only external vector, $\pp$, and  contribute terms of the form $\pp /|\pp| $, which do not correspond to local counterterms and are therefore not relevant for the RG equations.

Instead, the terms even in $q_l^0$ again lead to a $\delta$ function, $\left[v\cdot q_l+i\epsilon\right]^{-1} = \pi \delta(v\cdot q_l) +(\text{odd \,in} \,v\cdot q_l)$. After these simplifications, all terms in the sum over $l$ in Eq.~\eqref{eq:SeffSubleading} contribute equally.
Using this and the fact that $[j_W^\mu,{\mathcal Q}] = -j_W^\mu$ (due to $\big[\tau^+,\frac{\mathds{1}+\tau_3}{2}\big]=-\tau^+$), one finds
\begin{equation}
 S_\text{eff}^{(n)}\supset e^{2n} \int_{q_1\dots q_n}  \int_{y}\bar e_L  \slashed{v} \frac{\slashed p+\slashed q_1}{(p+q_1)^2}\dots \slashed{v} \frac{\slashed p+\slashed q_1+\dots +\slashed q_n}{(p+q_1+\dots +q_n)^2}  \gamma_\mu \nu_L \left(- \frac{n}{2}\right)\left[\prod_{k}^n\frac{2\pi \delta(q_k^0)}{q_k^2}\right]\, {\mathcal Q}^{n-1} j_W^\mu(y),
\end{equation}
which reproduces the same integrals as those encountered in the previous subsection. 

All in all, this then leads to new entries in Eq.~\eqref{eq:usoftRG}, which are similar to the $c_n\to c_m$  contributions of the previous section, but contribute to $c_{m-1}$ instead, with a relative factor of $-|m-n|/2$. Explicitly, these subleading terms give 
 \begin{equation}\label{eq:usoftRGnlo}
\gamma^{(1)}=  
\left(\begin{array}{*{20}c}
0&&&&\dots&&&&&&0\\
\frac{\alpha^2}{2}&0\\
0 & \frac{\alpha^2}{2} & 0\\
 \frac{\alpha^4}{4} & 0 &\frac{\alpha^2}{2} & 0\\
0& \frac{\alpha^4}{4}&0 &\frac{\alpha^2}{2} & 0&&&&&&\vdots\\
\frac{3\alpha^6}{16}&  0& \frac{\alpha^4}{4} & 0 &\frac{\alpha^2}{2} & 0\\
0&\frac{3\alpha^6}{16}&  0 & \frac{\alpha^4}{4}& 0&\frac{\alpha^2}{2} & 0\\
\frac{5\alpha^8}{32}&0&\frac{3\alpha^6}{16}&0 &\frac{\alpha^4}{4} & 0&\frac{\alpha^2}{2} & 0\\
0&\frac{5\alpha^8}{32}& 0&\frac{3\alpha^6}{16}& 0& \frac{\alpha^4}{4} &0 &\frac{\alpha^2}{2} & 0&\\
\vdots &&&&&&&&&\ddots
\end{array}\right),
 \end{equation}
which should be added to Eq.~\eqref{eq:usoftRG}.

\section{Renormalization group evolution kernels}\label{App:RGE}
We provide here a few more details on the solution of the RG equations for $C_\beta$, $g_V$, and $C_\text{eff}^{(g_V)}$, which resum large logarithms between $\muW$ and $\mue$.

\subsection{\texorpdfstring{$\boldsymbol{C_\beta}$}{} between \texorpdfstring{$\boldsymbol{\muW}$}{} and \texorpdfstring{$\boldsymbol{\muchi}$}{}}

The evolution matrix $U(\muchi,\muW)$ that appears in Eq.~\eqref{eq:gVM} captures the effect of the RG evolution of $C_\beta$, which, in the $\overline{\text{MS}}$ scheme, is given by~\cite{Sirlin:1981ie,Erler:2002mv,Hill:2019xqk}
\begin{align}
\frac{d C_\beta(\mu)}{d\log\mu} &=\left[\frac{\alpha}{\pi} \gamma_0+\left(\frac{\alpha}{\pi} \right)^2\gamma_1+\frac{\alpha}{\pi}\frac{\alpha_s}{4\pi} \gamma_{se}\right]C_\beta(\mu),\notag\\
\gamma_0&=-1, \qquad \gamma_1=\frac{\tilde n}{18} (2a+1), \qquad \gamma_{se}=1, \qquad \tilde n =\sum_f n_f Q_f^2,
\end{align}
where $n_f$ is the number of active fermions and $Q_f$ their charge. Further,  $a$ is a parameter related to the (arbitrary) choice of scheme used to treat evanescent operators, which drops out in observables.  
To NLL, this RG equation is solved by $C_\beta(\mu) = U(\mu,\muW)C_\beta(\muW)$, with 
\begin{align}
U(\mu,\muW)&=\left(\frac{\alpha(\mu)}{\alpha(\muW)}\right)^{-\frac{2\gamma_0}{\beta_0}} \left(\frac{\alpha_s(\mu)}{\alpha_s(\muW)}\right)^{-\frac{2\gamma_{se}}{\beta_{0,s}}\frac{\alpha(\mu)}{4\pi}}\left[1-\frac{2\gamma_1}{\beta_0}\frac{\alpha(\mu)-\alpha(\muW)}{\pi}\right],\notag\\
\beta_0 &= -4/3 \tilde n,\qquad \beta_{0,s} = \frac{11N_c-2n_f}{3}.
\end{align}
Control of terms $\simeq{\mathcal O}(\alpha^2 L)$ in principle requires the two-loop beta function, $\simeq \beta_1$, of $\alpha$. However, it turns out that the dependence on $\beta_1$ cancels in the solution of the RG equation, after expanding $\alpha$ in terms of $\beta_1$~\cite{Cirigliano:2023fnz}. The running couplings in the above expressions should therefore be evaluated using the solutions of the one-loop beta functions
\begin{align}
 \frac{d \alpha (\mu)}{d \log \mu} &= - \frac{\beta_0 (\mu)}{2 \pi}   \alpha^2 (\mu) ,\qquad 
  \frac{d \alpha_s (\mu)}{d \log \mu} = - \frac{\beta_{0,s} (\mu)}{2 \pi}   \alpha_s^2 (\mu),\notag\\
  \frac{1}{\alpha(\mu)}&=   \frac{1}{\alpha(\muW)}+\frac{\beta_0(\mu)}{2\pi}\log \frac{\mu}{\muW}, \qquad 
    \frac{1}{\alpha_s(\mu)}=   \frac{1}{\alpha_s(\muW)}+\frac{\beta_{0,s}(\mu)}{2\pi}\log \frac{\mu}{\muW},
\end{align}
with the boundary conditions $\alpha_s(M_Z)=0.1178$ and $\alpha(M_Z)=1/127.951$ \cite{ParticleDataGroup:2022pth}. 

At low energies, the combination that enters the matching for $g_V$ in Eq.~\eqref{eq:gVM}, can be written as
\begin{equation}\label{eq:Cbeta}
\bar C_\beta(\mu) = \frac{C_\beta(\mu )}{1+\frac{\alpha(\mu)}{\pi }B(a)}=U(\mu,\muW)\frac{C_\beta(\muW )}{1+\frac{\alpha(\mu)}{\pi }B(a)}.
\end{equation}
Although the evolution factor and the Wilson coefficient are separately $a$ dependent, one can show that the above combination is scheme independent by using the matching coefficient,
\begin{equation}
C_\beta(\muW)=1+\frac{\alpha}{\pi}\left[\log \frac{M_Z}{\muW}+B(a)\right],\qquad B(a)=\frac{a}{6}-\frac{3}{4}.
\end{equation}
Putting everything together, one finds $\bar C_\beta (\muchi) = 1.01092$ at $\muchi=\mN$.

\subsection{\texorpdfstring{$\boldsymbol{g_V}$}{} and \texorpdfstring{$\boldsymbol{C_\text{eff}^{(g_V)}}$}{} between \texorpdfstring{$\boldsymbol{\muchi}$}{} and \texorpdfstring{$\boldsymbol{\mue}$}{}}

At $\muchi$, the quark-level operator $O_\beta$ is matched onto the chiral theory, where we work in the $\overline{\text{MS}}_\chi$ scheme \cite{Gasser:1983yg}. This results in the matching of Eq.~\eqref{eq:gVM}, which requires  the non-perturbative input~\cite{Cirigliano:2023fnz,Seng:2018qru,Seng:2018yzq,Czarnecki:2019mwq,Shiells:2020fqp,Hayen:2020cxh,Seng:2020wjq,Cirigliano:2022yyo},
\begin{equation}\label{eq:box}
 \overline \Box^V_\text{had} (\mu_0) = \left[1.030(48)+0.49(11)+0.04(1)\right]\times 10^{-3} +\frac{\alpha}{8\pi}\left(1-\frac{\alpha_s}{\pi}\right)\log\frac{\mu_0^2}{Q_0^2}= 1.38(12)\times 10^{-3},
\end{equation}
for $Q_0^2 = 2\GeV^2$ and $\mu_0=1\GeV$. Here the first, second, and third numbers in square brackets arise from the elastic, Regge, and resonance contributions. Combining Eqs.~\eqref{eq:Cbeta} and \eqref{eq:box}  we obtain the boundary condition for $g_V$ in Eq.~\eqref{eq:gVmN},
\begin{equation}\label{eq:gVchi}
g_V(\mu=\mN) = 1.01153(12).
\end{equation}

In order to evolve $g_V$ to lower scales, we will also need the evolution of the QED coupling $\alpha_\chi$ in the $\overline{\text{MS}}_\chi$ scheme.
The relation between the fine-structure constant in this scheme and in the on-shell scheme, $\alpha^{-1}_\text{OS}= 137.036$, is discussed in detail in App.~A of Ref.~\cite{Cirigliano:2023fnz}.
$\alpha_\chi$ satisfies
\begin{align}
    \mu \frac{d \alpha_\chi (\mu)}{d \mu} &= - \frac{\beta_0 (\mu)}{2 \pi}   \alpha_\chi^2 (\mu) + \mathcal{O}(\alpha_\chi^3) ,\notag\\  
    \beta_0 (\mu) &=  - \frac{4}{3} \tilde n_\ell (\mu) -\frac{1}{3} \tilde n_\pi (\mu) , \qquad\tilde{n}_{\ell,\pi} (\mu) = \sum_{\ell,\pi} Q_{\ell,\pi}^2 n_{\ell,\pi}  \, \theta(\mu - m_{\ell,\pi}),
\end{align}
with $n_\ell = 1$, $Q_\ell =-1$ for leptons and $n_\pi = 1$, $Q_\pi = 1$ for pions. 
The matching relation at a given scale $\muchi$ is \cite{Cirigliano:2023fnz} 
\begin{equation}
    \frac{1}{\alpha_\chi(\muchi)}  = \frac{1}{\alpha_\text{OS}} + \frac{1}{3\pi} \sum_{\ell = e, \mu} \left(1 + \log \frac{m_\ell^2}{\mu^2_\chi}\right) \theta(\muchi - m_\ell) + \frac{1}{12 \pi} \left(1 + \log \frac{\mpi^2}{\muchi^2}\right) \theta(\muchi - \mpi).
\end{equation}
This formula accounts for the electron, muon, and pion thresholds.
The running of $\alpha_\chi$ between thresholds is then given by
\begin{equation}
    \frac{1}{\alpha_\chi(\mu)} = \frac{1}{\alpha_\chi(\muchi)} + \frac{\beta_0(\mu)}{2\pi} \log\frac{\mu}{\muchi}.
\end{equation}
In the following, and in the main text of the manuscript, we drop the subscript $\chi$ and $\alpha$ is always understood to be given in this scheme.
The RG kernel in Eq.~\eqref{eq:usoftU},
for the evolution of $C_\text{eff}^{(g_V)}$ between $\mupi$ and $\mue$,
is given by
\begin{equation}
\label{eq:kernel}
    U^{(g_V)}(\mue,\mupi) = \exp\left[- \frac{2}{\beta_0} \left( \tilde{\gamma}_0 \log r 
    + \tilde{\gamma}_1 \frac{\alpha(\mupi)}{\pi} (r - 1) + \pi \Big[  u\big(\alpha(\mue),Z\big) - u\big(\alpha(\mupi),Z\big) \Big]
    \right) \right],
\end{equation}
with
\begin{equation}
 \tilde{\gamma}_0 = -\frac{3}{4}, \qquad\tilde{\gamma}_1 = \frac{5}{24} \tilde{n} + \frac{5}{32} - \frac{\pi^2}{6}, \qquad    r = \frac{\alpha(\mue)}{\alpha(\mupi)},
\end{equation}
and
\begin{equation}
    u(\alpha,Z) =\frac{1}{\alpha} \left( 1 - \sqrt{1 - Z(1+Z) \alpha^2} \right)  - \sqrt{Z (1+Z)} \arcsin\big(\sqrt{Z (1+Z)} \alpha\big). 
\end{equation}
The running of $g_V$ between $\muchi$ and $\mupi$ is accomplished by a very similar kernel~\cite{Cirigliano:2023fnz}, without the $u$ functions, i.e.,
\begin{equation}
\label{eq:kernel2}
    \tilde{U}(\mupi,\muchi) = \exp\left[- \frac{2}{\beta_0} \left( \tilde\gamma_0 \log r_\pi 
    + \tilde\gamma_1 \frac{\alpha(\muchi)}{\pi} (r_\pi - 1) \right) \right],\qquad 
    r_\pi = \frac{\alpha(\mupi)}{\alpha(\muchi)}.
\end{equation}
Note that control of terms $\simeq{\mathcal O}(\alpha^2 L)$ would again require the two-loop beta function, $\simeq \beta_1$ of $\alpha$. Similarly to the $C_\beta$ case, however, this dependence drops out when expanding $\alpha$ in $\beta_1$,  justifying the use of the one-loop solution for $\alpha$ in the above equations.

\section{Toy model for the dispersive approach}
\label{app:toy}

Restoring the $i\eps$ prescriptions, the loop integral for our toy example becomes 
\beq
\Box_{\gamma W}^\text{toy}=\frac{-ig_A g_M}{M_\text{F}^{(0)}}\frac{ M}{\mN}\frac{\alpha}{\pi}\int\frac{\diff \nu}{2\pi}\int_{-1}^1\diff z\int_0^\Lambda \diff|\qq|\,\qq^2\frac{|\qq|(|\qq|-\nu z)}{[(p_e-q)^2+i\eps][q^2+i\eps][s-\bar M^2+i\eps ]},
\label{box_toy}
\eeq
with poles at
\begin{align}
\label{nu_i}
 \nu_\pm^{(1)}&=\pm |\qq|\mp i\eps,\notag\\
 \nu_\pm^{(2)}&=E_e\pm \sqrt{E_e^2+\qq^2-2E_e |\qq|z}\mp i\eps,\notag\\
 \nu_\pm^{(3)}&=-M\pm \sqrt{M^2+\qq^2-2M\Delta}\mp i\eps.
\end{align}
In writing Eq.~\eqref{box_toy}, we have  
set $\mW^2/(Q^2 + \mW^2) \to 1$ and regulated the UV divergence by a momentum cutoff $\Lambda$, which also makes the power divergences at intermediate steps visible. That is, the individual residue contributions for the three poles in Eq.~\eqref{nu_i} (upper plane) are
\begin{align}
 \Box_{\gamma W}^{(1)}&=\frac{g_A g_M}{M_\text{F}^{(0)}}\frac{M}{\mN}\frac{\alpha}{\pi}\frac{1}{8E_eM}\bigg(\Lambda^2+2\Lambda\Delta+2\Delta^2\log\frac{\Lambda}{\Delta}\bigg),\notag\\
 \Box_{\gamma W}^{(2)}&=\frac{g_A g_M}{M_\text{F}^{(0)}}\frac{M}{\mN}\frac{\alpha}{\pi}\bigg[-\frac{1}{8E_eM}\bigg(\Lambda^2+2\Lambda\Delta+2\Delta^2\log\frac{\Lambda}{\Delta}\bigg)+\frac{\Lambda^2}{8M^2}-\frac{(M-\Delta)\Lambda}{2M^2}+\frac{(M-\Delta)\Delta}{4M^2}\bigg(1-3\log\frac{\Lambda}{\Delta}\bigg)\bigg],\notag\\
 \Box_{\gamma W}^{(3)}&=\frac{g_A g_M}{M_\text{F}^{(0)}}\frac{M}{\mN}\frac{\alpha}{\pi}\bigg[-\frac{\Lambda^2}{8M^2}+\frac{(M-\Delta)\Lambda}{2M^2}-\frac{M^2-2M\Delta-4\Delta^2}{16M^2}\notag\\
 &+\frac{3}{16M^2}\bigg((M-2\Delta)^2\log\Big[1-\frac{2\Delta}{M}\Big]-2\big(M^2-2M\Delta+2\Delta^2\big)\log\frac{2\Lambda}{M}\bigg)\bigg],
\end{align}
leading to the sum
\begin{align}
\Box_{\gamma W}^{\text{toy}}&=-\frac{3g_A g_M}{16M_\text{F}^{(0)}M\mN}\frac{\alpha}{\pi}\bigg[2M^2\log\frac{2\Lambda}{M}+\frac{M}{3}(M-6\Delta)
-(M-2\Delta)^2\log\Big[1-\frac{2\Delta}{M}\Big]-4\Delta(M-\Delta)\log\frac{2\Delta}{M}\bigg]\notag\\
&=-\frac{g_A g_M}{16M_\text{F}^{(0)}}\frac{\alpha}{\pi}\frac{M}{\mN}\bigg(1+6\log\frac{2\Lambda}{M}\bigg)+\frac{3g_A g_M}{4}\frac{\alpha}{\pi}\frac{\Delta}{\mN}\log\frac{2\Delta}{M}+\Order\big(\Delta^2\big),
\label{Box_toy_app}
\end{align}
in which the power divergences in $\Lambda$ and the singularities in $1/E_e$ cancel. 

In the dispersive approach, there are two contributions that impede a straightforward Wick rotation, when $\nu^{(2)}_-$ moves into the first quadrant and when $\nu^{(3)}_+$ moves into the third. The former residue contribution vanishes for $E_e\to 0$, but the latter gives rise to
\begin{align}
\label{toy_res_app}
\Box_{\gamma W}^{\text{toy, res}}&=\frac{g_A g_M}{4M_\text{F}^{(0)}M\mN}\frac{\alpha}{\pi}\bigg[\frac{(5M-6\Delta)\sqrt{M\Delta}}{\sqrt{2}} -\frac{3}{4}(M-2\Delta)^2\log\frac{(\sqrt{M}-\sqrt{2\Delta})^2}{M-2\Delta}\bigg]\notag\\
&=\frac{g_A g_M}{M_\text{F}^{(0)}}\sqrt{\frac{M}{\mN}}\frac{\alpha}{\pi}\sqrt{\frac{2\Delta}{\mN}}+\Order\big(\Delta^{3/2}\big),
\end{align}
and therefore displays a scaling that differs from Eq.~\eqref{Box_toy_app}. However, to obtain the full result, Eq.~\eqref{toy_res} needs to be subtracted from the Wick-rotated integral, which can be brought into the following form
\beq
\label{toy_Wick_app}
\Box_{\gamma W}^\text{toy, Wick}=-\frac{2g_A g_M}{M_\text{F}^{(0)}}\frac{M}{\mN}\frac{\alpha}{\pi}\int\frac{\diff \nu}{2\pi}\int_0^\Lambda \diff|\qq|\,\qq^2\frac{\qq^2}{(\nu^2+\qq^2)^2[\nu^2+\qq^2-2M(\Delta + i\nu)]}.
\eeq
From the form of the denominator, one sees that the scale $\qq^2=2M\Delta$, which determines the maximum momentum up to which  $\nu^{(3)}_+$ lies in the third quadrant, again plays a role in the evaluation of the integral, and indeed the dependence on $\sqrt{\Delta}$ drops out in the difference of Eqs.~\eqref{toy_Wick_app} and~\eqref{toy_res_app}. By explicit evaluation one can show that $\Box_{\gamma W}^\text{toy}=\Box_{\gamma W}^\text{toy, Wick}-\Box_{\gamma W}^\text{toy, res}$.

\section{Corrections to the phase space}
\label{app:Cfactor}

We describe here the corrections to the phase space $\tilde C$ that enter the differential rate in Eq.~\eqref{eq:dGamma} and the half-life in Eq.~\eqref{eq:master2},
as well as the interplay between the EFT formulation and the corrections included in the $\beta$ decay literature, 
summarized in Refs.~\cite{Hardy:2004id,Hardy:2008gy,Hayen:2017pwg}.
In the standard framework, the most important corrections to the Fermi function arise from deviations of the nuclear charge distribution from a point charge, captured in the factors $L_0(Z,E_e)$ and $U(Z,E_e)$, from the momentum dependence of the weak form factor, $C(Z,E_e)$, and from atomic effects $S(Z,E_e)$ and $r(Z,E_e)$. Some of these corrections depend on nuclear parameters, such as the radius of the nuclear charge distribution or of the weak form factor. In an EFT approach, this dependence is reproduced by matrix element of one-, two- or higher-body operators, so that some pieces of the standard phase-space corrections need to be subtracted in order to avoid double counting. We first introduce the relevant correction factors, after which we combine them and discuss the issue of double counting.

\subsection{Atomic screening \texorpdfstring{$\boldsymbol{S}$}{}}
We start from atomic corrections, which are identical to the standard approach. To calculate the half-life in Sec.~\ref{sec:example}  we use the expressions from 
Refs.~\cite{Hardy:2004id,Hardy:2008gy}.
For the screening factor $S$, we have
\begin{equation}
    S(Z,E_e) = \frac{\tilde p\,  \tilde E_e \, F(Z,\tilde E_e)}{p\,  E_e\, F(Z,E_e)}.
\end{equation}
Here $F$ is the standard Fermi function
\begin{equation}\label{eq:Fermistandard}
    F(Z,E_e) =  \frac{2(1+\eta)}{\Gamma(2\eta+1)^2}|\Gamma(\eta+iy)|^2 e^{\pi y}\times \left(2 |\pp_e|R\right)^{2(\eta-1)},
\end{equation}
where $\eta = \sqrt{1-\alpha^2 Z^2}$ and $y =\mp Z\alpha/\beta$, while $\tilde E_e$ and $\tilde p$ are given by 
\begin{equation}
    \tilde E_e = E_e  - m_e V_0, \qquad \tilde p = (\tilde E_e^2-m_e^2)^{1/2},
\end{equation}
with $V_0$ 
\begin{equation}
    V_0 = \mp N(Z+1) \alpha^2  (Z+1)^{\frac{4}{3}},
\end{equation}
for positron and electron emission, respectively. $N(Z+1)$ is a slowly varying function of the charge of the parent nucleus, $N(8) = 1.42$, see Ref.~\cite{Hardy:2004id}. 

\subsection{Atomic overlap \texorpdfstring{$\boldsymbol{r}$}{}}
The factor $r$ takes into account the mismatch between the atomic states before and after $\beta$ decay. It was first considered in Ref.~\cite{Hardy:2008gy}, and it is given by
\begin{equation}
    r(Z,E_e) = 1 - \frac{1}{E_0 - E_e} \frac{\partial^2}{\partial Z^2} B(G),
\end{equation}
with
\begin{equation}
    B(G) = 13.080 (Z+1)^{2.42} \eV,
\end{equation}
for $5\leq Z\leq 9$.

\subsection{Finite-size correction \texorpdfstring{$\boldsymbol{L_0}$}{}}
$L_0$ encodes the effects of the nuclear charge distribution on the motion of the electron/positron emerging from the $\beta$ decay. References~\cite{Hardy:2004id,Hardy:2008gy,Hardy:2020qwl} computed this correction  numerically, by solving the Dirac equation with different nuclear charge distributions. Here we follow Ref.~\cite{Hayen:2017pwg}, which provides analytical expressions from which it is easier to identify possible double counting with the ab-initio approach. For $^{14}$O, we checked that the phase space $f$ obtained with the expressions from Ref.~\cite{Hayen:2017pwg}
agrees with Ref.~\cite{Hardy:2020qwl} within uncertainties. 
$L_0$ is given by~\cite{Hayen:2017pwg}
\begin{align}\label{eq:L0}
   L_0(Z,E_e) &= 1 + \frac{13}{60} (\alpha Z)^2 \pm 
  \alpha Z R E_e \left( 
 \frac{ (41-26 \gamma)}{15 (2\gamma -1)}  + 
 \frac{  \gamma (17 -2 \gamma)}{30  (2\gamma -1)} \frac{m_e^2}{E_e^2} \right) + a_{-1} \frac{m_e^2 R}{E_e} + \sum_{n = 0}^5 a_n (E_e R)^n \nonumber \\ &  + A ( m_e R - 0.0164) (\alpha Z)^{4.5}, 
\end{align}
with $A=0.41$ for electrons, $A=0.22$ for positrons. The coefficients $a_n$ have an expansion in $\alpha$
\begin{equation}
    a_n = \sum_{x=1}^6 b_{x, n} (\alpha Z)^x.
\end{equation}
The coefficients are tabulated in Tables I and II of Ref.~\cite{Hayen:2017pwg}.

\subsection{Shape factor \texorpdfstring{$\boldsymbol{C_0}$}{}}
A useful analytical expression for the shape factor $C$ is given by~\cite{Hayen:2017pwg}
\begin{align}\label{eq:C0}
    C(Z,E_e) = 1 + (E_0 R)^2 \left( - \frac{1}{5}   + 
    \frac{4}{15}  \frac{E_e}{E_0} + \frac{2}{15} \frac{m_e^2}{E_e E_0} - \frac{4}{15} \frac{E^2_e}{E_0^2} \right)
+ \alpha Z R \left( \pm\frac{6}{35} E_0 \pm \frac{13}{35} E_e  \mp \frac{1}{70} \frac{m_e^2}{E_e}\right)
   - \frac{233}{630} (\alpha Z)^2,
\end{align}
where we neglected an $\mathcal O (m_e^2 R^2)$ term, whose effect is numerically small since $E_0\gg m_e$~\cite{Seng:2022inj}. 
There are corrections to this approximate form, which, although small for light nuclei, can become relevant for heavier nuclei. For completeness we therefore include a more general expression, which can be separated into an isoscalar and isovector component. The former takes the form
\begin{align}\label{}
    C(Z,E_e)_0& = 1 + (E_0 R)^2F1110 \left( - \frac{1}{3}   + 
    \frac{4}{9}  \frac{E_e}{E_0} + \frac{2}{9} \frac{m_e^2}{E_e E_0} - \frac{4}{9} \frac{E^2_e}{E_0^2} \right)\notag\\
&+ \alpha Z R \left( \pm \frac{2}{9}E_0 F1111  \pm \frac{2}{3} E_e(F1221-F1111/3)  \pm \frac{F1211}{3} \frac{m_e^2}{E_e}\right)
-\frac{F1222}{3} (\alpha Z)^2,
\end{align}
with 
\begin{align}
F1111 &= 0.757+0.0069(1-e^{-A/1.008}),\qquad F1110=3/5,\notag\\
F1221 &= 0.844-0.0182(1-e^{-A/1.974}),\qquad F1222 = 1.219-0.0640(1-e^{-A/1.550}),
\end{align}
and $A$ a fit parameter related to the assumed charge distribution, ranging from $1.67$ for $^{14}$O to $3.00$ for $^{54}$Co \cite{Wilkinson:1993hx}. This should then be combined with the isovector correction,
\begin{equation}
C(Z,E)_I = 1-\frac{8}{5}\frac{w\xi R^2}{5A'+2},\qquad \xi = \frac{1}{6}\left[(E_0-E)^2+(E_e+V_0)^2-m_e^2\right],\qquad V_0 = \mp3 \alpha Z/(2 R),
\end{equation}
with $A'$ another fit parameter and $w$ a fraction that depends on the shell of the last nucleon, both of which are listed in Table 8 of Ref.~\cite{Wilkinson:1993hx}. As the difference between $C_0 C_I$ and Eq.~\eqref{eq:C0} is small for $^{14}$O, we use the simpler expression of Eq.~\eqref{eq:C0} in our numerical analysis for $^{14}$O, but note that the difference becomes sizable for larger nuclei.

\subsection{Nuclear recoil \texorpdfstring{$\boldsymbol{R}$}{}}
Finally, the correction due to recoil effects is given by
\begin{equation}
R(E_0) = 1-\frac{3E_0}{2M_A},
\end{equation}
with $M_A$ the mass of the nucleus.

\subsection{Combination and comparison to the EFT approach}

In an ab-initio set-up, the nuclear charge distribution 
and the weak form factor emerge 
from calculations with nucleon degrees of freedom. The $\mathcal O(\alpha^0)$ term in $C$ 
is captured by the momentum dependence of the LO weak form factor. Having an ab-initio calculation of the form factor, we can replace $R$ in the $\mathcal O(\alpha^0)$ terms in Eq.~\eqref{eq:C0} by the weak radius $R_W$ \cite{Seng:2022epj}, given by 
\begin{equation}
    R^2_W =  \frac{5}{3} \langle r^2_W \rangle,
\end{equation}
where $\langle r^2_W \rangle$ is defined in analogy to the charge form factor as
\begin{equation}
    M_\text{F}(\qq^2) = M_\text{F}(0) \left(1 - \langle r^2_W \rangle \frac{\qq^2}{6} + \ldots\right).
\end{equation}
The $\mathcal O(\alpha Z R)$ terms in Eqs.~\eqref{eq:L0} and \eqref{eq:C0}
are captured by 
matrix elements of the energy-dependent potentials  $\mathcal V_E$ and $\mathcal V_E^\pi$,
while terms of $\mathcal O(\alpha^2 Z^2)$
are captured in the matching coefficients $C_\delta$ and 
$C_{\delta}^\text{3b}$, and by the matrix elements of the potentials $\mathcal V_+$
and $\mathcal V_+^\text{3b}$. 
To avoid double counting, we therefore  define 
\begin{align}
    L^\text{sub}_0(Z,E_e) &= L_0(Z,E_e) \mp 
        \alpha Z R E_e \left(1 + \frac{m_e^2}{2 E_e}\right) -  \frac{13}{60} (\alpha Z)^2 - \alpha Z \left ( b_{1,-1} \frac{m_e^2 R}{E_e} + b_{1,0} + b_{1,1} E_e R \right) - b_{2,0} (\alpha Z)^2, \label{eq:L0sub} \\
    C^\text{sub}(Z,E_e) & =  1 + (E_0 R_W)^2 \left( - \frac{1}{5}   + 
    \frac{4}{15}  \frac{E_e}{E_0} + \frac{2}{15} \frac{m_e^2}{E_e E_0} - \frac{4}{15} \frac{E^2_e}{E_0^2} \right). \label{eq:C0sub}
\end{align}
$L^\text{sub}_0$ contains terms of 
$\mathcal O(\alpha^3)$, 
$\mathcal O(\alpha^2 \epsilon_{\slashed{\pi}})$, 
or higher. These are beyond the accuracy of our EFT calculation, and could be reproduced in the EFT by deriving two- and higher-body transition operators at higher order in $\epsilon_{\slashed{\pi}}$, and by calculating ultrasoft matrix elements at the same order. For $^{14}$O, the correction from $L^\text{sub}_0(Z,E_e)$
amounts to about an $\mathcal O(10^{-4})$ shift to $\bar{f}$, and gives us a sense of the size of subleading corrections. The shift is smaller than the effect of the scale variation, which we take as the theoretical error on $\bar f$. 
Concerning the $U$ correction, which takes into account deviations from a uniform charge distribution, we use the expression in Eq.~(29) of Ref.~\cite{Hayen:2017pwg}. 
In conclusion, our definition of $\tilde C$ is given by
\begin{align}
    \tilde C(E_e) = C^\text{sub}(Z,E_e) L_0^\text{sub}(Z,E_e) U(Z,E_e) S(Z,E_e) r(Z,E_e).
\end{align}

\begin{figure*}
    \centering
    \includegraphics[width=0.6\textwidth]{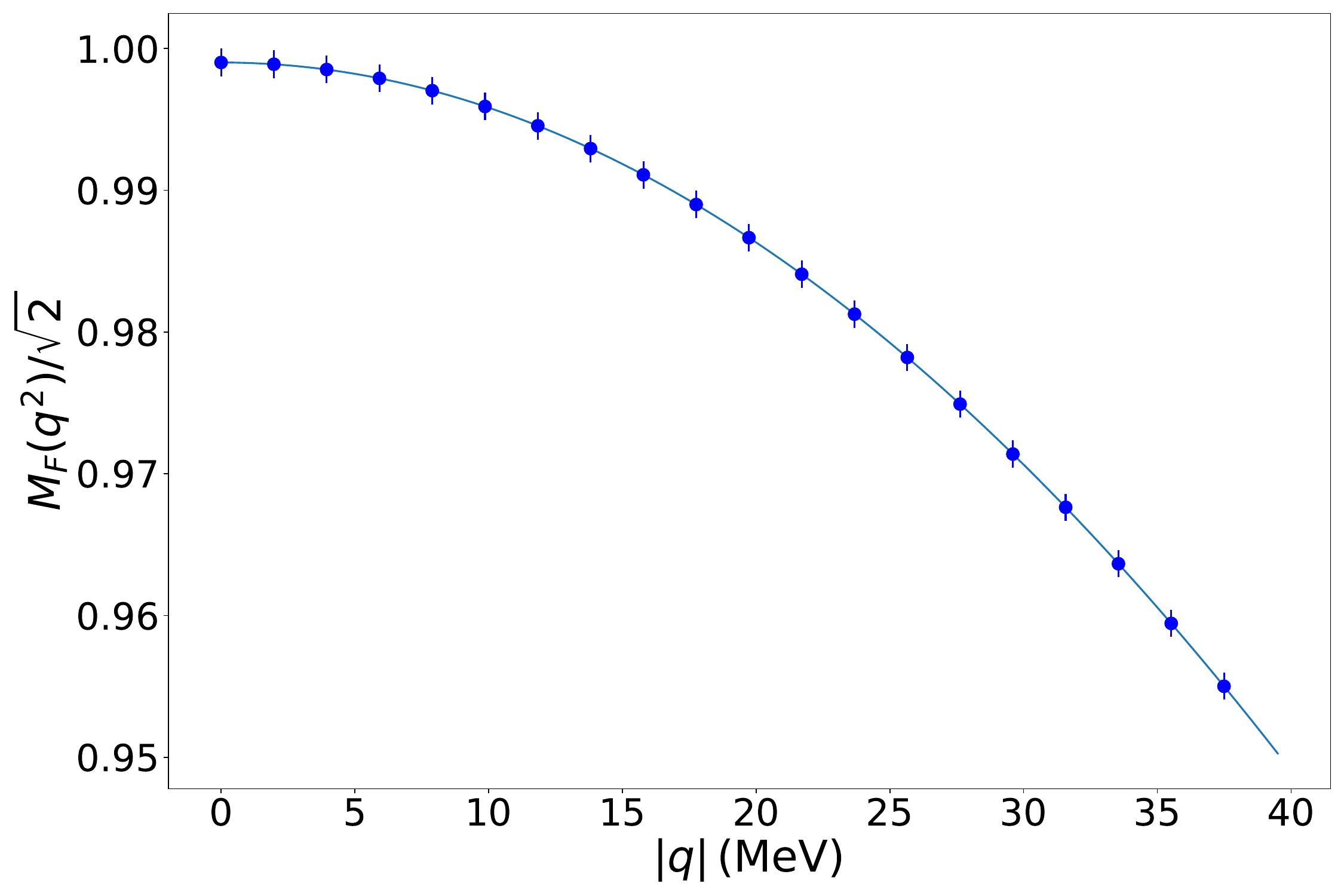}
    \caption{VMC calculation of the weak form factor for $^{14}\text{O}\rightarrow{}^{14}\text{N}$. The blue dots show the VMC calculation, with the error bar denoting the $2\sigma$ statistical error. The line is obtained by a fit to a polynomial function of $\qq^2$, including terms up to $\qq^4$.}
    \label{fig:VMC_FF}
\end{figure*}

Figure~\ref{fig:VMC_FF} shows a VMC calculation of the weak form factor for the transition $^{14}\text{O}\rightarrow{}^{14}\text{N}$, using the same chiral interaction  discussed in Sec.~\ref{sec:light_nuclei}.
By fitting the VMC results to the functional form
\begin{equation}
    M_\text{F}(\qq^2)  = M_\text{F}(0) \left(  1 -  \langle r^2_W \rangle \frac{\qq^2}{6} +  \langle r^4_W \rangle \frac{\qq^4}{5!}  \right),
\end{equation}
we obtain $\sqrt{\langle r^2_W \rangle} = 2.73(4)\fm$,   
where the error reflects only the statistical error of the VMC data points. 
Since $\sqrt{\langle r^2_W \rangle}$
differs from the charge radius of $^{14}$N by only about 10\%, in the numerical evaluations in Sec.~\ref{sec:example} we will keep using $R$ rather than $R_W$, as done in most of the superallowed-$\beta$-decay literature. For $^{14}$O, the difference amounts to a $\simeq 10^{-5}$ shift in the half-life $t$, much smaller than other theoretical uncertainties. For future refined studies and further cross checks, one could instead consider the weak radii, as at least for some nuclei they can be compared to experiment~\cite{Seng:2022inj}, while
the weak form factor is a prediction of the nuclear-structure calculations with which the wave functions are determined.

\end{widetext}

\bibliography{residue}

\end{document}